\documentclass[a4paper]{article}
\usepackage[margin=0.70in]{geometry}
\usepackage{amsmath,amssymb,empheq}
\usepackage{amsfonts} %Voor mathbb
\usepackage{graphicx}
\usepackage{parskip} %Zorgt dat er een enter is na elke paragraaf
\usepackage{subcaption} %Nodig voor subfigure
\usepackage{multirow}	%nodig om wat gesofisticeerdere tabellen te maken
\usepackage{float} %Nodig om [H] te kunnen gebruiken bij een figuur (zodat het op de juiste plek komt)
\usepackage{siunitx}%nuttig voor eenheden, vooral voor de mu in micrometer: \SI{10}{\micro\meter}
\usepackage{comment}
\usepackage{todonotes}%voor \todo{} en \todo[inline]{}
\usepackage{mathtools}

\usepackage{multirow}

\usepackage{pgfplotstable}
\usepackage{pgfplots}

\usepackage{algorithm}
\usepackage{algorithmic}

\usepackage{xcolor}

\usepackage{authblk}

\usepackage{adjustbox}

\usepackage{url}
\title{Predicting the statistical error of analog particle tracing Monte Carlo}

\author[1]{V. Maes}
\author[1]{I. Bossuyt}
\author[2]{H. Vandecasteele}
\author[3]{W. Dekeyser}
\author[4,5]{J. Koellermeier}
\author[3]{M. Baelmans}
\author[1]{G. Samaey}
\affil[1]{Department of Computer Science, KU Leuven, Belgium}
\affil[2]{Department of Chemical and Biomolecular Engineering, Johns Hopkins University, USA}
\affil[3]{Department of Mechanical Engineering, KU Leuven, Belgium}
\affil[4]{Bernoulli Institute for Mathematics, Computer Science and Artificial Intelligence, University of Groningen, Netherlands
}
\affil[5]{Groningen Cognitive Systems and Materials Center, University of Groningen, Netherlands
}
\date{}

\def \reponame {a public repository~\cite{zenodo}}

\begin{document}
%\pagenumbering{roman}
%\tableofcontents
%\newpage
\pagenumbering{arabic}
\setcounter{page}{1}
\maketitle

\section*{Abstract}
Large particle systems are often described by high-dimensional (linear) kinetic equations that are simulated using Monte Carlo methods for which the asymptotic convergence rate is independent of the dimensionality. Even though the asymptotic convergence rate is known, predicting the actual value of the statistical error remains a challenging problem. In this paper, we show how the statistical error of an analog particle tracing Monte Carlo method can be calculated (expensive) and predicted a priori (cheap) when estimating quantities of interest (QoI) on a histogram. We consider two types of QoI estimators: point estimators for which each particle provides one independent contribution to the QoI estimates, and analog estimators for which each particle provides multiple correlated contributions to the QoI estimates. The developed statistical error predictors can be applied to other QoI estimators and nonanalog simulation routines as well. The error analysis is based on interpreting the number of particle visits to a histogram bin as the result of a (correlated) binomial experiment. The resulting expressions can be used to optimize (non)analog particle tracing Monte Carlo methods and hybrid simulation methods involving a Monte Carlo component, as well as to select an optimal particle tracing Monte Carlo method from several available options. Additionally, the cheap statistical error predictors can be used to determine a priori the number of particles $N$ that is needed to reach a desired accuracy. We illustrate the theory using a linear kinetic equation describing neutral particles in the plasma edge of a fusion device and show numerical results. The code used to perform the numerical experiments is openly available.

Keywords: kinetic equation, variance prediction, statistical error, binomial experiment, correlated samples

\section{Introduction}
\label{sec:introduction}

The mesoscopic behaviour of large particle systems can be described by kinetic equations that have a high-dimensional phase space. A common approach to deal with the high-dimensionality of the problem is to employ Monte Carlo methods, consisting of a simulation and estimation routine, for which the statistical error converges with a rate that is independent of the dimensionality. For linear kinetic equations with a linear collision operator, modelling for instance collisions of the particles under consideration with a background medium, particle tracing Monte Carlo methods~\cite{spanier_monte_1969, mortier_advanced_2020, lapeyre_introduction_2003} allow for unbiased simulation and estimation of quantities of interest (QoI). Such particle tracing Monte Carlo methods do, however, introduce a statistical error (noise), which depends on the number of particles $N \in \mathbb{N}$ used in the simulation. Because of the linearity of the kinetic equation, the $N$ particles can be simulated independently, leading to the well-known result that the statistical error is inversely proportional to the square root of $N$~\cite{caflisch_monte_1998}. The precise value of the statistical error depends on the kinetic equation under consideration as well as on the choice of simulation and estimation method. For the illustration in this paper, we use the following linear kinetic equation:
\begin{equation}
\underbrace{\vphantom{\int}\partial_t f(x,v,t)}_{\substack{\text{transient}}} + \underbrace{\vphantom{\int} v \cdot \partial_x f(x,v,t)}_{\substack{\text{transport}}} = \underbrace{\vphantom{\int} S(x,v,t)}_{\substack{\text{particle source}}} - \underbrace{\vphantom{\int} R_i(x,t) f(x,v,t)}_{\substack{\text{ionization sink}}} +\underbrace{R_{cx}(x,t) \left( M(v \mid x,t) \int f(x,v',t)dv' - f(x,v,t) \right)}_{\substack{\text{charge-exchange collision operator}}},
\label{eq:kinetic_equation}
\end{equation}
which governs the neutral particle behaviour in the plasma edge of a fusion device~\cite{reiter_eirene_2005, horsten_fluid_2019, mortier_advanced_2020}. The particle velocity distribution $f(x,v,t)$ represents the density of neutral particles at time $t \geq t_0 = 0$ with a given position $x \in D \subset \mathbb{R}^d$ and velocity $v \in \mathbb{R}^d$; $S(x,v,t)$ is the particle source describing how many neutrals are created at a position $x$ with velocity $v$ per unit of time; the ionization rate $R_i(x,t)$ and charge-exchange rate $R_{cx}(x,t)$ are known positive functions that depend on the background plasma state; the sum of the two collision rates is called the total collision rate $R$:
\begin{equation}
R(x,t) = R_i(x,t) + R_{cx}(x,t);
\end{equation} 
$M(v \mid x,t)$ is the post-collisional velocity distribution describing the probability of a neutral acquiring velocity $v$ after a charge-exchange collision with a plasma particle. We here assume $M(v \mid x,t)$ to be a Maxwellian:
\begin{equation}
M(v \mid x,t) = \frac{1}{(2 \pi \sigma_p^2(x,t))^{d/2}} \exp \left(-\frac{1}{2}\frac{|v-u_p(x,t)|^2}{\sigma_p^2(x,t)} \right).
\label{eq:Maxwellian}
\end{equation}
The mean velocity $u_p(x,t)$ and variance $\sigma_p^2(x,t)$ (which we assume to be isotropic, i.e., a scalar) of the post-collisional velocity distribution are known from the background plasma state, rendering the kinetic equation~\eqref{eq:kinetic_equation} linear. The QoI $Q(x,t)$ that have to be estimated using Monte Carlo estimation routines are typically moments of the particle velocity distribution $f(x,v,t)$, i.e., integrals over velocity space of the following form:
\begin{equation}
Q(x,t) = \int q(v)f(x,v,t)dv,
\label{eq:QoI_Generic}
\end{equation} 
where $q(v)$ is a function of velocity. We focus on the neutral particle density $\rho(x,t)$, momentum density $m(x,t)$, and energy density $E(x,t)$, the three lowest order moments of the particle velocity distribution, which are defined as follows:
\begin{equation}
\rho(x,t) = \int f(x,v,t)dv,\quad m(x,t) = \int v f(x,v,t)dv,\quad 
E(x,t) = \int \frac{|v|^2}{2}f(x,v,t)dv.
\label{eq:QoIs}
\end{equation}
Using particle tracing Monte Carlo methods, we can obtain unbiased estimates $\hat{Q}(x,t)$ that approximate the exact QoI $Q(x,t)$. In this work, we search for expressions that predict a priori the statistical error on these estimates $\hat{Q}(x,t)$ as a function of two types of parameters: (1) model parameters such as the collision rates $R_i(x,t)$, $R_{cx}(x,t)$ and the post-collisional velocity distribution $M(v \mid x,t)$; and (2) the number of particles $N$ without the need for asymptotic arguments that only hold in the limit for $N \rightarrow \infty$. As error measure, we consider the mean squared error, i.e., the variance on the QoI estimates (this actually represents the statistical error squared and therefore has an asymptotic convergence rate of $\frac{1}{N}$~\cite{caflisch_monte_1998}). Predicting the variance allows to determine the number of particles $N$ needed to achieve a desired accuracy a priori and gives a good indication of the computational cost of a Monte Carlo simulation. Note that there might be multiple particle sources (so-called strata) in a simulation that each require a different number of particles~\cite{reiter_time_1995}. Additionally, the variance predictors can be used to select the best simulation and estimation combinations for a given problem~\cite{macmillan_comparison_1966, lux_systematic_1978, indira_analytical_1988, mortier_advanced_2020} and to optimize particle tracing Monte Carlo methods~\cite{sarkar_prediction_1979, indira_optimization_1989} and hybrid simulation methods, e.g., domain decomposition methods~\cite{boyd_hybrid_2011,valentinuzzi_two-phases_2019} that encompass a Monte Carlo solver. Finally, there are some other applications such as in multilevel Monte Carlo~\cite{giles_multilevel_2015} where the variance predictors can be used to determine the required number of particles per level, and in Bayesian inversion settings~\cite{law_data_2015} where the uncertainty of the forward Monte Carlo simulations has to be taken into account~\cite{bal_bayesian_2013}.
%, and in the Monte Carlo simulation driven generation of high-fidelity data for training surrogate models~\cite{miles_radiation_2021,ThesisWillem}.
% Meerdere strata ander aantal deeltjes --> vooral in transient: deeltjes die later in de simulatie terecht komen zijn minder belangrijk dus daar kan je wrs minder deeltjes per massa nemen om eenzelfde fout te krijgen

In the literature, there is some work on analytical variance calculations, based on invariant imbedding~\cite{spanier_analytic_1970, bellman_introduction_1975, mortier_advanced_2020, mortier_accepted_2023,mortier_technical_2020} and on the construction of so-called history-score moment equations~\cite{amster_prediction_1976, lux_systematic_1978, sarkar_prediction_1979, solomon_verification_2011}. These analytical calculations build on derivations that only seem tractable in simple (academic) test cases, e.g., one-group equations in infinite domains. In these simple cases, closed form expressions for the variance can be obtained that allow for selection and optimization of particle tracing Monte Carlo methods. However, in more complex (realistic) cases such as equation~\eqref{eq:kinetic_equation}, the expressions are more involved such that they can only be solved numerically. In those cases, and especially for higher dimensional problems, the numerical variance calculations rapidly become impractical, which renders these methods unattractive for a priori error assessment.

Most of the remaining literature focuses on data-driven on-the-fly variance estimation, typically based on the assumption that the number of particles $N$ is high enough such that asymptotic arguments hold~\cite{talay_stochastic_2013} (this is not always easy to verify). Then the following asymptotic expression for the variance on a QoI is valid:
\begin{equation}
\mathbb{V}[\hat{Q}] = \frac{C}{N},
\label{eq:asymptotic_variance_expression}
\end{equation}
where estimation of the factor $C \geq 0$ leads to a quantitative variance expression~\cite{koehler_assessment_2009, ghoos_accuracy_2016}. Empirical simulation driven variance estimation along these lines is also performed for DSMC methods~\cite{chen_statistical_1996, myong_review_2019}. Recently, error analysis has also been performed for kernel density estimators for particle-in-cell methods~\cite{evstatiev_noise_2021}. 

When sampling stationary stochastic processes, where consecutive samples are correlated, the correlation can be taken into account in the on-the-fly variance estimation by replacing $N$ in~\eqref{eq:asymptotic_variance_expression} by an effective number of independent samples $N_{\text{eff}} \leq N$ based on an estimate of the integrated autocorrelation time~\cite{wolff_monte_2007, kalugin_statistical_2011}.  Similarly, for Markov chain Monte Carlo methods sampling an invariant distribution, there is a vast literature on methods for on-the-fly variance estimation that incorporate correlations between consecutive samples, such as the batch means approach and spectral methods~\cite{geyer_practical_1992, damerdji_strong_1994, hobert_applicability_2002, jones_markov_2004, goodman_ensemble_2010, Flegal_batch_2010}. These methods typically require ergodicity properties and asymptotic arguments, such as the existence of a central limit theorem. Note, however, that stochastic processes underlying linear kinetic equations with sources and sinks typically are not ergodic. An advantage of the variance expressions derived below is that they do not require ergodicity and are valid for any number of samples $N$ without relying on asymptotic arguments or a central limit theorem.

Data-driven on-the-fly variance estimation techniques cannot be used for a priori variance prediction and do not provide insight in the dependence of the variance on model parameters, unless when repeating a lot of simulations for different parameter settings and comparing the results, but such an approach quickly becomes an expensive endeavour. Additionally, in complex codes (such as B2.5-EIRENE for plasma edge modelling, where the Monte Carlo code is coupled to a finite volume code) it might be unclear which error is dominant, e.g., when also a finite sampling bias and discretization bias are present~\cite{ghoos_accuracy_2016}. For such codes, the conclusions of on-the-fly variance estimation might be unreliable. In these situations, theory-based a priori statistical error predictors that reveal how the error depends on model parameters are more desirable for tuning numerical simulation parameters, as they do not require numerous initial experiments to tune and optimize each separate production run.

In this paper, we show how the variance on the QoI estimates can be calculated (expensive) and predicted (cheap) in the setting of an analog particle tracing Monte Carlo method on a spatial grid, where each grid cell is used as a bin to construct a histogram for the QoI. This setting is common, e.g., in plasma edge modelling where equation~\eqref{eq:kinetic_equation} is solved to model the neutral particle behaviour, and the grid (histogram) on which the QoI are estimated is also used to simulate the background plasma using a finite volume method~\cite{reiter_eirene_2005, horsten_fluid_2019, mortier_advanced_2020}. In Ref.~\cite{scott_optimal_1979}, the idea is introduced that building a histogram can be interpreted as performing a binomial experiment, where for each bin the Bernoulli trial consists of a particle (or more generally speaking a sample) being present in that bin (success) or not (failure). In that paper, only independent samples are considered. However, for particle tracing Monte Carlo methods, a particle can contribute multiple times to the histogram, leading to multiple correlated samples. Therefore, we extend the ideas introduced in Ref.~\cite{scott_optimal_1979} to the setting of particle tracing Monte Carlo, where the construction of a histogram can be interpreted as performing a correlated binomial experiment. 

The literature on correlated binomial experiments (binary data) is focussed on proposing a probability model (e.g., the Bahadur model) for the joint probability of $N$ Bernoulli trials, and then fitting the parameters in that model by regression using data (e.g., by using a maximum-likelihood estimator)~\cite{cox_analysis_1972, lipsitz_estimation_1995, pendergast_survey_1996, zhang_use_2019}. This probability model for the joint probability distribution then specifies the variance on the outcome of such a binomial experiment. These probability models are, however, data-driven, which in our simulation context means that it requires (expensive) simulations to generate the required data. In this paper, we aim at providing an a priori variance predictor that does not require (expensive) simulations. As we will see below, several theory-based variance expressions exist for correlated binomial experiments, some of which can be used as a cheap a priori variance predictor. One variance expression (based on hidden Markov model theory) allows to calculate the actual variance, but at a high cost. Another variance expression only accounts for the Markov dependence between consecutive samples and provides a cheap approximate predictor for the actual variance. For each variance expression, we verify the correctness and accuracy.

% Overview paper
The rest of this paper is structured as follows. Section~\ref{sec: particle dynamics} introduces the concepts underlying analog particle tracing Monte Carlo and QoI estimation. In Section~\ref{sec: binomial experiments}, different expressions to describe the variance on (correlated) binomial experiments in an exact and approximate way are introduced. In Section~\ref{sec: point estimators}, we use the obtained expressions to calculate and predict the variance on an analog particle tracing Monte Carlo method with point estimators, in which each particle provides one independent contribution to the QoI estimates. In Section~\ref{sec: analog estimators}, we use the obtained expressions to calculate and predict the variance on an analog particle tracing Monte Carlo method with analog estimators, for which each particle provides multiple correlated contributions to the QoI estimates. Section~\ref{sec: discussion} discusses some important remarks with respect to the topic of statistical error prediction. Finally, the conclusions and some pointers for future work are summarized in Section~\ref{sec: conclusion}. The code accompanying this paper is openly available in \reponame.

\section{Particle dynamics and QoI estimators}
\label{sec: particle dynamics}
In this section, we describe how kinetic equation~\eqref{eq:kinetic_equation} can be solved using an analog particle tracing Monte Carlo scheme and introduce some statistical assumptions on the particle dynamics that we will use to derive the a priori variance predictors. Next, we show how QoI can be estimated on a grid and how this estimation can be interpreted as the execution of a binomial experiment. Finally, we illustrate how the variance on such a binomial experiment depends on the model parameters of the kinetic equation.
\subsection{Analog particle tracing Monte Carlo}
\label{sec: binomial experiments dynamics}
Analog particle tracing Monte Carlo builds on the insight that the kinetic equation~\eqref{eq:kinetic_equation} can be interpreted as the population dynamics of particles that execute the following velocity-jump process~\cite{lapeyre_introduction_2003}:
\begin{equation}
\begin{split}
&x_{n,k+1} = x_{n,k} + v_{n,k} \tau_{n,k}, \quad x_{n,k} \in D \subset \mathbb{R}^d\\
&v_{n,k} \sim M(v_{n,k}) = \mathcal{N}(v_{n,k} \mid u_p,\sigma_p^2) = \frac{1}{(2 \pi \sigma_p^2)^{d/2}} \exp \left( - \frac{1}{2} \frac{|v_{n,k}-u_p|^2}{\sigma_p^2} \right), \quad v_{n,k} \in \mathbb{R}^d\\
&\tau_{n,k} \sim \mathcal{E}(\tau_{n,k} \mid R) = R \exp(-R \tau_{n,k} ), \quad \tau_{n,k} \in \mathbb{R}_+.
\end{split}
\label{eq:particle_dynamics}
\end{equation}
In this velocity-jump process, $x_{n,k}$ represents the position where particle $1 \leq n \leq N$ has its $k$-th collision ($x_{n,0}$ represents the initial particle position at time $t_0 = 0$), $v_{n,k}$ is the velocity of the particle after that collision, and $\tau_{n,k}$ is the flight time until the next collision. At each collision point $x_{n,k}$, the trajectory is terminated with probability
\begin{equation}
\mathcal{P}_t = \frac{R_i}{R_i + R_{cx}} = \frac{R_i}{R}.
\end{equation}
The transition kernel corresponding to the velocity-jump process is given by
\begin{equation}
\mathcal{K}(x_{n,k} \rightarrow x_{n,k+1}) = \int_0^\infty \frac{1}{(\tau_{n,k})^d} \cdot \mathcal{E} \left( \tau_{n,k} \mid R \right) \mathcal{N}\left(\frac{x_{n,k+1} - x_{n,k}}{\tau_{n,k}} \mid u_p, \sigma_p^2 \right) d\tau_{n,k}.
\label{eq:transition_kernel}
\end{equation}
The transition kernel~\eqref{eq:transition_kernel} represents the probability density for having a transition from $x_{n,k}$ to $x_{n,k+1}$ and intuitively follows from integrating over all combinations of $\tau_{n,k}$ and $v_{n,k}$ that result in this transition. Mathematically, the transition kernel is obtained by differentiating the cumulative distribution function 
\begin{equation}
\begin{split}
\mathcal{P}(v_{n,k} \tau_{n,k} \leq x_{n,k+1}-x_{n,k}) &= \int_0^\infty \mathcal{P}\left(v_{n,k} \leq \frac{x_{n,k+1}-x_{n,k}}{\tau_{n,k}} \mid \tau_{n,k} \right)\mathcal{E}(\tau_{n,k} \mid R)d\tau_{n,k}\\
&= \int_0^\infty \int_{-\infty}^\frac{x_{n,k+1} - x_{n,k}}{\tau_{n,k}} \mathcal{N}(v_{n,k} \mid u_p, \sigma_p^2 )dv_{n,k} \cdot \mathcal{E}(\tau_{n,k} \mid R)d\tau_{n,k}, 
\end{split}
\end{equation}
where we used the probability densities defined in~\eqref{eq:particle_dynamics}.

\paragraph{Remark 1} The dynamics~\eqref{eq:particle_dynamics} and the transition kernel~\eqref{eq:transition_kernel} are correct for unbounded domains, i.e., they do not yet take boundary conditions into account. There are many boundary conditions that can be imposed on a linear kinetic equation, see for example Ref.~\cite{horsten_fluid_2019}.
In the numerical experiments, we will only consider periodic boundary conditions, leaving more general boundary conditions for future work.

When simulating kinetic equation~\eqref{eq:kinetic_equation} with velocity-jump process~\eqref{eq:particle_dynamics}, the total mass that has to be simulated,
\begin{equation}
M = \int \int f(x,v,t=0) dxdv + \int \int \int S(x,v,t) dx dv dt,
\end{equation}
originates from the initial condition and/or the particle source and has to be represented by $N$ particles. In an analog simulation, each particle gets the same constant weight:
\begin{equation}
w=\frac{M}{N}.
\label{eq:particle_weight}
\end{equation}
There are two ways to treat the particle source $S(x,v,t)$. The first way works for transient sources and deterministically discretizes the particle source in time using $I$ time steps of size $\Delta t_s$. We then have $N_0$ particles at the start of the simulation (at $t_0 = 0$) and generate $N_{s,i}$ additional particles in each time step during the simulation. For variance prediction, we can then simply treat each group (stratum) of particles ($N_0$, $N_{s,i}$ $i=1..I$) separately. 

The second way only works if the particle source is stationary ($S(x,v,t) = S(x,v)$) over a time interval $t \in [t_1,t_2]$, meaning that the times at which particles are generated are distributed uniformly in that time interval. We can then split up the particles in a group (stratum) of $N_0$ particles that is present at the start of the simulation (at $t_0 = 0$) and a group (stratum) of $N_s$ particles that is generated by the particle source during the simulation. For each particle $n$ of those $N_s$ particles, we then sample a generation time $t_n$ at which that particle enters the simulation. The generation times are uniformly distributed in $[t_1,t_2]$:
\begin{equation}
\begin{split}
t_n &\sim U[t_1,t_2].
\end{split}
\label{eq:stationary_source_statistics}
\end{equation}
How to take such a stationary particle source into account for variance prediction will be discussed in sections~\ref{sec: point estimators} and~\ref{sec: analog estimators}.

\subsection{Statistical assumptions}
Note in~\eqref{eq:particle_dynamics} that a particle $n$ requires no information about other particles, i.e., each particle is simulated independently from the others. This stems from the fact that the kinetic equation~\eqref{eq:kinetic_equation} is linear. Furthermore, the particles all follow the same dynamics, making them independent and identically distributed (iid). For the variance prediction, we also assume that the simulation parameters $u_p$, $\sigma_p^2$, $R_{cx}$, and $R_i$ are homogeneous (space independent) and stationary (time independent), such that $\tau_{n,k}$, $v_{n,k}$, and $x_{n,k}$ are independent random variables (probability distributions $\mathcal{N}(v_{n,k} \mid u_p, \sigma_p^2)$ and $\mathcal{E}(\tau_{n,k} \mid R)$ in~\eqref{eq:particle_dynamics} do not depend on $x$, $v$, $t$). From these probability distributions, it follows that both the velocities $v_{n,k}$ and the flight times $\Delta t_{n,k}$ are iid. The positions $x_{n,k}$, however, are not iid, because position $x_{n,k+1}$ depends on $x_{n,k}$. Note that the assumption of constant simulation parameters is a typical finite volume assumption within one grid cell of the simulation domain. In realistic simulation cases, however, the simulation parameters do vary throughout the simulation domain and in time. We neglect these space and time dependencies below in the construction of the variance predictors.

\subsection{QoI estimation on a grid}
\label{subsec: QoI estimation}
We estimate the QoI~\eqref{eq:QoI_Generic} on a grid, which consists of $J$ cells $D_j \subset D$, interpreting each grid cell as a bin of a histogram. We define an indicator function $I_j(x)$ over each of the grid cells that equals 1 for $x \in D_j$ and 0 elsewhere. Note that a random evaluation of such an indicator function is a Bernoulli random variable where 1 can be interpreted as a success and 0 as a failure. We can then define the QoI in cell $j$ as follows:
\begin{equation}
Q_j(t) = \int_{D_j} Q(x,t)dx = \int_D Q(x,t) I_j(x)dx = \int_D \int q(v)f(x,v,t)I_j(x)dvdx.
\end{equation}
A particle can only contribute to the estimate $\hat{Q}_j(t)$ of $Q_j(t)$ if its position is in cell $D_j$. We consider two QoI estimators: point estimators and analog estimators. For a point estimator, we are interested in the QoI at one specific point in time $t=T$. If we have $N$ particles, the estimator is of the form
\begin{equation}
Q_j(T) \approx \hat{Q}_j(T) = w \sum_{n=1}^N q(v_n(T)) I_j(x_n(T)),
\label{eq:QoI_Point_Estimator}
\end{equation} 
where $x_n(T)$, $v_n(T)$ denote the position and velocity of particle $n$ at time $T$ without the constraint that it has to be a collision point. Note that the estimate consists of $N$ independent contributions, as each particle is simulated independently and only contributes once.
For an analog estimator, we are interested in the average QoI over a time interval $t \in [t_1, t_2]$, where we count contributions to the QoI at collision points $k \in [0, K(n)]$, where $k=0$ represents the first collision in the time interval and $K(n)$ represents the last collision of particle $n$ in the time interval. If we have $N$ particles, the estimator is of the form
\begin{equation}
Q_j = \frac{1}{\Delta t}\int_{t_1}^{t_2} Q_j(t)dt \approx \hat{Q}_j = \frac{w}{R\Delta t} \sum_{n=1}^N \sum_{k=0}^{K(n)} q(v_{n,k})I_j(x_{n,k}),
\label{eq:QoI_Analog_Estimator}
\end{equation}
where we represent the duration of the time interval by $\Delta t = t_2 - t_1$. In this estimator, a particle $n$ delivers $K(n)$ contributions to the estimate, where the particle dynamics introduce correlations between the positions $x_{n,k}$ in these contributions.

Using expression~\eqref{eq:particle_weight}, we see that both estimators are sample averages $S_n$ of the form
\begin{equation}
S_N = \frac{1}{N}\sum_{n=1}^{N} F_n,
\label{eq:iid_estimators}
\end{equation}
where for the point estimator~\eqref{eq:QoI_Point_Estimator} we have
\begin{equation}
F_n = Mq(v_n(T))I_j(x_n(T)),
\end{equation}
and for the analog estimator~\eqref{eq:QoI_Analog_Estimator} we have
\begin{equation}
F_n = \frac{M}{R\Delta t} \sum_{k=0}^{K(n)} q(v_{n,k})I_j(x_{n,k}).
\end{equation}
Since the particles $1 \leq n \leq N$ are iid, it follows from the variance sum law~\cite{hodges_basic_2005} that the variance on sample averages~\eqref{eq:iid_estimators} is of the form
\begin{equation}
\mathbb{V}[S_N] = \frac{C}{N},
\end{equation}
like the asymptotic expression~\eqref{eq:asymptotic_variance_expression}, where in this case $C = \mathbb{V}[F_n]$. In what follows, we search for good theory-based approximations of the constant $C$ to get a quantitative prediction of the statistical error, valid both for low and high values of $N$. To achieve this goal, we interpret the sums of indicator functions as binomial experiments.

\subsection{(Correlated) binomial experiments}
In both estimators~\eqref{eq:QoI_Point_Estimator} and~\eqref{eq:QoI_Analog_Estimator}, we have a sum of indicator function evaluations. We now write a generic sum of $L$ indicator function evaluations at arbitrary positions $x_l \in D$ as
\begin{equation}
S_{L,j} = \sum_{l=1}^L I_j(x_l).
\label{eq:general_binomial_experiment}
\end{equation}
This sum can be interpreted as a binomial experiment, in which each evaluation of the indicator function $I_j(x_l)$ represents a Bernoulli trial. We denote the probability for success in Bernoulli trial $l$ by $p_{l,j}$. The probability for success can be related to the expected value $\langle I_j(x_l) \rangle$ of the indicator function evaluation $I_j(x_l)$ by using the fact that an expected value of a discrete random variable can be written as the sum of each possible outcome times the probability for that outcome:
\begin{equation}
p_{l,j} = \mathcal{P}(I_j(x_l) = 1) = 1 \cdot \mathcal{P}(I_j(x_l) = 1) + 0 \cdot \mathcal{P}(I_j(x_l) = 0) = \langle I_j(x_l) \rangle.
\label{eq:binomial_success_probability_singletrial}
\end{equation}
The variance on the sum $S_{L,j}$ can be written as
\begin{equation}
\begin{split}
\mathbb{V}[S_{L,j}] = \mathbb{V}[\sum_{l=1}^L I_j(x_l)] &= \langle (\sum_{l=1}^L I_j(x_l) - \langle \sum_{l=1}^L I_j(x_l) \rangle )^2 \rangle\\
&= \langle (\sum_{l=1}^L I_j(x_l))^2 \rangle - \langle \sum_{l=1}^L I_j(x_l) \rangle^2\\
&= \langle \sum_{l=1}^L I_j(x_l)^2 + 2\sum_{l=1}^L \sum_{m=1}^{l-1} I_j(x_l) I_j(x_m) \rangle - \langle \sum_{l=1}^L I_j(x_l) \rangle^2 \\
&= \sum_{l=1}^L \langle I_j(x_l)^2 \rangle + 2 \sum_{l=1}^L \sum_{m=1}^{l-1} \langle I_j(x_l) I_j(x_m) \rangle - (\sum_{l=1}^L \langle I_j(x_l) \rangle)^2\\
&= \sum_{l=1}^L \langle I_j(x_l) \rangle + 2 \sum_{l=1}^L \sum_{m=1}^{l-1} \langle I_j(x_l) I_j(x_m) \rangle - (\sum_{l=1}^L \langle I_j(x_l) \rangle)^2\\
&= \sum_{l=1}^L p_{l,j} + 2 \sum_{l=1}^L \sum_{m=1}^{l-1} \langle I_j(x_l) I_j(x_m) \rangle - (\sum_{l=1}^L p_{l,j})^2,
\end{split}
\label{eq:variance_on_nj}
\end{equation}
where in the penultimate equality, we used the following identity for indicator functions: 
\begin{equation}
I_j(x_l)^2 = I_j(x_l).
\label{eq:indicator_function_property}
\end{equation}
The goal of Section~\ref{sec: binomial experiments} is to find useful expressions for the variance $\mathbb{V}[S_{L,j}]$, which we can then use to derive variance predictors for the QoI estimators~\eqref{eq:QoI_Point_Estimator} and~\eqref{eq:QoI_Analog_Estimator} in Section~\ref{sec: point estimators} and Section~\ref{sec: analog estimators}, respectively.

\subsection{Variance as a function of model parameters}
\label{sec:comparison_variance_experiment}
\begin{figure}
\centering
\includegraphics[width=0.8\linewidth]{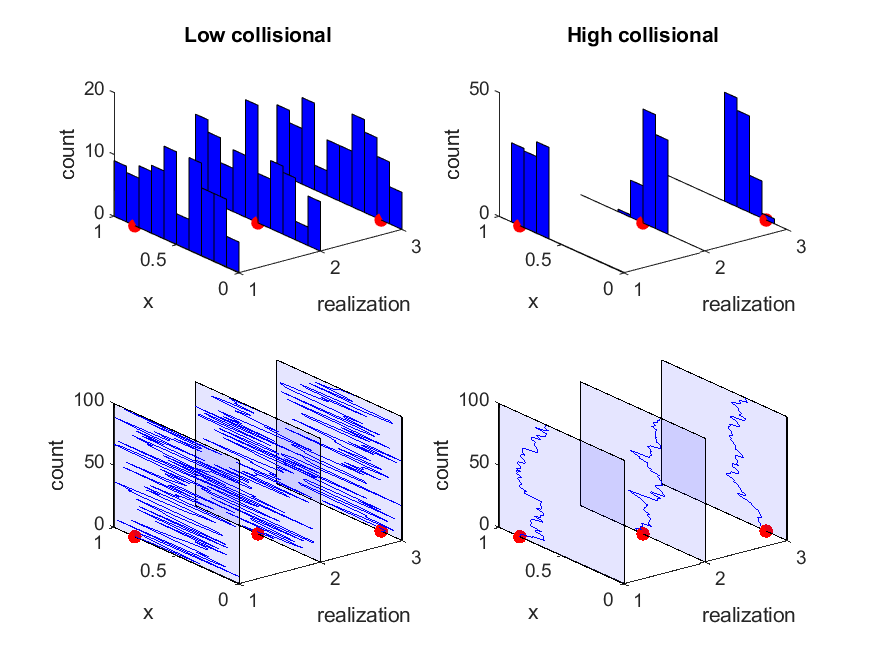}
\caption{Particle paths (bottom) and corresponding histograms counting the number of collisions happening inside each bin (top). Left: three realizations with $u_p = 0$, $\sigma_p^2 = 1$, $R_i = 0$, $R_{cx} = 1$. Right: three realizations with $u_p = 0$, $\sigma_p^2 = 1$, $R_i = 0$, $R_{cx} = 25$. The red dots indicate the initial particle positions. All simulations are terminated after 100 collisions.}
\label{fig:comparison_variance}
\end{figure}

The variance on the outcome of a correlated binomial experiment representing QoI estimation on a grid depends on model parameters such as $u_p$, $\sigma_p^2$, $R_i$, and $R_{cx}$. To illustrate this, we compare three realizations of two simulation set-ups with different simulation parameters in Figure~\ref{fig:comparison_variance}. The simulation parameters are chosen to be homogeneous and stationary such that particles are distributed uniformly in space and time throughout the simulation. In each realization, we build a histogram that counts the number of collisions happening inside each bin for one simulated particle, which for each bin results in a sum of the form~\eqref{eq:general_binomial_experiment} resembling an analog estimator~\eqref{eq:QoI_Analog_Estimator} with $N=1$. Each contribution to the histogram is a location where that particle has a collision. The histograms have $J=10$ bins and the particle is allowed to execute $L=100$ collisions. For both simulation set-ups, the expected number of collisions in each bin is the same, namely $\frac{L}{J} = 10$. 

The simulation set-up in the left panel has a low value for the charge-exchange rate $R_{cx}$, meaning that particles can travel a long distance in between collisions. As a result, the histograms look as if the contributions (the collision points) are independent of each other, leading to a small variability between different realizations, which in turn leads to a low variance on the realized histograms. The simulation set-up in the right panel has a high value for the charge-exchange rate $R_{cx}$, meaning that particles can only travel a short distance in between collisions. As a result, the realized histograms are tightly centered around their initial collision point, because consecutive collision points are strongly correlated. The histograms therefore depend heavily on the initial position of the particle. The variability between different realizations, with different initial particle positions, is large, which in turn leads to a high variance on the realized histograms. 

The experiment supports the following intuition about the variance of histogram based QoI estimators in which each particle can deliver an equal number of correlated contributions to an estimate: (1) model parameters that result in particles only being able to travel short distances lead to a high variance; (2) model parameters that result in particles being able to travel long distances lead to a low variance. This intuition is confirmed mathematically in the subsequent sections of this paper.

%\newpage
\section{Variance expressions for (correlated) binomial experiments}

In this section, we derive four expressions for the variance $\mathbb{V}[S_{L,j}]$~\eqref{eq:variance_on_nj} on a (correlated) binomial experiment consisting of $L$ Bernoulli trials. First, we derive an upper bound on $\mathbb{V}[S_{L,j}]$ by searching for the worst-case relation between consecutive Bernoulli trials $x_l$; this upper bound is independent of any model parameters of the kinetic equation~\eqref{eq:kinetic_equation}. Second, we derive an expression for a binomial experiment in which consecutive Bernoulli trials are assumed to be independent; this expression is also independent of any model parameters. Third, we assume that consecutive Bernoulli trials have a Markov dependence, i.e., we assume that there is a 2-state Markov process that determines if a particle is inside bin $D_j$ or not. The probabilities involved in the assumed Markov process can be related to the bin size and model parameters of the kinetic equation~\eqref{eq:kinetic_equation}. Finally, we assume that the binomial experiment is the observed process of a hidden Markov model, where the underlying (hidden) process is the particle dynamics~\eqref{eq:particle_dynamics}. The involved probabilities can again be related to the bin size and model parameters of the kinetic equation.
\label{sec: binomial experiments}
\subsection{Upper bound}
Let us start by providing a simple upper bound for the variance on $S_{L,j}$. We look for stationary worst-case values $p$, $z$ for the expected values in the variance expression~\eqref{eq:variance_on_nj}:
\begin{equation}
\begin{split}
 &p \coloneq \langle I_j(x_l) \rangle = p_{l,j},\\
 &z \coloneq \langle I_j(x_l) I_j(x_m) \rangle.
 \end{split}
\end{equation}
Because the indicator functions $I_j(x_l)$ can only take values 0 or 1, we have that $ 0 \leq p \leq 1$, and $0 \leq z \leq 1$. We insert $p$, $z$ in~\eqref{eq:variance_on_nj} and solve the following optimization problem to find an upper bound for the variance on $S_{L,j}$:
\begin{equation}
\max_{p,z} Lp + L(L-1)z - L^2 p^2 \quad \text{s.t.} \quad 0 \leq p \leq 1, \hspace{2pt} 0 \leq z \leq 1,
\end{equation}
where $L \geq 1$ is the number of Bernoulli trials in the binomial experiment.
It can be verified that the solution for this maximization problem is $p=\frac{1}{2L}$, $z=1$, resulting in the following upper bound for the variance:
\begin{equation}
\mathbb{V}[S_{L,j}] = \mathbb{V}[\sum_{l=1}^L I_j(x_l)] \leq L(L-1) + 0.25 \sim \mathcal{O}(L^2),
\label{eq:binomial_variance_simple_upperbound}
\end{equation}
which is quadratic in the number of Bernoulli trials $L$. As we will see in the numerical experiments below, the upper bound~\eqref{eq:binomial_variance_simple_upperbound} typically provides a huge overestimation of the actual variance of a binomial experiment.

\paragraph{Remark 2} A quadratic variance $\sim \mathcal{O}(L^2)$ on a binomial experiment~\eqref{eq:general_binomial_experiment} means that for a sample mean of the form $\frac{1}{L} \sum_{l=1}^L I_j(x_l)$, the variance is independent of $L$, therefore, the variance does not decrease for $L \rightarrow \infty$. A non-decreasing variance corresponds to taking $L$ times the same sample in each realization of the sum (perfect correlation, i.e., $z=1$). More practically, the equivalent in the QoI estimation context described in Section~\ref{subsec: QoI estimation} is an analog estimator~\eqref{eq:QoI_Analog_Estimator} with one particle ($N=1$) that only moves within one bin, which is a limiting case of the high-collisional regime displayed in Figure~\ref{fig:comparison_variance} in which each particle still samples three bins. In this limiting case, we would have $I_j(x_{1,k}) = I_j(x_{1,0})$ $\forall k$, indeed resulting in the worst case estimator for the QoI as the particle does not sample different bins $D_j$ in the domain $D$. This means that increasing the number of samples (number of collisions) does not improve the QoI estimate.

\subsection{Binomial experiment with independent Bernoulli trials}
Let us assume that the Bernoulli trials are independent and that each trial occurs with the same (stationary) success probability, meaning that the samples of the random variable $I_j(x_l)$ are iid:
%de tweede gelijkheid hier volgt uit dat <I_n I_m> = P(I_n = 1, I_m = 1) * 1 + (P(I_n = 0, I_m = 0) + P(I_n = 1, I_m = 0) + P(I_n = 0, I_m = 1)) * 0 = P(I_n = 1, I_m = 1) = P(I_n = 1 \mid I_m = 1) * P(I_m = 1) = (independent) = P(I_n = 1) * P(I_m = 1) = p * p = p^2
\begin{equation}
\begin{split}
&\langle I_j(x_l) \rangle = p_{l,j} = p_j,\\
&\langle I_j(x_l) I_j(x_m) \rangle = \langle I_j(x_l) \rangle \langle I_j(x_m) \rangle = p_{l,j} p_{m,j} = p_j^2.
\end{split}
\label{eq:binomial_statistics_independent}
\end{equation} 
Variance expression~\eqref{eq:variance_on_nj} then simplifies to the form
\begin{equation}
\mathbb{V}[S_{L,j}] = \mathbb{V}[\sum_{l=1}^L I_j(x_l)] =  L p_j + L(L-1)p_j^2 - L^2 p_j^2 = L p_j (1-p_j),
\label{eq:binomial_variance_independent_stationary}
\end{equation}

which is the well known expression for the variance of a binomial experiment with independent trials. Note that the variance expression for independent trials~\eqref{eq:binomial_variance_independent_stationary} is linear in $L$, while the upper bound~\eqref{eq:binomial_variance_simple_upperbound} derived above is quadratic. Information on how to obtain a value for $p_j$ can be found in Section~\ref{subsec:pandlambda}.

\subsection{Binomial experiment driven by a Markov process}
\label{subsec:markov_process}
In this section, we assume that the binomial experiment for each bin $D_j$ is driven by a 2-state Markov process in which the current Bernoulli trial depends on the outcome of the previous trial (but not on outcomes of older trials). Binomial experiments with a Markov dependence have been investigated in a stationary setting in Ref.~\cite{klotz_statistical_1973} and in a transient setting in Ref.~\cite{xekalaki_binomial_2004}. The Markov dependence of the current trial on the previous trial is encoded by conditional probabilities. In the transient setting, the (conditional) probabilities are allowed to have a different value for each Bernoulli trial, while in the stationary setting, the same probabilities are assumed for each trial. The (trial dependent) probability for success is defined above as $p_{l,j}$ in~\eqref{eq:binomial_success_probability_singletrial}. The probability for failure follows logically as $\mathcal{P}(I_j(x_l) = 0) = 1 - \mathcal{P}(I_j(x_l) = 1) = 1 - p_{l,j}$. To fully characterize the Markov dependence, we define the following conditional probabilities encoding the dependence on the outcome of the previous trial:
\begin{equation}
\begin{split}
&\mathcal{P}(I_j(x_l) = 1 \mid I_j(x_{l-1}) = 1) = \lambda_{l-1,j},\\
&\mathcal{P}(I_j(x_l) = 0 \mid I_j(x_{l-1}) = 1) = 1 - \mathcal{P}(I_j(x_l) = 1 \mid I_j(x_{l-1}) = 1) = 1-\lambda_{l-1,j},\\
&\mathcal{P}(I_j(x_l) = 0 \mid I_j(x_{l-1}) = 0) = \nu_{l-1,j},\\
&\mathcal{P}(I_j(x_l) = 1 \mid I_j(x_{l-1}) = 0) = 1 - \mathcal{P}(I_j(x_l) = 0 \mid I_j(x_{l-1}) = 0) = 1 - \nu_{l-1,j}.
\end{split}
\label{eq:binomial_markovdependence_conditionalprobabilities}
\end{equation}
The conditional probabilities serve as transition probabilities in the Markov process that drives the binomial experiment. We can construct the transition matrix $T_j(P_{l-1,j} \rightarrow P_{l,j})$ as follows:
\begin{equation}
T_j(P_{l-1,j} \rightarrow P_{l,j}) = 
\begin{bmatrix}
\nu_{l-1,j} & 1-\lambda_{l-1,j} \\ 1 - \nu_{l-1,j} &
\lambda_{l-1,j}
\end{bmatrix},
\end{equation}
defining the vector $P_{l,j} = \begin{bmatrix}
1-p_{l,j} \\
p_{l,j}
\end{bmatrix}$ containing the failure and success probability for trial $l$. We then recover the following Markov process:
\begin{equation}
P_{l,j} = T_j(P_{l-1,j} \rightarrow P_{l,j})P_{l-1,j}.
\label{eq:binomial_discret_markov_process}
\end{equation}
Having defined the Markov process and the probabilities involved, we can elaborate an expression for the variance on $S_{L,j}$~\eqref{eq:variance_on_nj}. Following Ref.~\cite{xekalaki_binomial_2004}, we observe that in the second term in the variance expression~\eqref{eq:variance_on_nj}, we can rewrite $\langle I_j(x_l) I_j(x_m) \rangle$ with $m<l$ as follows:
\begin{equation}
\begin{split}
\langle I_j(x_l) I_j(x_m) \rangle &= \langle I_j(x_l) I_j(x_m) \mid I_j(x_l) = 0, I_j(x_m) = 0 \rangle \cdot \mathcal{P}(I_j(x_l) = 0, I_j(x_m) = 0)\\ 
&+ \langle I_j(x_l) I_j(x_m) \mid I_j(x_l) = 1, I_j(x_m) = 0 \rangle \cdot \mathcal{P}(I_j(x_l) = 1, I_j(x_m) = 0)\\ 
&+ \langle I_j(x_l) I_j(x_m) \mid I_j(x_l) = 0, I_j(x_m) = 1 \rangle \cdot \mathcal{P}(I_j(x_l) = 0, I_j(x_m) = 1)\\ 
&+ \langle I_j(x_l) I_j(x_m) \mid I_j(x_l) = 1, I_j(x_m) = 1 \rangle \cdot \mathcal{P}(I_j(x_l) = 1, I_j(x_m) = 1)\\
&= 0 \times \mathcal{P}(I_j(x_l) = 0, I_j(x_m) = 0) + 0 \cdot \mathcal{P}(I_j(x_l) = 1, I_j(x_m) = 0)\\ &+ 0 \cdot \mathcal{P}(I_j(x_l) = 0, I_j(x_m) = 1) + 1 \cdot \mathcal{P}(I_j(x_l) = 1, I_j(x_m) = 1) \\
&= \mathcal{P}(I_j(x_l) = 1, I_j(x_m) = 1)\\
&= \begin{bmatrix} 0 & 1 \end{bmatrix} \left(\prod_{i=m+1}^{l} T_j(P_{i-1,j} \rightarrow P_{i,j})\right) \begin{bmatrix}
0 \\ p_{m,j}
\end{bmatrix}.
\end{split}
\label{eq:binomial_markovdependence_secondterm_variance_of_nj}
\end{equation}
This result can be interpreted as follows (by reading the final expression from the right to the left). We take the fraction $p_{m,j}$ of the Markov chains that have success as outcome in trial $m$. Then, we look at how this fraction redistributes its probability in each of the next $l-m$ trials, to find which fraction of those chains ends up with a success as outcome in trial $l$ and which fraction has a failure as outcome in trial $l$. Finally, we are only interested in the fraction of Markov chains that have a success in trial $l$. Inserting~\eqref{eq:binomial_markovdependence_secondterm_variance_of_nj} into the variance expression~\eqref{eq:variance_on_nj} results in
\begin{equation}
\mathbb{V}[S_{L,j}] = \mathbb{V}[\sum_{l=1}^L I_j(x_l)] = \sum_{l=1}^L p_{l,j} + 2 \sum_{l=1}^L \sum_{m=1}^{l-1} \begin{bmatrix} 0 & 1 \end{bmatrix} \left(\prod_{i=m+1}^{l} T_j(P_{i-1,j} \rightarrow P_{i,j})\right) \begin{bmatrix}
0 \\ p_{m,j}
\end{bmatrix} - \left(\sum_{l=1}^L p_{l,j}\right)^2,
\end{equation}
which can also be written in the following form~\cite{xekalaki_binomial_2004}:
\begin{equation}
\begin{split}
\mathbb{V}[S_{L,j}] = \mathbb{V}[\sum_{l=1}^L I_j(x_l)] &= \sum_{l=1}^L p_{l,j}(1-p_{l,j})\\ 
&+ 2 \sum_{l=1}^L \sum_{m=1}^{l-1} \left( \begin{bmatrix} 0 & 1 \end{bmatrix} \left(\prod_{i=m+1}^{l} T_j(P_{i-1,j} \rightarrow P_{i,j})\right) \begin{bmatrix}
0 \\ p_{m,j}
\end{bmatrix} - p_{l,j} p_{m,j} \right).
\end{split}
\label{eq:binomial_variance_markovdependence_xekalaki}
\end{equation}
In this last expression, we recognize the variance for a transient independent binomial experiment (compare the first term to the expression for the variance of an independent stationary binomial experiment~\eqref{eq:binomial_variance_independent_stationary}). The second term can therefore be considered as a correction term that corrects for the presence of a Markov dependence between consecutive trials.

For stationary processes, we can drop the subscript $l$ on the probabilities $p_j$, $\lambda_j$ and $\nu_j$. As the probabilities should remain the same in each trial, we obtain the following detailed balance condition:
\begin{equation}
\mathcal{P}(I_j(x_l) = 1, I_j(x_{l-1}) = 0) = \mathcal{P}(I_j(x_l) = 0, I_j(x_{l-1}) = 1),
\label{eq:MarkovDependence_DetailedBalance}
\end{equation}
from which it follows that
\begin{equation}
1 - \nu_j = \mathcal{P}(I_j(x_l) = 1 \mid I_j(x_{l-1}) = 0) = \frac{\mathcal{P}(I_j(x_l) = 0 \mid I_j(x_{l-1}) = 1) \cdot \mathcal{P}(I_j(x_{l-1}) = 1)}{\mathcal{P}(I_j(x_{l-1}) = 0)} = \frac{(1-\lambda_j)p_j}{1-p_j},
\label{eq:MarkovDependence_DetailedBalance_corollary}
\end{equation}
i.e.,  $\nu_j$ becomes a function of $p_j$ and $\lambda_j$, meaning that the last two now fully characterize the 2-state Markov process~\cite{klotz_statistical_1973}. Information on how to obtain values for $p_j$ and $\lambda_j$ for a given problem can be found in Ref.~\cite{ingelaere_prep_2023} and in Section~\ref{subsec:pandlambda}. We now obtain the stationary Markov process $P_{l,j} = T_j P_{l-1,j}$ with stationary transition matrix $T_j$:
\begin{equation}
T_j = \begin{bmatrix}
1 - \frac{(1-\lambda_{j})p_{j}}{1-p_{j}} & 1-\lambda_{j} \\ \frac{(1-\lambda_{j})p_{j}}{1-p_{j}} &
\lambda_{j}  
\end{bmatrix}.
\label{eq:binomial_markov_transitionmatrix_klotz}
\end{equation}
The stationary equivalent of~\eqref{eq:binomial_markovdependence_secondterm_variance_of_nj} is as follows:
% Remark: see (5.4) in Klotz 1973
\begin{equation}
\langle I_j(x_l) I_j(x_m) \rangle = \begin{bmatrix} 0 & 1 \end{bmatrix} T_j^{l-m} \begin{bmatrix}
0 \\ p_{m,j}
\end{bmatrix} = p_j \left( p_j + (1-p_j) \left(\frac{\lambda_j - p_j}{1 - p_j} \right)^{l-m} \right).
\label{eq:binomial_statistics_Klotz}
\end{equation}
The expression for the variance on $S_{L,j}$ then simplifies from~\eqref{eq:binomial_variance_markovdependence_xekalaki} to
% Remark: the double sum in the variance expression becomes a geometric series for which sum formulas exist!
\begin{equation}
\begin{split}
\mathbb{V}[S_{L,j}] = \mathbb{V}[\sum_{l=1}^L I_j(x_l)] &= L p_j (1-p_j)\\ 
& + \frac{2 p_j (1-p_j) (\lambda_j - p_j)}{1-\lambda_j} \left( L - \frac{1 - p_j}{1 - \lambda_j} \left(1 - \left( \frac{\lambda_j - p_j}{1 - p_j} \right)^L \right) \right).
\end{split}
\label{eq:binomial_variance_markovdependence_klotz}
\end{equation}
Note that this is a slightly different result than expression (5.5) in Ref.~\cite{klotz_statistical_1973}, which contained an error.
The variance expression once again splits up into a term representing the variance for an uncorrelated binomial experiment with independent trials and a correction term for the Markov dependence. Expression~\eqref{eq:binomial_variance_markovdependence_klotz} can be used as a cheap-to-compute a priori predictor for the variance on QoI estimates. In the remainder, we refer to~\eqref{eq:binomial_variance_markovdependence_klotz} as the Markov process based variance predictor.

We draw attention to three important points in variance expression~\eqref{eq:binomial_variance_markovdependence_klotz}. First, all the conditional probabilities in the transition matrix~\eqref{eq:binomial_markov_transitionmatrix_klotz} must be values between 0 and 1. This leads to the following constraint:
\begin{equation}
\max \left( \frac{2 p_j - 1}{p_j},0 \right) \leq \lambda_j \leq 1,
\label{eq:Klotz_feasibility_border}
\end{equation}
which naturally constitutes a feasibility border that separates feasible and infeasible combinations of $p_j$ and $\lambda_j$. Under this constraint it follows that
\begin{equation}
\left|\frac{\lambda_j-p_j}{1-p_j} \right| \leq 1,
\end{equation}
meaning that the exponential part $\left( \frac{\lambda_j - p_j}{1 - p_j} \right)^L$ in expression~\eqref{eq:binomial_variance_markovdependence_klotz} cannot blow up. Second, the correction term scales with $\frac{1}{1-\lambda_j}$ and for $\lambda_j \rightarrow 1$, with $0 < p_j < 1$, the variance seems to blow up to infinity, which is not realistic for the variance of a finite sum. However, working out the limit $\lambda_j \rightarrow 1$ using l'H\^opital's rule shows that the actual limit of the variance is
\begin{equation}
\lim_{\lambda_j \rightarrow 1} \mathbb{V}[\sum_{l=1}^L I_j(x_l)] = L^2 p_j(1-p_j),
\label{eq:binomial_variance_markovdependence_klotz_largeL}
\end{equation}
which is a quadratic expression in $L$, just like the upper bound~\eqref{eq:binomial_variance_simple_upperbound}. It can be verified that expression~\eqref{eq:binomial_variance_markovdependence_klotz_largeL} respects the upper bound~\eqref{eq:binomial_variance_simple_upperbound} for all $ L \geq 1$.
The limit $\lambda_j \rightarrow 1$ corresponds to transitions between success and failure becoming rare events, leading to a sharp increase in the variance. Third, for $\lambda_j < 1$, we observe that the expression for the variance grows linearly with $L$, just as the variance of a binomial experiment with independent trials~\eqref{eq:binomial_variance_independent_stationary} (the quadratic behaviour~\eqref{eq:binomial_variance_markovdependence_klotz_largeL} is only present for $\lambda_j \rightarrow 1$). For large values of $L$ and $\lambda_j<1$, the variance expression~\eqref{eq:binomial_variance_markovdependence_klotz} simplifies approximately to
\begin{equation}
\mathbb{V}[\sum_{l=1}^L I_j(x_l)] \approx L \cdot \left( p_j (1-p_j) + \frac{2 p_j (1-p_j) (\lambda_j - p_j)}{1-\lambda_j} \right).\label{eq:binomial_variance_markovdependence_klotz_largeN}
\end{equation}

\begin{figure}
\centering
\begin{subfigure}{0.49\linewidth}
\centering
\includegraphics[width=\linewidth]{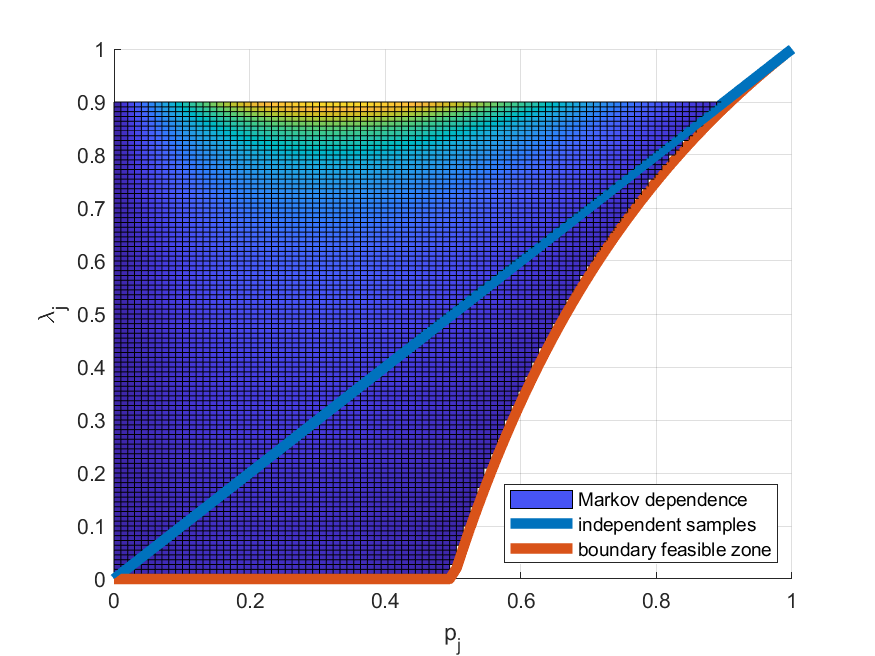}
\end{subfigure}
\begin{subfigure}{0.49\linewidth}
\centering
\includegraphics[width=\linewidth]{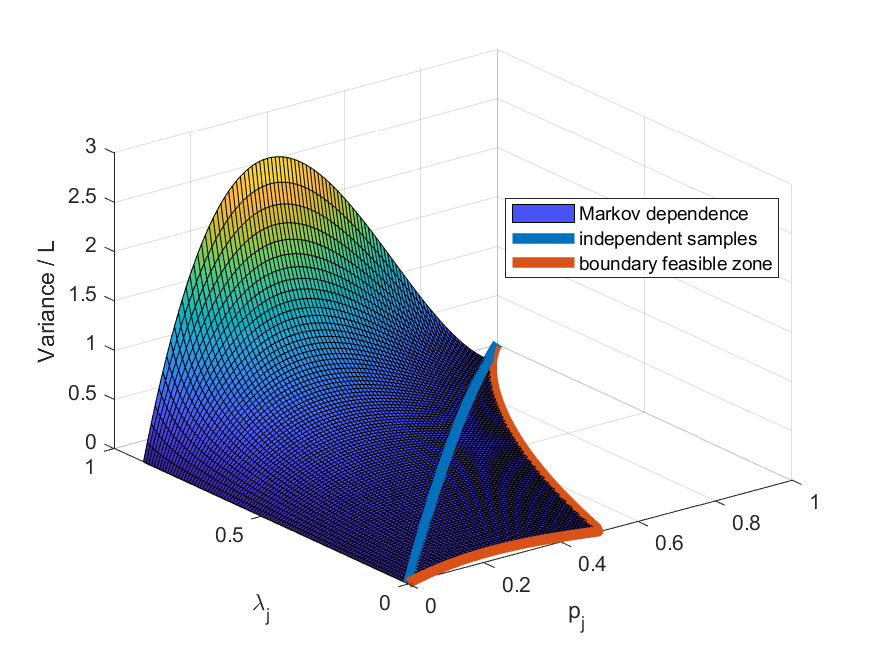}
\end{subfigure}
\caption{Markov dependence based variance expression for large $L$~\eqref{eq:binomial_variance_markovdependence_klotz_largeN} divided by $L$ for $0 \leq p_j \leq 1$ and $0 \leq \lambda_j \leq 0.9$. For large values of $\lambda_j$, transitions between success and failure become rare events, inflating the variance. The blue line represents the variance of a binomial experiment with $\lambda_j = p_j$, i.e., independent Bernoulli trials~\eqref{eq:binomial_variance_independent_stationary} divided by $L$. The red line represents the feasibility border~\eqref{eq:Klotz_feasibility_border}. Left: top view. Right: 3D view.}
\label{fig:Klotz_Variance_per_largeN}
\end{figure}

In Figure~\ref{fig:Klotz_Variance_per_largeN}, we plot the variance expression~\eqref{eq:binomial_variance_markovdependence_klotz_largeN} divided by $L$ for $\lambda_j \leq 0.9$. If $\lambda_j = p_j$, then the variance expression reduces to the expression for independent trials. This is logical, because it means that knowledge of the previous outcome does not affect the probability for success. If $\lambda_j < p_j$, which corresponds to more mixing in the Markov chain compared to when doing independent trials, then the correction term is negative, meaning that the Markov dependence lowers the variance on the binomial experiment. If $\lambda_j > p_j$, which corresponds to more clustering in the Markov chain compared to when doing independent trials, then the correction term is positive, meaning that the Markov dependence increases the variance on the binomial experiment. As discussed above, the variance increases sharply for $\lambda_j \rightarrow 1$ and $0 < p_j < 1$. E.g., for $\lambda_j = 0.9$, $p_j = 0.5$, we see that the variance per trial is inflated from $0.25$ for the independent case to about $2.5$ for the Markov dependence case, i.e., we obtain a factor 10 variance inflation due to the Markov dependence.

\paragraph{Remark 3} The conditional probability $\lambda_j$ of staying within bin $j$ increases if particles move over shorter distances between collisions~\cite{ingelaere_prep_2023}. The observed variance increase for increasing $\lambda_j$ therefore corresponds to the experimental observations from Section~\ref{sec:comparison_variance_experiment}.

%\paragraph{Remark x} One can try to locally reduce the variance on a QoI estimate by biasing the simulation procedure in such a way that the value of $\lambda_j$ is lowered. The bias then has to be accounted for in the QoI estimator, which also affects the variance. The net result on the variance is, however, not immediately clear.

\newpage
\subsection{Binomial experiment driven by a hidden Markov process}
\label{sec:HMP}
In this section, we assume that the binomial experiment is driven by a hidden Markov process, which corresponds exactly to the setting of estimating QoI on a histogram. When we build estimators on a histogram, we are actually coarse-graining a (hidden) continuous Markov process for the position of a particle $x_l$ to a binomial experiment monitoring the presence of that particle in a bin $j$, represented by random variable $I_j(x_l)$ (an evaluation of an indicator function). This corresponds to the description of a hidden Markov model, where the presence of particles in a given bin, i.e., the binomial experiment, can be interpreted as the observed process, and the stochastic particle dynamics can be interpreted as the underlying hidden Markov process. In this case, however, the hidden Markov process is not really `hidden' as we know the particle dynamics. If the particle dynamics are governed by the transition kernel $\mathcal{K}(x_{l-1} \rightarrow x_l)$ (e.g.,~\eqref{eq:transition_kernel} for particle dynamics~\eqref{eq:particle_dynamics}), then this transition kernel can be identified to be the conditional probability density $\mathcal{P}(x_l \mid x_{l-1})$. If we then have the probability density $P_l(x_l) = \mathcal{P}(x_l = x)$, we can write the following continuous Markov process to go from a probability density on state (particle position) $x_{l-1}$ to a probability density on state (particle position) $x_l$~\cite{spanier_monte_1969}:
%cfr eq 2.4.4 in Spanier, Gelbard 1969
\begin{equation}
P_{l}(x_l) = \int \mathcal{K}(x_{l-1} \rightarrow  x_l) P_{l-1}(x_{l-1})dx_{l-1}.
\label{eq:binomial_continuous_markov_process}
\end{equation}
This Markov process (a recursion relation that only depends on state $x_{l-1}$, not on older states) is a continuous equivalent of the discrete Markov process~\eqref{eq:binomial_discret_markov_process}. 
We can use expression~\eqref{eq:binomial_continuous_markov_process} to calculate the different terms in the expression for the variance of $S_{L,j}$~\eqref{eq:variance_on_nj}. The first term in that expression becomes
\begin{equation}
\begin{split}
\sum_{l=1}^L \langle I_j(x_l) \rangle &= \sum_{l=1}^L \mathcal{P}(I_j(x_l) = 1)\\
&= \sum_{l=1}^L \int \hdots \int I_j(x_l) \mathcal{K}(x_{l-1} \rightarrow x_l) \hdots \mathcal{K}(x_1 \rightarrow x_2)P_1(x_1)dx_1 \hdots dx_l,
\end{split}
\label{eq:binomial_hmm_firstterm_variance_on_nj}
\end{equation} 
i.e., for each value of $l$, we take the probability density $P_1(x_1)$ on the initial state, propagate it using the transition kernels to the probability density on the $l$-th state, and then integrate that probability density over bin $j$, indicated by indicator function $I_j(x_l)$, to count the expected fraction of Markov chains that have state $x_l$ in bin $j$. Following a similar reasoning, we find the analog for~\eqref{eq:binomial_markovdependence_secondterm_variance_of_nj} to be
\begin{equation}
\begin{split}
\langle I_j(x_l) I_j(x_m) \rangle &= \mathcal{P}(I_j(x_l) = 1, I_j(x_m) = 1)\\
&= \int \hdots \int I_j(x_l)I_j(x_m)\mathcal{K}(x_{l-1} \rightarrow x_l) \hdots \mathcal{K}(x_1 \rightarrow x_2)P_1(x_1)dx_1 \hdots dx_l,
\end{split}
\label{eq:binomial_hmm_secondterm_variance_on_nj}
\end{equation}
with $m < l$.
The main difference between~\eqref{eq:binomial_markovdependence_secondterm_variance_of_nj} and~\eqref{eq:binomial_hmm_secondterm_variance_on_nj} is that in the former, because of the Markov dependence, the probabilities for the current state only depend on the previous state, while in the latter, the coarse-grained (hidden) Markov process, the random variables $I_j(x_l)$ do not necessarily have the Markov property anymore, meaning that we need to take all the transitions of the (hidden) continuous Markov process into account starting from the initial state~\cite{macdonald_coarse-grained_2020}. 

The high-dimensional integrals~\eqref{eq:binomial_hmm_firstterm_variance_on_nj},~\eqref{eq:binomial_hmm_secondterm_variance_on_nj} are typically not computable analytically. Instead, we discretize the (hidden) continuous Markov process by subdividing the continuous phase space into small cells and using these cells as discrete states. We define the vector $P_l$ containing the probability of position $x_l$ being in a certain cell. We assume that particles are distributed uniformly in each of the cells, which enables the computation of transition probabilities between the different cells, as explained in Ref.~\cite{ingelaere_prep_2023}. These transition probabilities are stored in a transition matrix $K_d(P_{l-1} \rightarrow P_l)$, a discretized counterpart to the transition kernel $\mathcal{K}(x_{l-1} \rightarrow  x_l)$. The indicator function $I_j(x_l)$ becomes a diagonal matrix $\mathbb{I}_{l,j}$ with the value 1 on the diagonal elements corresponding to the cells that belong to bin $j$, and the value 0 on the diagonal elements corresponding to the cells that are outside bin $j$. The probability density $P_1(x)$ becomes a vector $P_1$ with at each entry the probability of the initial state being in the cell corresponding to that element. We obtain the following discrete Markov process:
\begin{equation}
P_l = K_d(P_{l-1} \rightarrow P_l) P_{l-1}.
\label{eq:discretized_transition_kernel}
\end{equation}
The discretized versions of expressions~\eqref{eq:binomial_hmm_firstterm_variance_on_nj},~\eqref{eq:binomial_hmm_secondterm_variance_on_nj}  are as follows~\cite{macdonald_coarse-grained_2020}:
\begin{equation}
\begin{split}
\sum_{l=1}^L \langle I_j(x_l) \rangle &= \sum_{l=1}^L \mathcal{P}(I_j(x_l) = 1)\\
&= \sum_{l=1}^L \left( \boldsymbol{1}' \mathbb{I}_{l,j} \left( \prod_{i=2}^l K_d(P_{i-1} \rightarrow P_i) \right) P_1 \right),
\end{split}
\label{eq:binomial_hmmdiscretized_firstterm_variance_on_nj}
\end{equation} 

\begin{equation}
\begin{split}
\langle I_j(x_l) I_j(x_m) \rangle &= \mathcal{P}(I_j(x_l) = 1, I_j(x_m) = 1)\\
&= \boldsymbol{1}' \mathbb{I}_{l,j} \left( \prod_{i=m+1}^l K_d(P_{i-1} \rightarrow P_i) \right)\mathbb{I}_{l,j} \left( \prod_{i=2}^m K_d(P_{i-1} \rightarrow P_i) \right) P_1,
\end{split}
\label{eq:binomial_hmmdiscretized_secondterm_variance_on_nj}
\end{equation}
where $\boldsymbol{1}$ is a vector of ones. To fully appreciate the difference with the Markov dependence case from Section~\ref{subsec:markov_process}, equation~\eqref{eq:binomial_hmmdiscretized_secondterm_variance_on_nj} should be compared to~\eqref{eq:binomial_markovdependence_secondterm_variance_of_nj}.

The high-dimensional integrals~\eqref{eq:binomial_hmm_firstterm_variance_on_nj},~\eqref{eq:binomial_hmm_secondterm_variance_on_nj} and their discretizations~\eqref{eq:binomial_hmmdiscretized_firstterm_variance_on_nj},~\eqref{eq:binomial_hmmdiscretized_secondterm_variance_on_nj} are expensive to compute and therefore unfit for a priori variance prediction. There is, however, a way to obtain a cheap approximate expression for the variance on $S_{L,j}$. A typical simplification for a coarse-grained (hidden) Markov process, is to only consider the Markovian component of the observed dynamics~\cite{andrieux_bounding_2012}. We then retrieve the case of a binomial experiment driven by a Markov process discussed in Section~\ref{subsec:markov_process}, which means that we can use the expression~\eqref{eq:binomial_variance_markovdependence_klotz} as a cheap predictor of the variance on $S_{L,j}$, provided that we have expressions for $p_j$ and $\lambda_j$.

\paragraph{Remark 4} The discretized transition kernel $K_d(P_{l-1} \rightarrow P_l)$ can incorporate information about boundary conditions. Our implementation for the numerical experiments, for instance, incorporates periodic boundary conditions. For details, we refer to the code~\cite{zenodo}.

\subsection{Numerical experiments}
\label{sec:verification_binomial_experiments}
We test the derived variance expressions by applying them to the velocity-jump process~\eqref{eq:particle_dynamics} for a range of values for $R$ and $\sigma_p^2$, while setting $u_p, R_i = 0$. For each set of parameters, we simulate a single particle for $L=10^3$ collisions and count how many collision points (Bernoulli trials) are in a bin $j$, which corresponds to a sum of the form~\eqref{eq:general_binomial_experiment}. The test cases are all executed on a 1D ($d=1$) domain $D = [0,1]$ with length 1 and periodic boundary conditions. All particles are launched at time $t=0$, with an initial position distributed uniformly over the domain and a velocity drawn from the post-collisional velocity distribution given in~\eqref{eq:particle_dynamics}. The space is subdivided in $J=10$ bins, such that the probability for success equals $p_j = \frac{1}{J} = 0.1$. The subdomain $D_j = [0,0.1]$ is considered as the bin of interest. We compare the upper bound~\eqref{eq:binomial_variance_simple_upperbound} for the variance, the variance expression for independent trials~\eqref{eq:binomial_variance_independent_stationary}, the Markov process based variance expression~\eqref{eq:binomial_variance_markovdependence_klotz}, and the discretized hidden Markov process based variance expression. The hidden Markov process is discretized on an equidistant grid consisting of 100 cells. The conditional probability $\lambda_j$ for the Markov process and transition probabilities for the discretized hidden Markov process can be calculated as a function of the model parameters as described in Ref.~\cite{ingelaere_prep_2023}. The code is openly available in \reponame.

\begin{figure}
\centering
\begin{subfigure}{0.32\linewidth}
\centering
\includegraphics[width=\linewidth]{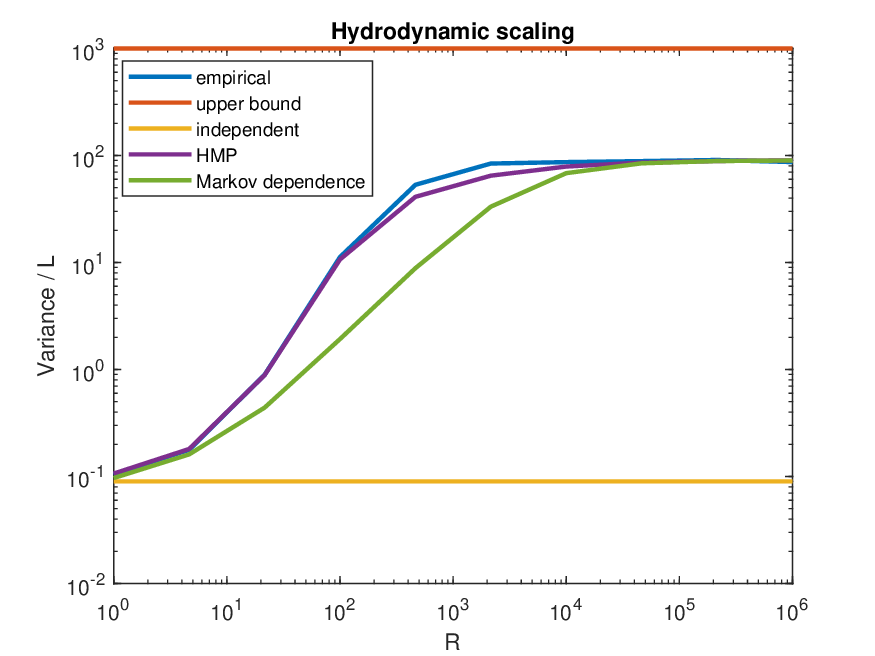}
\end{subfigure}
\begin{subfigure}{0.32\linewidth}
\centering
\includegraphics[width=\linewidth]{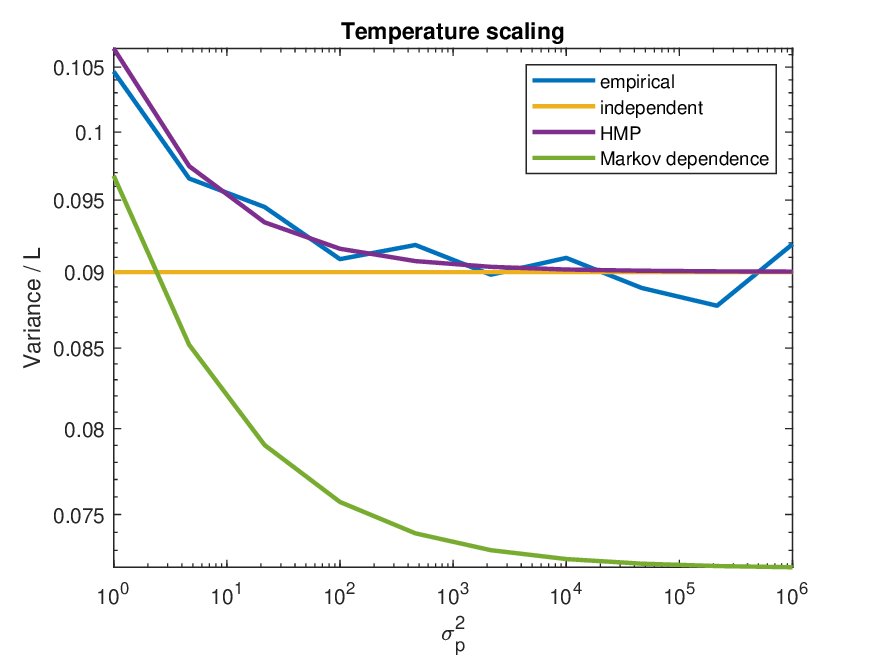}
\end{subfigure}
\begin{subfigure}{0.32\linewidth}
\centering
\includegraphics[width=\linewidth]{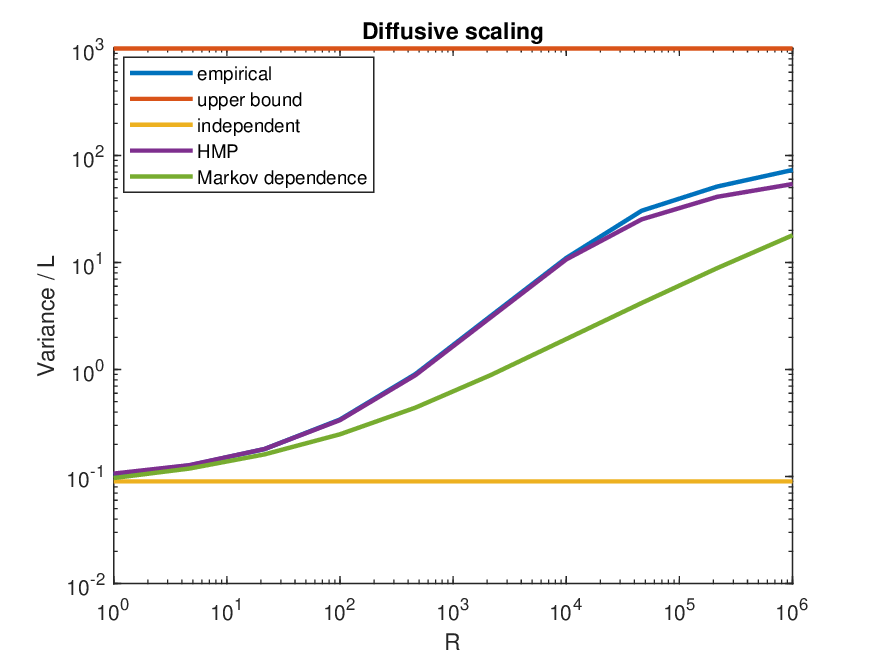}
\end{subfigure}
\caption{Variance divided by $L$ for $L=10^3$ Bernoulli trials. Left: hydrodynamic scaling. Middle: temperature scaling. Right: diffusive scaling. The empirical variance (blue) is calculated based on $10^4$ realizations. The upper bound (red) is given by~\eqref{eq:binomial_variance_simple_upperbound}. The expression for independent Bernoulli trials (yellow) is given by~\eqref{eq:binomial_variance_independent_stationary}. The used Markov dependence approximation (green) is given by~\eqref{eq:binomial_variance_markovdependence_klotz}. The variance expression based on a discretized hidden Markov process (purple) is explained in Section~\ref{sec:HMP}.}
\label{fig:NeutralBinomial_scalings}
\end{figure}

We consider three experiments: the first experiment is in the so-called hydrodynamic scaling (vary $R$ over orders of magnitude while keeping $\sigma_p^2 \sim \mathcal{O}(1)$); the second experiment is in the so-called temperature scaling (vary $\sigma_p^2$ over orders of magnitude while keeping $R \sim \mathcal{O}(1)$); and the third experiment is in the so-called diffusive scaling (choose $\sigma_p^2 = R$ and vary over orders of magnitude). As a reference solution, we determine for each set of parameters the empirical variance based on $10^4$ realizations of the experiment. We plot the different variance predictors divided by $L$ for these three scalings in Figure~\ref{fig:NeutralBinomial_scalings}, where we each time use $\sigma_p^2 = 1$ in the hydrodynamic scaling and $R=1$ in the temperature scaling. The upper bound for the variance is omitted in the temperature scaling plot, to zoom in on the other variance expressions.

As expected, parameter settings that result in particles being able to travel longer distances (low $R$, high $\sigma_p^2$) correspond to lower variances, while parameter settings that result in particles being able to travel shorter distances (high $R$, low $\sigma_p^2$) correspond to higher variances. The discretized hidden Markov process based expression correctly calculates the empirical variance, except for a small discretization error. In each plot, we can see that using the Markov process based expression for the variance~\eqref{eq:binomial_variance_markovdependence_klotz} is an enormous improvement over the assumption of independent trials, except in the temperature scaling. 

The peculiar behaviour in the temperature scaling can be explained by a boundary condition effect as follows. Due to increasing $\sigma_p^2$ for constant $R$, particles can travel further and further in one step. Because of the periodic boundary conditions, this means that at some point particles can fly around the whole domain returning to the bin where they left. In the limit for very large step sizes, the periodic boundary conditions make that the variance on the binomial experiment evolves to the case with independent samples, i.e., to $\lambda_j = p_j$. The conditional probability $\lambda_j$, however, is calculated as described in Ref.~\cite{ingelaere_prep_2023} and does not take the periodic boundary conditions into account. The calculated $\lambda_j$ therefore corresponds to the case in which a particle that escapes from a cell cannot end up in that cell again, leading to $\lambda_j < p_j$ and an underestimation of the variance. Taking boundary conditions into account in the variance predictors is left for future work.

In the hydrodynamic scaling for $R \rightarrow \infty$, the variance on the binomial experiment reaches a plateau, which is also predicted by the Markov process based variance expression~\eqref{eq:binomial_variance_markovdependence_klotz}. When $R$ is large enough, the particles can only move within one bin, i.e., we reach the limit $\lambda_j \rightarrow 1$, meaning that expression~\eqref{eq:binomial_variance_markovdependence_klotz_largeL} holds for the variance on the binomial experiment. This experiment indicates that~\eqref{eq:binomial_variance_markovdependence_klotz_largeL} constitutes a more practical worst-case variance predictor for a binomial experiment than the upper bound~\eqref{eq:binomial_variance_simple_upperbound}, given a good estimate for $p_j$.

\subsection{Towards variance expressions for QoI estimators}
The expressions obtained above deal with the variance on sums $S_{L,j}$ of the form~\eqref{eq:general_binomial_experiment}.
The QoI estimators~\eqref{eq:QoI_Point_Estimator} and~\eqref{eq:QoI_Analog_Estimator}, however, contain more general sums of the form
\begin{equation}
S_{L,x,j} = w \sum_{l=1}^L q_x(v_l) I_j(x_l),
\end{equation}
where for the lowest order moments~\eqref{eq:QoIs}, we have $q_\rho(v_l) = 1$, $q_m(v_l) =  v_l$, and $q_E(v_l) = |v_l|^2/2$. We compute the variance on such a sum for a fixed value of $L$ by performing a similar derivation as in~\eqref{eq:variance_on_nj}. Using the fact that the velocities $v_l$ are iid, and that $v_l$ and $x_l$ are independent and therefore functions of $x_l$ and functions of $v_l$ are also independent, we obtain
\begin{equation}
\mathbb{V}[S_{L,x,j}] = w^2 \mathbb{V}[q_x(v_l)] \langle S_{L,j} \rangle + w^2 \langle q_x(v_l) \rangle^2 \mathbb{V}[S_{L,j}].
\label{eq:QoI_Generic_Variance_Formula}
\end{equation}
The statistics on $q_x(v_l)$ can be calculated given the probability density $M(v_l)$ in~\eqref{eq:particle_dynamics}. For the three lowest order moments~\eqref{eq:QoIs}, we obtain
\begin{equation}
\begin{split}
&\langle q_\rho(v_l) \rangle = 1, \quad\hspace{2mm} \langle q_m(v_l) \rangle = |u_p|, \quad\hspace{2mm} \langle q_E(v_l) \rangle = \frac{1}{2} (|u_p|^2 + \sigma_p^2),\\
&\mathbb{V}[q_\rho(v_l)] = 0, \quad \mathbb{V}[q_m(v_l)] = \sigma_p^2, \quad \mathbb{V}[q_E(v_l)] = |u_p|^2 \sigma_p^2 + \frac{1}{2} \sigma_p^4.
\end{split}
\label{eq:statistics_QoIs}
\end{equation}
From above, we have an upper bound~\eqref{eq:binomial_variance_simple_upperbound}, an expression for independent trials~\eqref{eq:binomial_variance_independent_stationary}, Markov process based expressions, and hidden Markov process based expressions for the variance $\mathbb{V}[S_{L,j}]$. We therefore have all the required ingredients to construct variance predictors for analog particle tracing Monte Carlo methods with point estimators and analog estimators.

\section{Predicting the variance of analog particle tracing Monte Carlo with point estimators}
\label{sec: point estimators}
For point estimators~\eqref{eq:QoI_Point_Estimator}, the contributions (Bernoulli trials) are independent. The variance on the point estimator $\hat{Q}_j(T)$ can be written in the form~\eqref{eq:QoI_Generic_Variance_Formula} as follows:
\begin{equation}
\begin{split}
\mathbb{V}[\hat{Q}_j(T)] &= w^2 \mathbb{V}[q_x(v_n(T))] \langle \sum_{n=1}^N I_j(x_n(T)) \rangle + w^2 \langle q_x(v_n(T))\rangle^2 \mathbb{V}[\sum_{n=1}^N I_j(x_n(T))]\\
&= w^2 N p_j \mathbb{V}[q_x(v_n(T))] + w^2 \langle q_x(v_n(T))\rangle^2 \mathbb{V}[\sum_{n=1}^N I_j(x_n(T))],
\end{split}
\end{equation}
where the last equality only holds for a stationary probability for success $p_j$. Because of the independent contributions, we can use expression~\eqref{eq:binomial_variance_independent_stationary} for the variance on the binomial experiment with independent trials:
\begin{equation}
\mathbb{V}[\hat{Q}_j(T)] = w^2 N p_j \mathbb{V}[q_x(v_n(T))] + w^2 N p_j(1-p_j) \langle q_x(v_n(T))\rangle^2.
\label{eq:var_predictor_point_estimators_general}
\end{equation}
Working out this expression explicitly for the three lowest order moments~\eqref{eq:QoIs} by inserting~\eqref{eq:statistics_QoIs} results in
\begin{equation}
\begin{split}
\mathbb{V}[\hat{\rho}_j(T)] &= w^2 N p_j (1-p_j),\\
\mathbb{V}[\hat{m}_j(T)] &= w^2 N p_j \sigma_p^2 + w^2 N p_j (1-p_j) |u_p|^2,\\
\mathbb{V}[\hat{E}_j(T)] &= w^2 N p_j (|u_p|^2 \sigma_p^2 +  \frac{1}{2} \sigma_p^4) + \frac{1}{4} w^2 N p_j (1-p_j) (|u_p|^2 + \sigma_p^2)^2.
\end{split}
\label{eq:var_predictor_point_estimators}
\end{equation}
Recall that the particle weight $w$ depends on the number of particles $N$.
Inserting expression~\eqref{eq:particle_weight} for the particle weight $w$ shows that the variances scale with $\frac{1}{N}$, as they should (see Section~\ref{subsec: QoI estimation}).

\subsection{Influence of sources and sinks on probability for success $p_j$}
If there are no sinks ($R_i = 0$), then we can select a probability for success $p_j = \bar{p}_j$, e.g., by assuming that a success in each bin $j$ occurs with the same probability: $\bar{p}_j = \frac{1}{J}$, such that $\sum_{j=1}^J p_j = \sum_{j=1}^J \bar{p}_j = 1$.
If there is an ionization sink, then we have that $\sum_{j=1}^J p_j < 1$, because a particle can disappear in the sink before reaching time $T$. We can incorporate the sink in the probability for success as follows:
\begin{equation}
\begin{split}
p_j &= \mathcal{P}(\text{particle ends up in cell $j$ and is not ionized})\\
&= \mathcal{P}(\text{particle ends up in cell $j$}) \cdot \mathcal{P}(\text{particle is not ionized})\\
&= \bar{p}_j \cdot \exp(-R_i T).
\end{split}
\label{eq:point_estimator_success_Probability_sinks}
\end{equation}
Note that each particle still has the same probability for success $p_j$. When the particles originate from a stationary particle source, as described in Section~\ref{sec: binomial experiments dynamics}, then a particle $n$ that survives (does not get ionized) up to time $T$ spends a time $\Delta t_n \sim U[0,T]$ in the simulation. In that case, the probability for success $p_{n,j} = \bar{p}_j \cdot \exp(-R_i \Delta t_n)$ becomes particle dependent, because each particle has to survive for a different duration $\Delta t_n$, which means that they have a different probability of being ionized before time $T$. As the probability distribution for $\Delta t_n$ is known, we can calculate the following stationary probability for success:
\begin{equation}
\begin{split}
p_j = \langle p_{n,j} \rangle &= \int_0^T \bar{p}_j \cdot \exp(-R_i \delta) \cdot \mathcal{P}(\Delta t_n = \delta) d \delta\\ 
&= \int_0^T \bar{p}_j \cdot \exp(-R_i \delta) \cdot \frac{1}{T} d\delta\\
&= \bar{p}_j \cdot \frac{1-\exp(-R_i T)}{R_i T}.
\end{split}
\label{eq:point_estimator_success_Probability_combi}
\end{equation}
This stationary probability for success $p_j$ can be used in variance expression~\eqref{eq:var_predictor_point_estimators_general} for simulations including sources and sinks.

\subsection{Numerical experiments}
We test the point estimator variance predictors for the three lowest order moments~\eqref{eq:var_predictor_point_estimators} for a range of values for $R$ and $\sigma_p^2$. For each set of parameters, we simulate $N=10^2$ particles and count one contribution to each of the point estimators for each particle (Bernoulli trial) that ends up in a bin $j$ at a time $T$, where $T$ is chosen such that $RT = 10$. We repeat the experiments once with $R_i = 0$, and once with $R_i = 1$. The test cases are all executed on a 1D ($d=1$) domain $D = [0,1]$ with length 1 and periodic boundary conditions. The particles are launched from a stationary particle source, with an initial position distributed uniformly over the domain and a velocity drawn from the post-collisional velocity distribution as given in~\eqref{eq:particle_dynamics}. The space is subdivided in $J=10$ bins, such that the probability for success~\eqref{eq:point_estimator_success_Probability_combi} equals
\begin{equation}
p_j = \frac{1}{J} \cdot \frac{1-\exp(-R_i T)}{R_i T} = 0.1 \cdot \frac{1-\exp(-R_i T)}{R_i T}.
\end{equation}  
The subdomain $D_j = [0,0.1]$ is considered as the bin of interest. The code is openly available in \reponame.

\begin{figure}
\centering
\begin{subfigure}{0.32\linewidth}
\centering
\includegraphics[width=\linewidth]{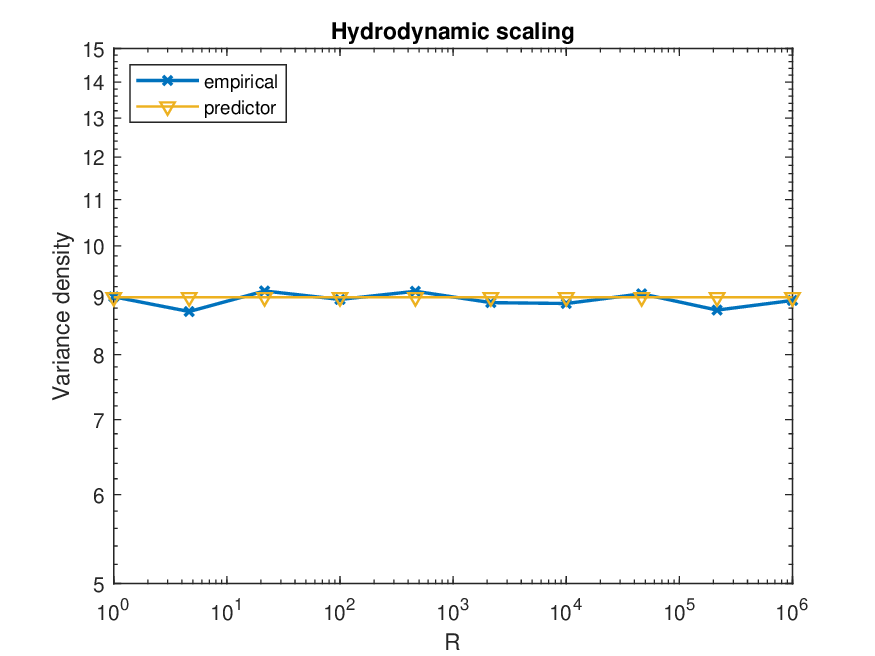}
\end{subfigure}
\begin{subfigure}{0.32\linewidth}
\centering
\includegraphics[width=\linewidth]{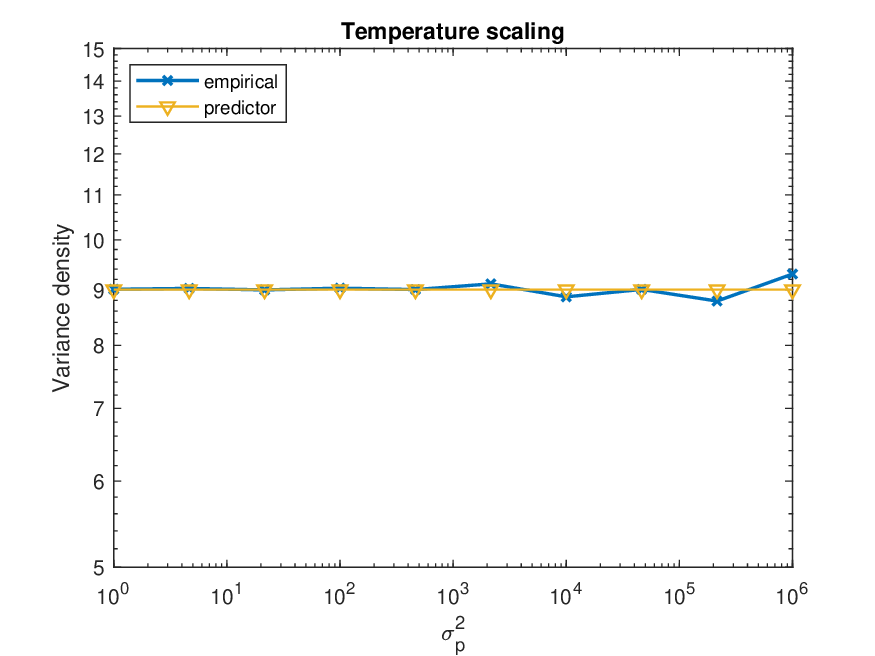}
\end{subfigure}
\begin{subfigure}{0.32\linewidth}
\centering
\includegraphics[width=\linewidth]{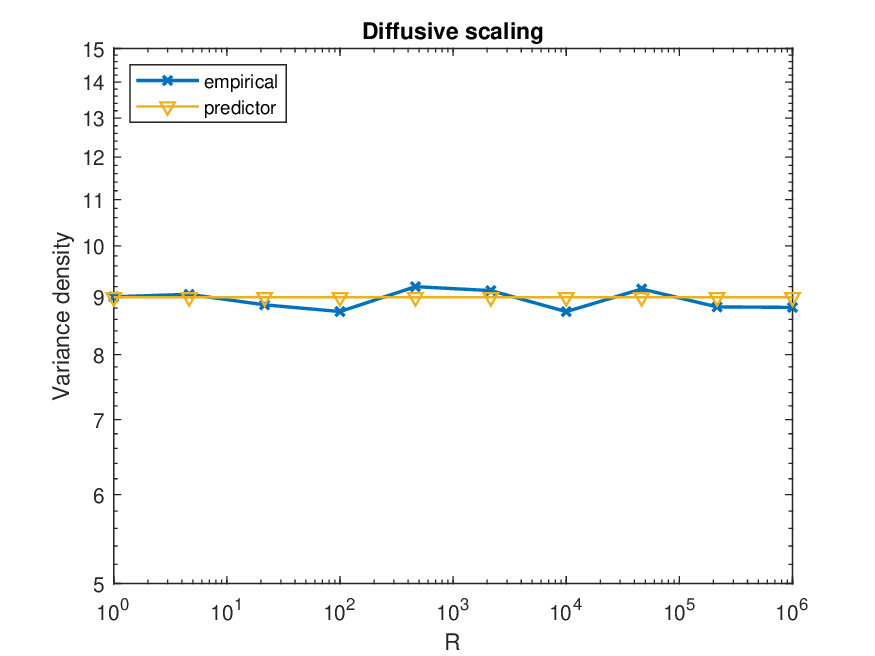}
\end{subfigure}
\begin{subfigure}{0.32\linewidth}
\centering
\includegraphics[width=\linewidth]{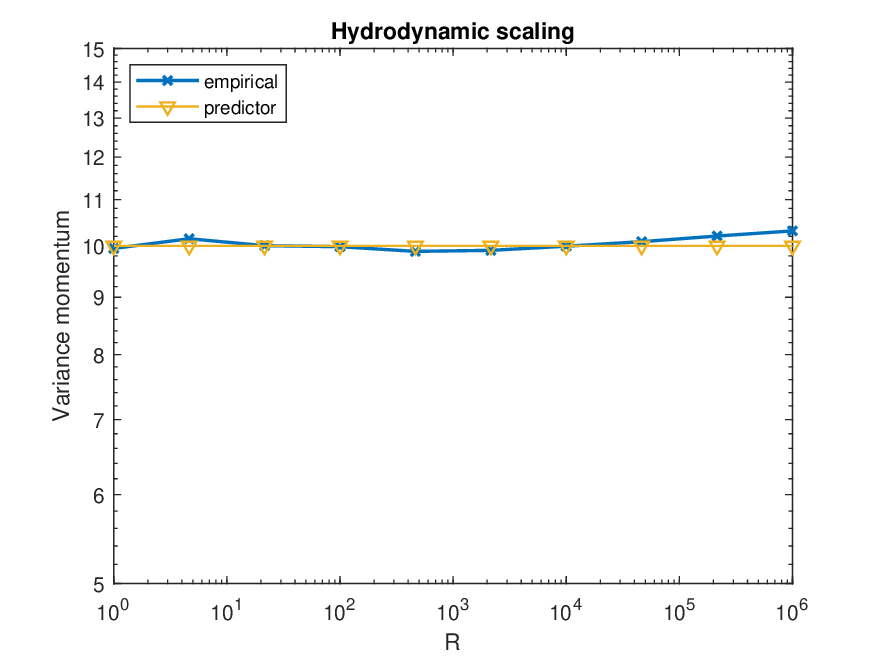}
\end{subfigure}
\begin{subfigure}{0.32\linewidth}
\centering
\includegraphics[width=\linewidth]{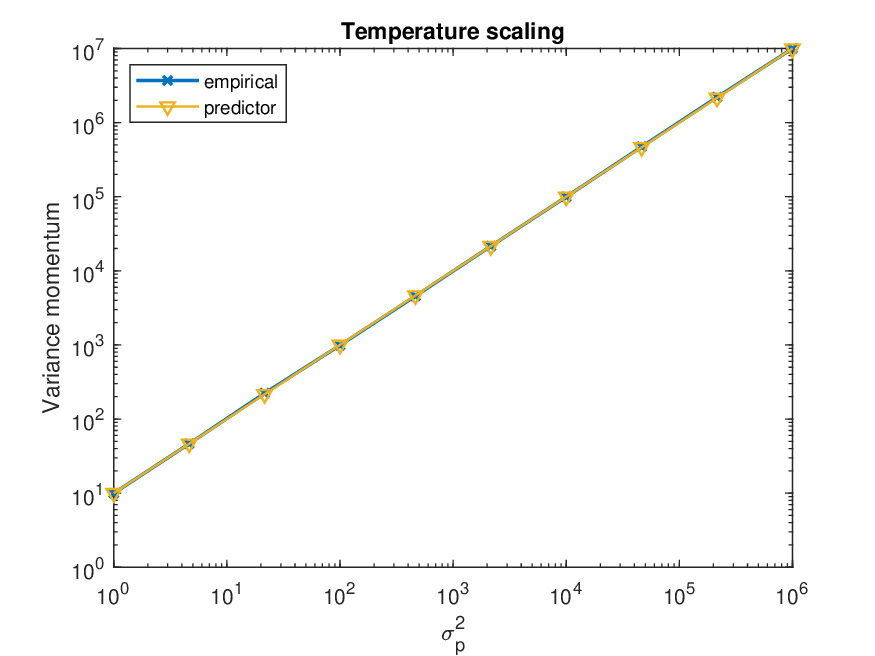}
\end{subfigure}
\begin{subfigure}{0.32\linewidth}
\centering
\includegraphics[width=\linewidth]{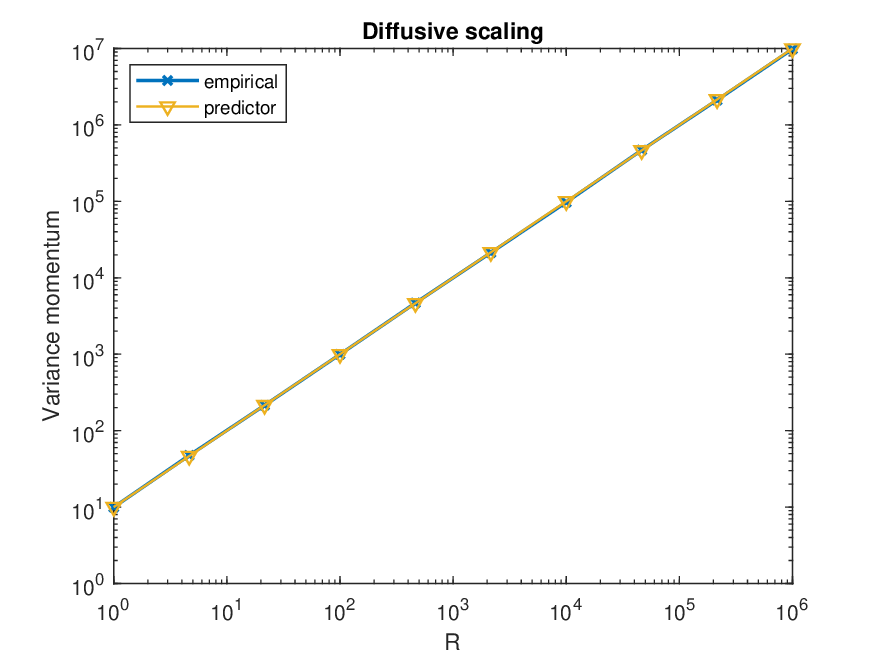}
\end{subfigure}
\begin{subfigure}{0.32\linewidth}
\centering
\includegraphics[width=\linewidth]{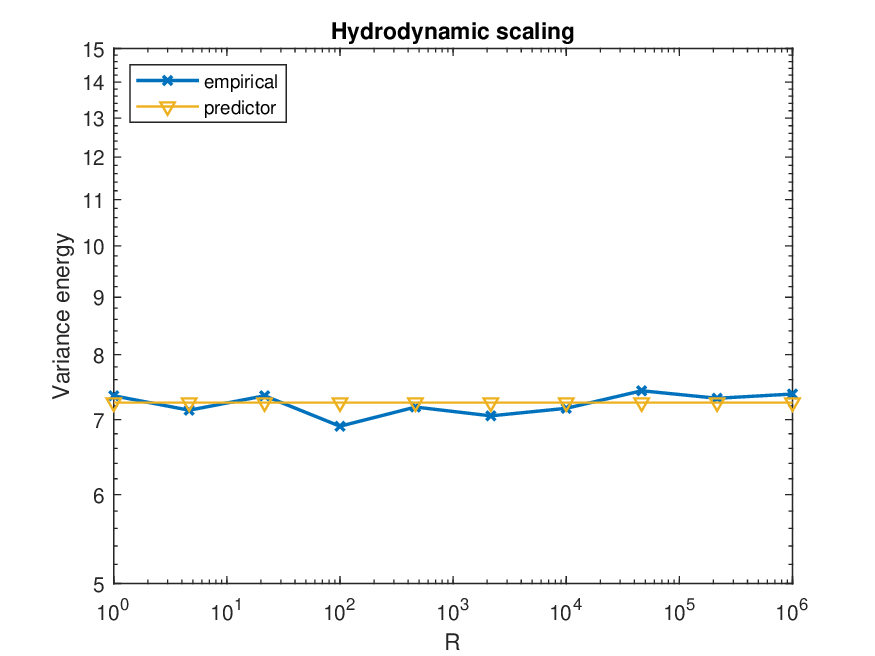}
\end{subfigure}
\begin{subfigure}{0.32\linewidth}
\centering
\includegraphics[width=\linewidth]{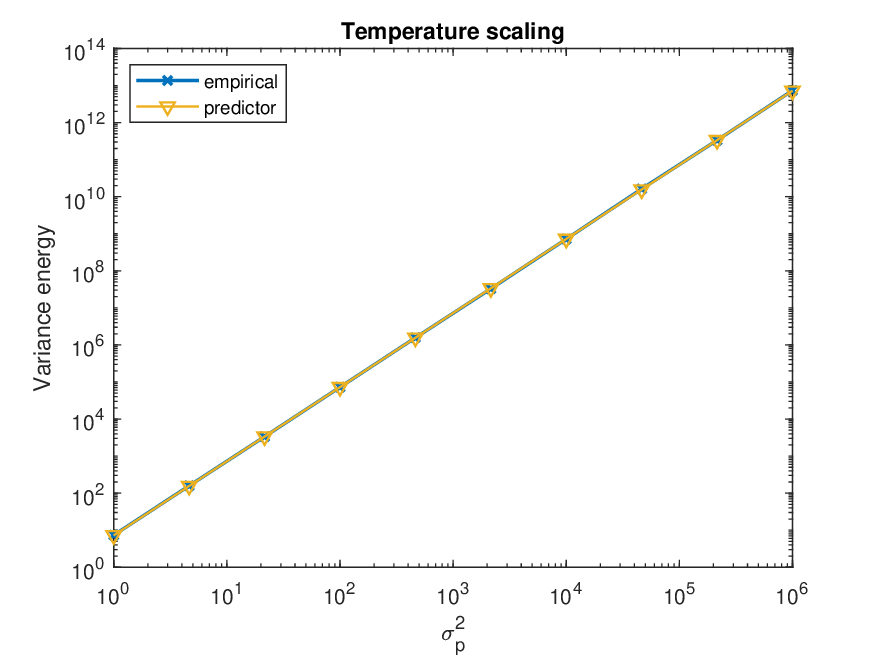}
\end{subfigure}
\begin{subfigure}{0.32\linewidth}
\centering
\includegraphics[width=\linewidth]{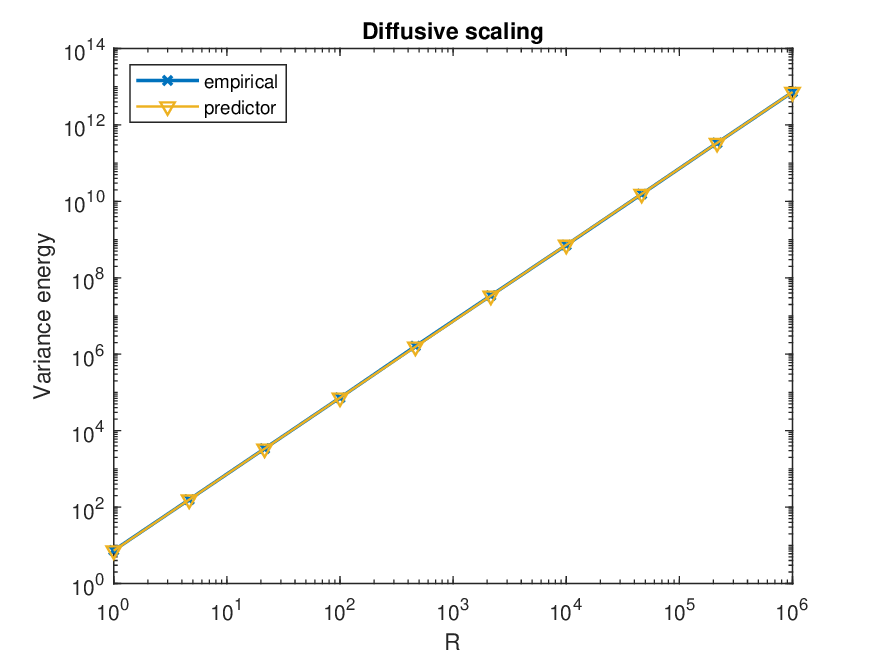}
\end{subfigure}
\caption{Analog particle tracing Monte Carlo with point estimator. Particles are generated from a random source that is uniform in time, there is no ionization sink ($R_i=0$). Top to bottom: variance on density, momentum and energy. Left to right: hydrodynamic, temperature and diffusive scaling.}
\label{fig:point_estimators_u0}
\end{figure}

\begin{figure}
\centering
\begin{subfigure}{0.32\linewidth}
\centering
\includegraphics[width=\linewidth]{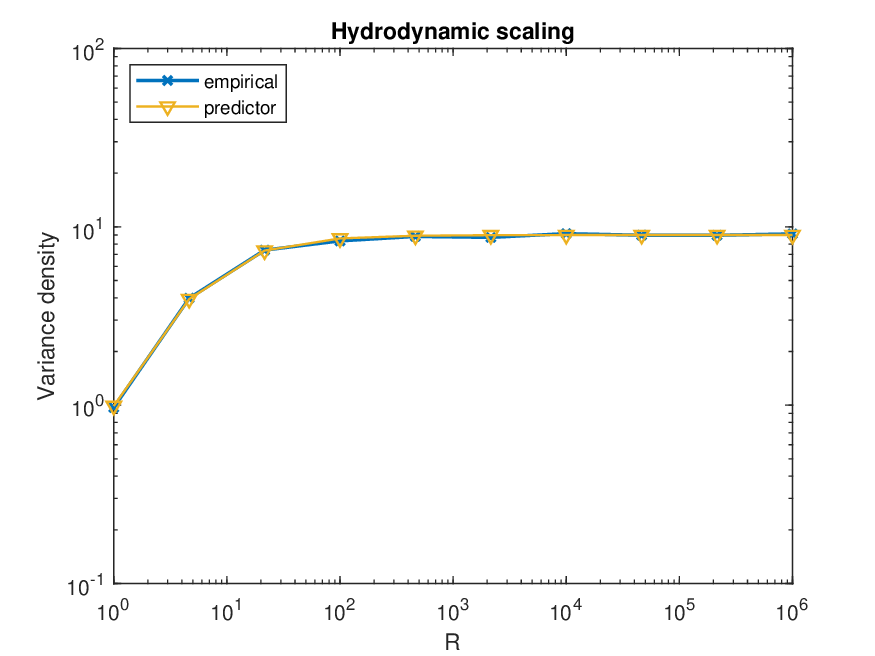}
\end{subfigure}
\begin{subfigure}{0.32\linewidth}
\centering
\includegraphics[width=\linewidth]{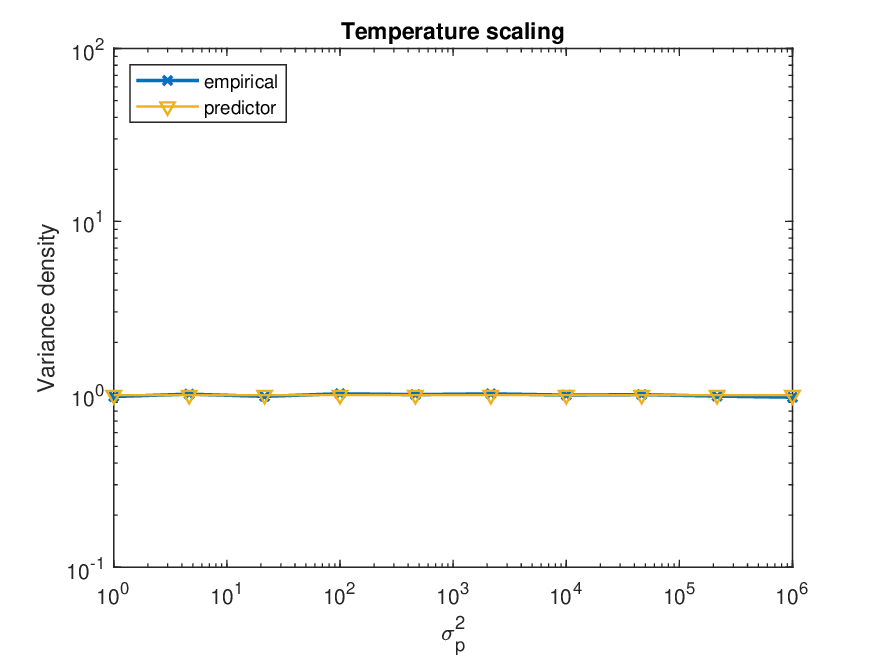}
\end{subfigure}
\begin{subfigure}{0.32\linewidth}
\centering
\includegraphics[width=\linewidth]{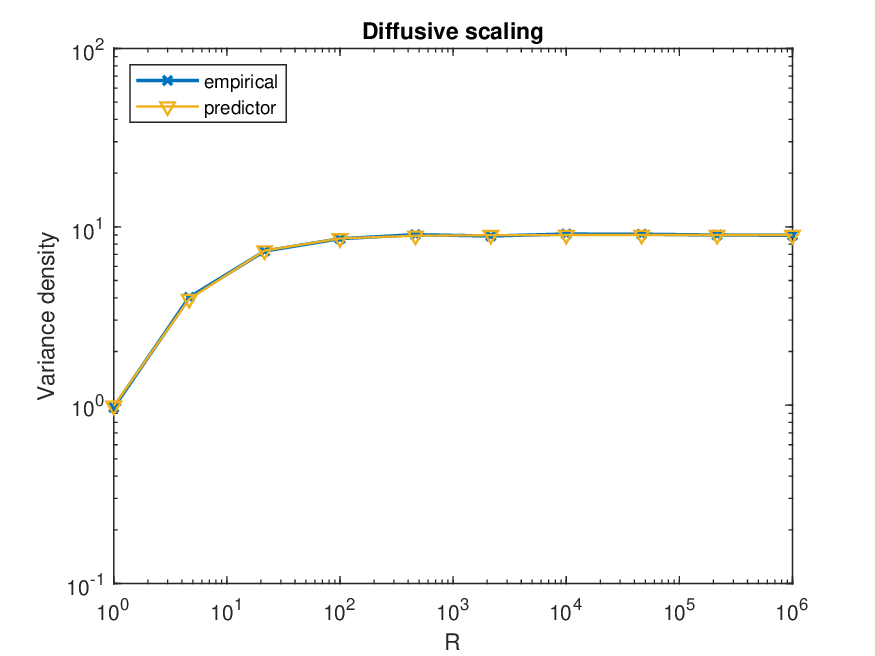}
\end{subfigure}
\begin{subfigure}{0.32\linewidth}
\centering
\includegraphics[width=\linewidth]{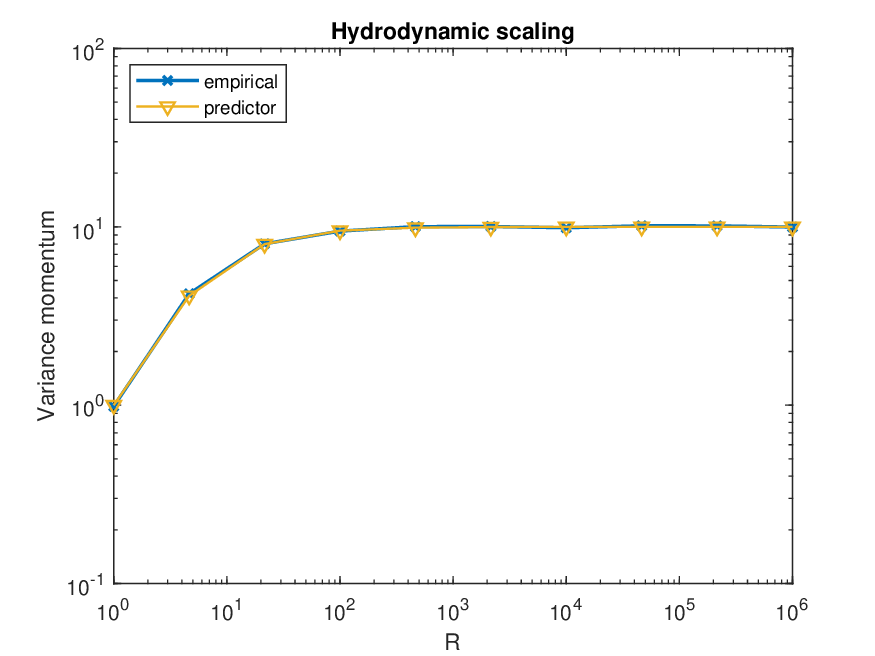}
\end{subfigure}
\begin{subfigure}{0.32\linewidth}
\centering
\includegraphics[width=\linewidth]{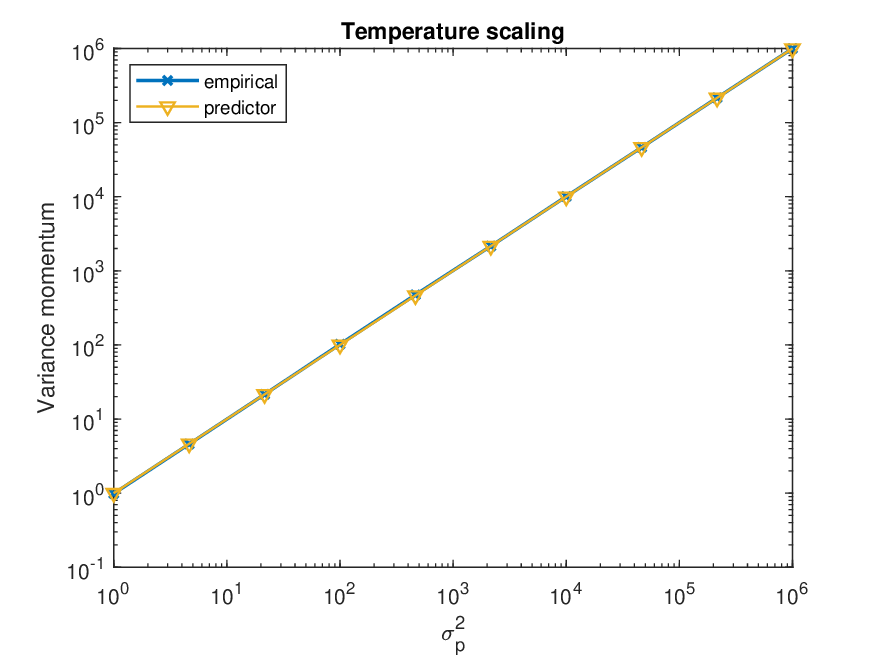}
\end{subfigure}
\begin{subfigure}{0.32\linewidth}
\centering
\includegraphics[width=\linewidth]{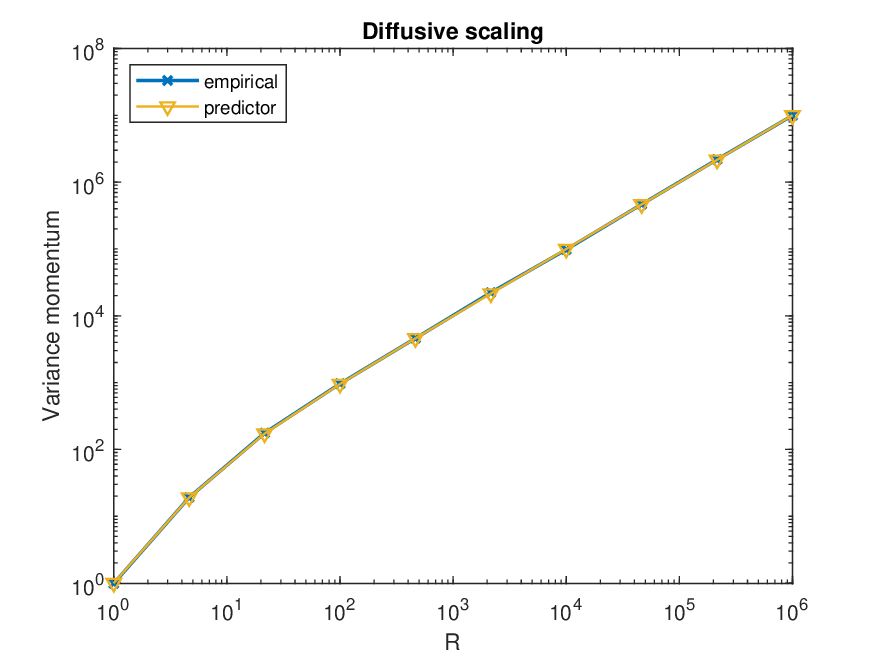}
\end{subfigure}
\begin{subfigure}{0.32\linewidth}
\centering
\includegraphics[width=\linewidth]{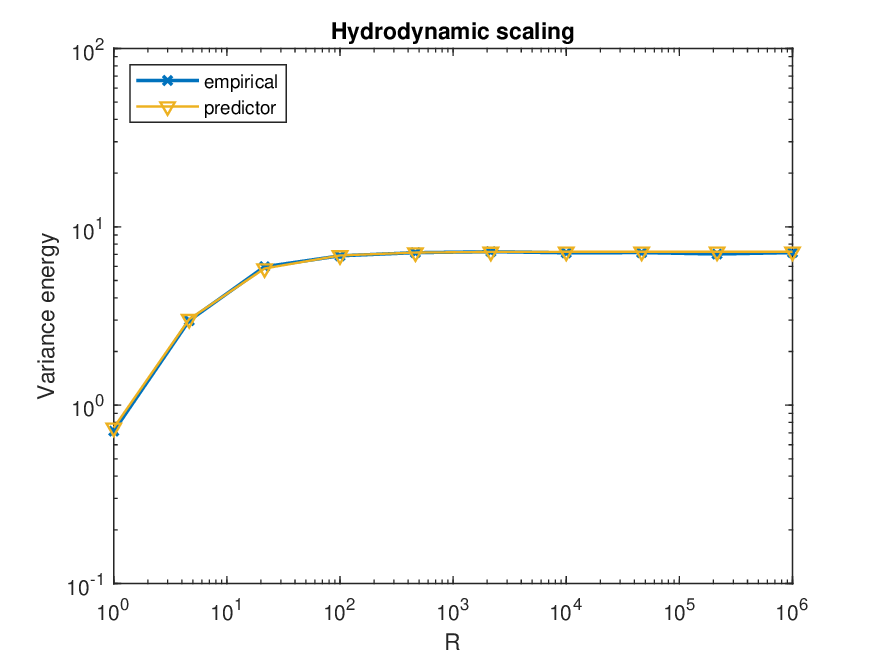}
\end{subfigure}
\begin{subfigure}{0.32\linewidth}
\centering
\includegraphics[width=\linewidth]{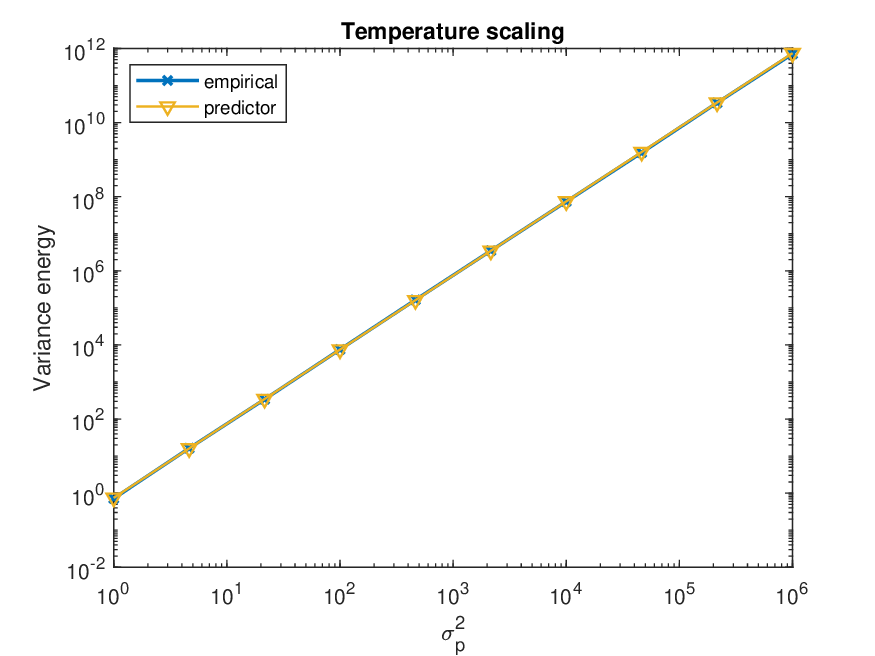}
\end{subfigure}
\begin{subfigure}{0.32\linewidth}
\centering
\includegraphics[width=\linewidth]{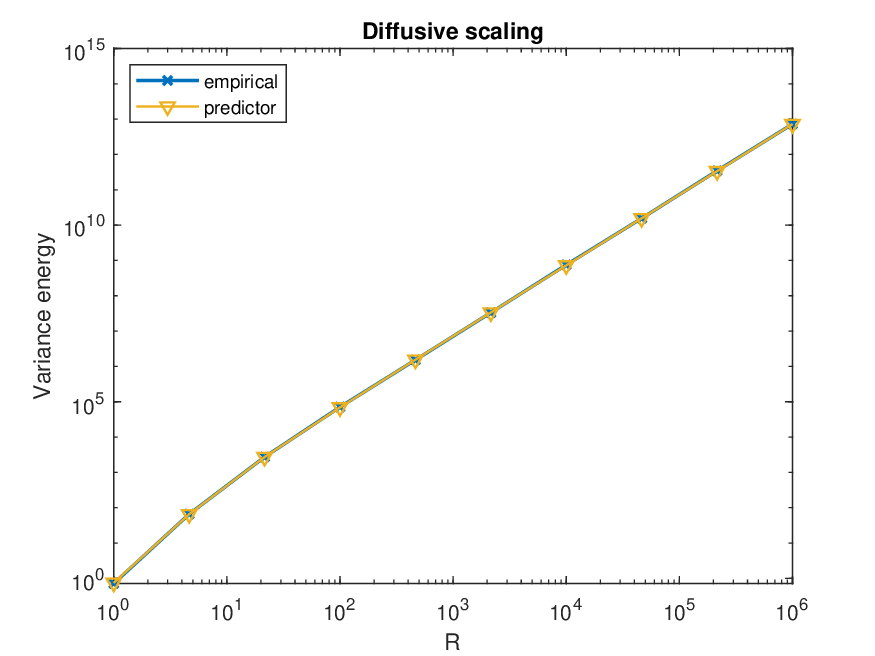}
\end{subfigure}
\caption{Analog particle tracing Monte Carlo with point estimator. Particles are generated from a random source that is uniform in time, there is an ionization sink with $R_i = 1$. Top to bottom: variance on density, momentum and energy. Left to right: hydrodynamic, temperature and diffusive scaling.}
\label{fig:point_estimators_Ri_u0}
\end{figure}

We consider the same three experiments as in Section~\ref{sec:verification_binomial_experiments}: the hydrodynamic scaling experiment, the temperature scaling experiment, and the diffusive scaling experiment. In each scaling, we estimate the variance on the three lowest order moments: particle density, momentum, and energy. As a reference solution, we determine for each set of parameters the empirical variance based on $10^4$ realizations of the experiment. We plot the variance predictors divided by $N$ for these three scalings in Figure~\ref{fig:point_estimators_u0} (with $R_i=0$) and Figure~\ref{fig:point_estimators_Ri_u0} (with $R_i=1$), where we each time choose $u_p = 0$, and use $\sigma_p^2 = 1$ in the hydrodynamic scaling and $R=1$ in the temperature scaling. In the plots where the variance remains constant for different values of $R$ and $\sigma_p^2$, the statistical noise on the empirical variance is visible. In each plot, the variance predictor corresponds well to the empirical values for each QoI.

\newpage
\section{Predicting the variance of analog particle tracing Monte Carlo with analog estimators}
\label{sec: analog estimators}
For analog estimators~\eqref{eq:QoI_Analog_Estimator}, the different particles are iid, but the contributions (Bernoulli trials) delivered by each particle are correlated. Using the knowledge that the correlation follows from the positions $x_{n,k}$ (the current position $x_{n,k}$ depends on previous positions $x_{n,m}$ with $0 \leq m < k$), while the velocities $v_{n,k}$ are iid and position independent, the variance on an analog estimator can be written in a similar form as~\eqref{eq:QoI_Generic_Variance_Formula} as follows:
\begin{equation}
\mathbb{V}[\hat{Q}_j] = \frac{w^2}{(R\Delta t)^2} \mathbb{V}[q_x(v_{n,k})] \langle \sum_{n=1}^N \sum_{k=0}^{K(n)} I_j(x_{n,k}) \rangle + \frac{w^2}{(R\Delta t)^2} \langle q_x(v_{n,k}) \rangle^2 \mathbb{V}[\sum_{n=1}^N \sum_{k=0}^{K(n)} I_j(x_{n,k})].
\label{eq:analog_estimator_generic_expression}
\end{equation}
Note that the number of particles $N$ that reaches time $t_1$ (the time at which the analog estimator starts recording) is a random variable, as is the number of collisions $K(n)$ during the time interval of duration $\Delta t = t_2 - t_1$. In the following, we first derive the variance expression~\eqref{eq:analog_estimator_generic_expression} under the assumption that $N$ is known deterministically, such that only $K(n)$ is a random variable. We begin with the case without a particle source and subsequently adjust the derivations for the case where there is an active stationary particle source. Finally, we let $N$ be a random variable as well, where we again make the distinction between having no particle source and having an active stationary particle source.

\subsection{Deterministic particle number $N$}
\label{subsec:knownN}
Let us first assume that the number of particles $N$ at time $t_1$ is known, e.g., because $t_1 = t_0 = 0$ such that the time interval $\Delta t$ starts at the known initial condition or because there are no sources or sinks. Then~\eqref{eq:analog_estimator_generic_expression} becomes:
\begin{equation}
\mathbb{V}[\hat{Q}_j] = \frac{w^2N}{(R\Delta t)^2} \mathbb{V}[q_x(v_{n,k})] \langle \sum_{k=0}^{K(n)} I_j(x_{n,k}) \rangle + \frac{w^2N}{(R\Delta t)^2} \langle q_x(v_{n,k}) \rangle^2 \mathbb{V}[\sum_{k=0}^{K(n)} I_j(x_{n,k})].
\label{eq:analog_estimator_generic_expression_knownN}
\end{equation}
Assuming that $N$ is known allows us to focus on the number of collisions $K(n)$ occurring during the time interval $\Delta t$. The collisions happen at exponentially distributed times, so the number of collision events in time interval $\Delta t$ is Poisson distributed. A collision is either a charge-exchange collision or an ionization collision.
If there is an ionization sink ($R_i > 0$), then we have the following probability density for having $K(n) = \kappa$ collisions: 
% Note here that we do not count over all possible chains of length k that have k-1 cx and 1 ionization events in the sum with index l, (cfr binomial distribution, we would need to take a binomial factor into account to get that probability); we count the probability for one unique chain (each chain with k-1 cx events and 1 ionization event is equally likely) with length k that has those properties, of which the chain in which the first k-1 are cx and the last one is ionization is one.
\begin{equation}
\begin{split}
\mathcal{P}(K(n) = 0) &= \exp(-R \Delta t),\\
\mathcal{P}(K(n) = \kappa >0) &= \left(\frac{R_{cx}}{R} \right)^{\kappa-1}  \cdot \frac{(R\Delta t)^\kappa \exp(-R\Delta t)}{\kappa!}\\
&+ \sum_{l = \kappa+1}^\infty \left(\frac{R_{cx}}{R} \right)^{\kappa-1} \left(1 - \frac{R_{cx}}{R} \right) \cdot \frac{(R\Delta t)^l \exp(-R\Delta t)}{l!}.
\end{split}
\label{eq:probability:collisions_sink_case}
\end{equation}
For $K(n) = 0$, we have the probability that no collision happens during time interval $\Delta t$. For $K(n) > 0$, the probability density~\eqref{eq:probability:collisions_sink_case} has two contributions. First, a sequence of $\kappa$ events counts as a sequence of $\kappa$ collisions if the first $\kappa-1$ events are charge-exchange collisions. Second, a sequence with $l > \kappa$ events counts as a sequence of $\kappa$ collisions if the first $\kappa-1$ events are charge-exchange collisions and the $\kappa$-th event is an ionization collision. The expected value and variance of distribution~\eqref{eq:probability:collisions_sink_case} are
\begin{equation}
\begin{split}
\langle K(n) \rangle &= \frac{R}{R_i} \left(1 - \exp(-R_i \Delta t) \right),\\
\mathbb{V}[K(n)] &= \frac{2 R^2 - R_i R -(2R_i R_{cx} R \Delta t + 2 R^2 - R_i R) \exp(-R_i \Delta t)}{R_i^2}\\ 
&- \left(\frac{R}{R_i}(1-\exp(-R_i \Delta t)) \right)^2.
\end{split}
\label{eq:probability:collisions_sink_case_MV}
\end{equation}
A derivation of these results is provided in Appendix~\ref{app:meanvark}.
%NOTE: the limit $\Delta t \rightarrow \infty$ gives the probability on the amount of collisions before the particle is ionized (because it then gets infinite time to be ionized). We then also obtain $lim_{\Delta t \rightarrow \infty} \langle K(n) \rangle = \frac{R}{R_i}$, which is the expected amount of collisions before a particle gets ionized.
Taking the limit $R_i \rightarrow 0$ results in the standard Poisson distribution
% probability for k collisions in time dt is Poisson distributed with expected value lambda = R * dt
\begin{equation}
\lim_{R_i \rightarrow 0} \mathcal{P}(K(n) = \kappa) = \frac{(R \Delta t)^\kappa \exp(-R \Delta t)}{\kappa!},
\label{eq:probability:collisions_no_sink_case}
\end{equation}
with as expected value and variance
\begin{equation}
\begin{split}
\lim_{R_i \rightarrow 0} \langle K(n) \rangle &= R \Delta t,\\
\lim_{R_i \rightarrow 0} \mathbb{V}[K(n)] &= R \Delta t.
\end{split}
\end{equation}
Using the law of total expectation (see Appendix~\ref{app:total_exp_var}), we can write the expected value of a sum of indicator function evaluations with a random number of terms $K(n)$ as follows:
\begin{equation}
\langle {\sum_{k=0}^{K(n)} I_j(x_{n,k})} \rangle = \sum_{\kappa=0}^\infty \langle \sum_{k=0}^{\kappa} I_j(x_{n,k}) \rangle \cdot \mathcal{P}(K(n)=\kappa).
\label{eq:expected_value_with_random_k_collision}
\end{equation}
Using the law of total variance (see Appendix~\ref{app:total_exp_var}), we can write the variance of such a sum as
\begin{equation}
\mathbb{V}[\sum_{k=0}^{K(n)} I_j(x_{n,k})] = \sum_{\kappa=0}^\infty \left( \mathbb{V}[\sum_{k=0}^{\kappa} I_j(x_{n,k})] + \left( \langle \sum_{k=0}^{\kappa} I_j(x_{n,k}) \rangle - \langle \sum_{k=0}^{K(n)} I_j(x_{n,k}) \rangle \right)^2 \right) \cdot \mathcal{P}(K(n)=\kappa).
\label{eq:general_variance_with_random_k_collision}
\end{equation}
Expression~\eqref{eq:general_variance_with_random_k_collision} is completely general and any of the expressions for binomial experiments with a fixed number of trials $\kappa$, e.g., the upper bound~\eqref{eq:binomial_variance_simple_upperbound}, the expression for independent trials~\eqref{eq:binomial_variance_independent_stationary}, the Markov process based expressions, or the hidden Markov process based expressions, can be inserted to obtain an (approximate) expression for the variance on the binomial experiment with a random number of trials. 

As an example, we insert the Markov process based expression for large $K(n)$~\eqref{eq:binomial_variance_markovdependence_klotz_largeN} for the variance on the binomial experiment into~\eqref{eq:expected_value_with_random_k_collision} and~\eqref{eq:general_variance_with_random_k_collision} to obtain
\begin{equation}
\begin{split}
\langle {\sum_{k=0}^{K(n)} I_j(x_{n,k})} \rangle &\approx \sum_{\kappa=0}^\infty \kappa p_j \cdot \mathcal{P}(K(n)=\kappa) = \langle K(n) \rangle p_j,\\
\mathbb{V}[\sum_{k=0}^{K(n)} I_j(x_{n,k})] &\approx \langle K(n) \rangle \left( p_j (1-p_j) + \frac{2 p_j (1-p_j) (\lambda_j - p_j)}{1-\lambda_j} \right) + \mathbb{V}[K(n)] p_j^2.
\end{split}
\end{equation}
Inserting these expressions in~\eqref{eq:analog_estimator_generic_expression_knownN} results in:
\begin{equation}
\begin{split}
\mathbb{V}[\hat{Q}_j] &\approx \frac{w^2 N}{(R\Delta t)^2} \mathbb{V}[q_x(v_{n,k})] \langle K(n) \rangle p_j\\ 
&+ \frac{w^2 N}{(R\Delta t)^2} \langle q_x(v_{n,k}) \rangle^2 \left( \langle K(n) \rangle \left( p_j (1-p_j) + \frac{2 p_j (1-p_j) (\lambda_j - p_j)}{1-\lambda_j} \right) + \mathbb{V}[K(n)] p_j^2 \right),
\end{split}
\label{eq:Analog_Variance_LargeK}
\end{equation}
where we can substitute~\eqref{eq:statistics_QoIs} to obtain variance predictors for the three lowest order moments. Inserting expression~\eqref{eq:particle_weight} for the particle weight $w$, we see that the variance expression scales with $\frac{1}{N}$, as it should (see Section~\ref{subsec: QoI estimation}).

\subsection{Treatment of a stationary particle source}
\label{subsec:knownNSource}
When there is a stationary source as described in Section~\ref{sec: binomial experiments dynamics}, the time that a particle $n$ can effectively spend in the time interval $[t_1, t_2]$ is a uniformly distributed random variable $\Delta t_n \sim U[0,t_2-t_1]$, see~\eqref{eq:stationary_source_statistics}. This directly affects the probability distribution of the number of collisions (Bernoulli trials) that the particle contributes to the analog estimator. We can write the probability of having an effective time interval of length $\delta$ and having $\kappa$ collisions in that time interval as follows:
\begin{equation}
\begin{split}
\mathcal{P}(\Delta t_n = \delta, K(n) = \kappa) &= \mathcal{P}(K(n) = \kappa \mid \Delta t_n = \delta) \cdot \mathcal{P}(\Delta t_n = \delta)\\
&= \mathcal{P}(K(n) = \kappa \mid \Delta t_n = \delta) \cdot \frac{1}{(t_2-t_1)},
\end{split}
\label{eq:5.2-joint_probability}
\end{equation}
where $\mathcal{P}(K(n) = \kappa \mid \Delta t_n = \delta)$ is given by~\eqref{eq:probability:collisions_sink_case} with $\Delta t = \delta$. In variance expression~\eqref{eq:general_variance_with_random_k_collision}, we need the marginal probability distribution $\mathcal{P}(K(n)=\kappa)$.  Integrating out the dependence on $\delta$ in~\eqref{eq:5.2-joint_probability}, we obtain the following expression for the marginal probability distribution for having $\kappa$ collisions:
\begin{equation}
\begin{split}
\mathcal{P}(K(n)=\kappa) = & \int_0^{(t_2-t_1)} \mathcal{P}(\Delta t_n = \delta, K(n) = \kappa) d\delta \\
= &\left( \frac{R_{cx}}{R} \right)^{\kappa-1} \Bigg[ \frac{1}{R (t_2-t_1)} - \frac{\Gamma(\kappa+1,R(t_2-t_1))}{R (t_2-t_1) \cdot \kappa!} + \\ 
&\left( 1 - \frac{R_{cx}}{R} \right) \left(1 - \sum_{l=0}^\kappa \left( \frac{1}{R (t_2-t_1)} - \frac{\Gamma(l+1, R(t_2-t_1))}{R (t_2-t_1) \cdot l!} \right) \right) \Bigg],
\end{split}
\label{eq:sec5_marginal_dist}
\end{equation}
where we make use of incomplete Gamma functions (see Appendix~\ref{app:gamma}). Additionally, we can derive the following expression for the expected number of collisions:
\begin{equation}
\begin{split}
\langle K(n) \rangle &= \int_0^{(t_2-t_1)} \sum_{\kappa=0}^\infty \kappa \cdot \mathcal{P}(\Delta t_n = \delta, K(n) = \kappa) d\delta\\
&= \frac{R(R_i (t_2-t_1) + \exp(-R_i(t_2-t_1)) - 1)}{R_i^2 (t_2-t_1)}.
\end{split}
\end{equation}
Using l'H\^opital's rule, we can verify that in the limit $R_i \rightarrow 0$, we obtain
\begin{equation}
\lim_{R_i \rightarrow 0} \langle K(n) \rangle = \frac{R (t_2-t_1)}{2}.
\end{equation} 
In this limit, the expected number of collisions is the expected number of collisions for a fixed time interval of duration $\frac{(t_2-t_1)}{2}$, i.e., for the expected value of the effective time interval $\Delta t_n$.

The variance on the number of collisions is
\begin{equation}
\begin{split}
\mathbb{V}[K(n)] &= \int_0^{(t_2-t_1)} \sum_{\kappa=0}^\infty (\kappa-\langle K(n) \rangle)^2 \cdot \mathcal{P}(\Delta t_n = \delta, K(n) = \kappa) d\delta\\
&= \frac{2R^2 \exp(-R_i(t_2-t_1)) (R_i(t_2-t_1) + \exp(R_i(t_2-t_1))(R_i(t_2-t_1)-2)+2)}{R_i^3(t_2-t_1)}\\
&- \frac{R R_i \exp(-R_i(t_2-t_1))  (2R_i(t_2-t_1) + \exp(R_i(t_2-t_1))(R_i(t_2-t_1)-3)+3)}{R_i^3(t_2-t_1)}\\
&- \left( \frac{R(R_i (t_2-t_1) + \exp(-R_i(t_2-t_1)) - 1)}{R_i^2 (t_2-t_1)} \right)^2.
\end{split}
\end{equation}
Again, using l'H\^opital's rule, we can verify that in the limit $R_i \rightarrow 0$ we obtain
\begin{equation}
\lim_{R_i \rightarrow 0} \mathbb{V}[K(n)] = \frac{R (t_2-t_1)}{2} + \frac{(R (t_2-t_1))^2}{12}.
\end{equation}
Note in this limit of the variance expression that $\frac{(t_2-t_1)^2}{12}$ is the variance of the uniform distribution from which we draw $\Delta t_n$. So the new variance is the variance on the number of collisions in the expected effective time interval (first term), augmented with a term that scales with the variance on the effective time interval duration.

To account for the stationary particle source, we can insert the marginal probability distribution~\eqref{eq:sec5_marginal_dist} in the variance expression~\eqref{eq:general_variance_with_random_k_collision}. The expressions for the mean and variance on $K(n)$ can, e.g., be used in the variance predictor for the analog estimator based on the approximate Markov process expression for large $K(n)$~\eqref{eq:Analog_Variance_LargeK}.

\subsection{Random particle number $N$}
\label{subsec:randomN}

Let us now consider the case of equation~\eqref{eq:analog_estimator_generic_expression} where the number of particles $N$ at time $t_1$, the time at which the analog estimator starts recording, is uncertain. We denote the number of particles in the initial condition by $N(t=t_0=0) = N_0$, the particles generated by a stationary particle source by $N_S$, and the particles present at time $t_1$ by $N(t=t_1) = N$. 

If we know the probability distribution on $N$, then it follows from the law of total expectation (see Appendix~\ref{app:total_exp_var}) that
\begin{equation}
\langle \sum_{n=1}^N \sum_{k=0}^{K(n)} I_j(x_{n,k}) \rangle = \sum_{m=1}^{N_i} m \langle \sum_{k=0}^{K(n)} I_j(x_{n,k}) \rangle \cdot \mathcal{P}(N=m) = \langle N \rangle \langle \sum_{k=0}^{K(n)} I_j(x_{n,k}) \rangle,
\label{eq:analog_estimators_expectedvalue_doublesum}
\end{equation}
where $N_i = N_0$ for particles originating from the initial condition and $N_i = N_S$ for particles originating from the stationary particle source that is  active during the time interval $[t_0,t_1]$. Additionally, we can use the law of total variance (see Appendix~\ref{app:total_exp_var}) in the following way:
\begin{equation}
\begin{split}
\mathbb{V}[\sum_{n=1}^{N} \sum_{k=0}^{K(n)} I_j(x_{n,k})] &= \mathbb{V}[\langle \sum_{n=1}^{N} \sum_{k=0}^{K(n)} I_j(x_{n,k}) \mid N \rangle ] + \langle \mathbb{V}[ \sum_{n=1}^{N} \sum_{k=0}^{K(n)} I_j(x_{n,k}) \mid N]\rangle\\ 
&= \mathbb{V}[N \langle\sum_{k=0}^{K(n)} I_j(x_{n,k})\rangle] + \langle N \mathbb{V}[ \sum_{k=0}^{K(n)} I_j(x_{n,k}) ] \rangle\\ 
&= \mathbb{V}[N] \langle \sum_{k=0}^{K(n)} I_j(x_{n,k}) \rangle^2 + \langle N \rangle \mathbb{V}[\sum_{k=0}^{K(n)} I_j(x_{n,k})].
\end{split}
\label{eq:analog_estimators_variance_doublesum}
\end{equation}

Expressions for the statistics of the sum over $k$ have been derived above in sections~\ref{subsec:knownN} and~\ref{subsec:knownNSource}. Therefore, the variance expressions~\eqref{eq:analog_estimators_variance_doublesum} and~\eqref{eq:analog_estimator_generic_expression} are completely determined if we derive expressions for $\mathbb{V}[N]$ and $\langle N \rangle$. The randomness in $N$ is introduced by the particle source and ionization sink. We first consider the case where there is an ionization sink, but no particle source. Then, we will look at the case where there is also a stationary particle source.

\textbf{Case 1: no stationary particle source}\\
We are interested in the probability $\mathcal{P}(N=m)$, i.e., we want to know how many particles there are at time $t_1$, the start of the estimation time interval. Each of the $N_0$ particles at the start of the simulation ionizes with probability $P_{ion} = 1-\exp(-R_i t_1)$ or survives with probability $P_{survive} = 1 - P_{ion} = \exp(-R_i t_1)$. If we then count all the different realizations in which $m$ particles survive up to time $t_1$, we get the following binomial probability distribution:
\begin{equation}
\begin{split}
\mathcal{P}(N=m) &= \frac{N_0!}{m!(N_0-m)!}P_{survive}^m P_{ion}^{N_0-m}\\ 
&= \frac{N_0!}{m!(N_0-m)!} \exp(-R_i t_1)^m (1-\exp(-R_i t_1))^{N_0-m}.
\end{split}
\end{equation}

Recall that the binomial distribution describes binomial experiments with independent trials. Using the properties of this distribution, we find the following expressions for the expected number of particles $N$ and the variance on $N$:
\begin{equation}
\begin{split}
\langle N \rangle &= N_0 \exp(-R_i t_1),\\
\mathbb{V}[N] &= N_0 \exp(-R_i t_1) (1 - \exp(-R_i t_1)).
\end{split}
\end{equation}
Note that we can also find these expressions by defining a binomial experiment
$N = \sum_{n=1}^{N_0} I_n$, where this time $I_n$ represents the Bernoulli trial of a particle surviving the time interval $[0,t_1]$ without being ionized, which happens with success probability $p_n = p = \exp(-R_i t_1)$, and then considering the expected value and variance of this sum under the assumption of independent trials. Also note that in the limit of $R_i \rightarrow 0$ we obtain $\langle N \rangle = N_0$ and $\mathbb{V}[N] = 0$, which both are to be expected as all the initial particles then surely reach time $t_1$.

\textbf{Case 2: active stationary particle source}\\
When $N_S$ particles originate from a stationary particle source, the survival probability becomes particle dependent: $p_n = \exp(-R_i \Delta t_n)$ with $\Delta t_n \sim \mathcal{P}(\Delta t_n = \delta) = U[0,t_1]$. This time, obtaining an expression for $\mathcal{P}(N=m)$ is not trivial. We therefore immediately write the binomial experiment $N = \sum_{n=1}^{N_S} I_n$ with Bernoulli trials $I_n$ as explained in the previous paragraph. The expected value of a binomial experiment with independent trials, but trial dependent success probabilities $p_n$, is as follows:
\begin{equation}
\begin{split}
\langle N \rangle &= \sum_{n=1}^{N_S} \langle I_n \rangle = \sum_{n=1}^{N_S} \int_0^{t_1} (p_n \cdot 1 + (1-p_n) \cdot 0) \cdot \frac{1}{t_1} d\delta = N_S \int_0^{t_1} \exp(-R_i \delta) \cdot \frac{1}{t_1} d \delta\\ 
&= N_S \cdot \frac{1 - \exp(-R_i t_1)}{R_i t_1}.
\end{split}
\end{equation}
In the limit for $R_i \rightarrow 0$, this expression reduces using l'H\^opital's rule to $\langle N \rangle = N_S$. For the variance, we find
\begin{equation}
\begin{split}
\mathbb{V}[N] &= \mathbb{V}[\sum_{n=1}^{N_S} I_n ] = \sum_{n=1}^{N_S} \langle I_n \rangle (1-\langle I_n \rangle)\\
&= N_S \cdot \left( \frac{1 - \exp(-R_i t_1)}{R_i t_1} \right) \left( 1 - \left( \frac{1 - \exp(-R_i t_1)}{R_i t_1} \right) \right).
\end{split}
\end{equation}
In the limit for $R_i \rightarrow 0$, this expression reduces to a variance of $0$, as expected because then we have $N \equiv N_S$.

\subsection{Numerical experiments}
We test the analog estimator variance predictors of the form~\eqref{eq:analog_estimator_generic_expression} for the three lowest order moments~\eqref{eq:statistics_QoIs} for a range of values for $R$ and $\sigma_p^2$. For each set of parameters, we simulate a single particle and count one contribution to each of the analog estimators for each collision point (Bernoulli trial) that is inside a bin $j$ during a time interval $[t_1, t_2]$. All the experiments use $t_1 = 0$ and $t_2 = \frac{100}{R}$ such that the expected number of collisions during the time interval equals 100. We repeat the experiments two times: once with particles being launched from an initial condition without sink ($R_i = 0$); and once with particles being launched from a stationary particle source with an active sink ($R_i = 1$). The test cases are all executed on a 1D ($d=1$) domain $D = [0,1]$ with length 1 and periodic boundary conditions. The particles have an initial position distributed uniformly over the domain and a velocity drawn from the post-collisional velocity distribution as given in~\eqref{eq:particle_dynamics}. The space is subdivided in $J=10$ bins, such that the probability for success equals $p_j = \frac{1}{J} = 0.1$. The subdomain $D_j = [0,0.1]$ is considered as the bin of interest. The code is openly available in \reponame.

\begin{figure}
\centering
\begin{subfigure}{0.32\linewidth}
\centering
\includegraphics[width=\linewidth]{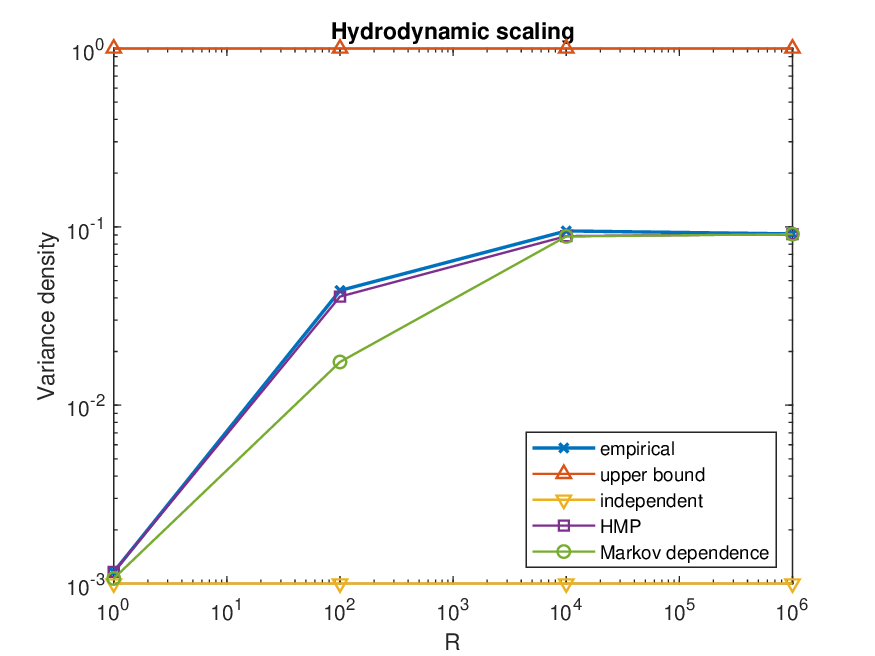}
\end{subfigure}
\begin{subfigure}{0.32\linewidth}
\centering
\includegraphics[width=\linewidth]{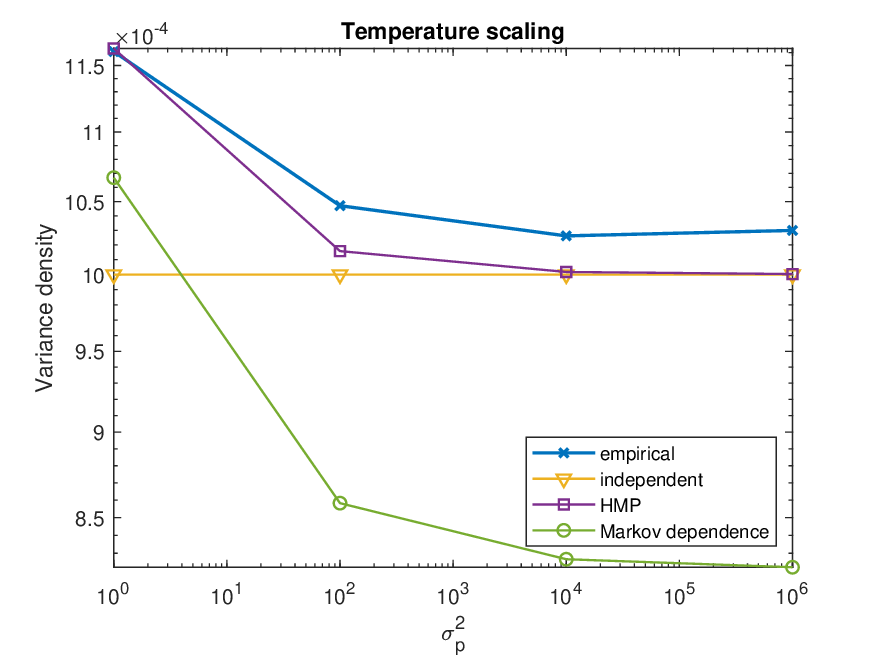}
\end{subfigure}
\begin{subfigure}{0.32\linewidth}
\centering
\includegraphics[width=\linewidth]{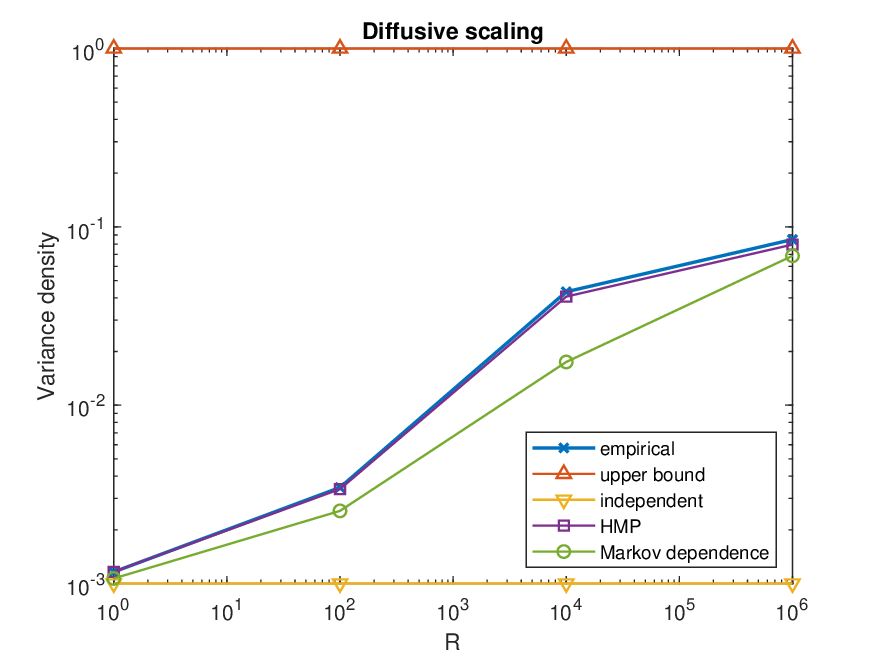}
\end{subfigure}
\begin{subfigure}{0.32\linewidth}
\centering
\includegraphics[width=\linewidth]{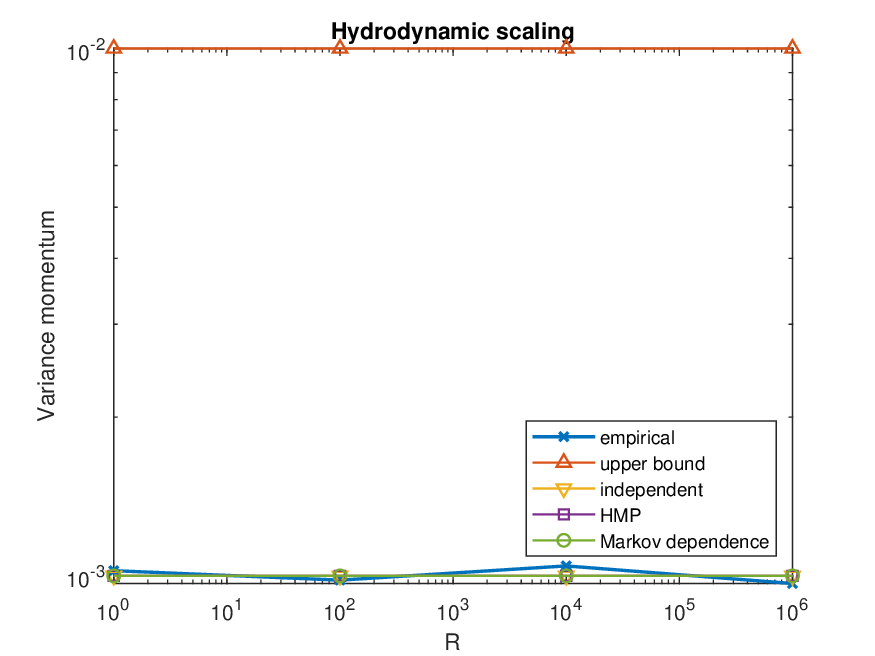}
\end{subfigure}
\begin{subfigure}{0.32\linewidth}
\centering
\includegraphics[width=\linewidth]{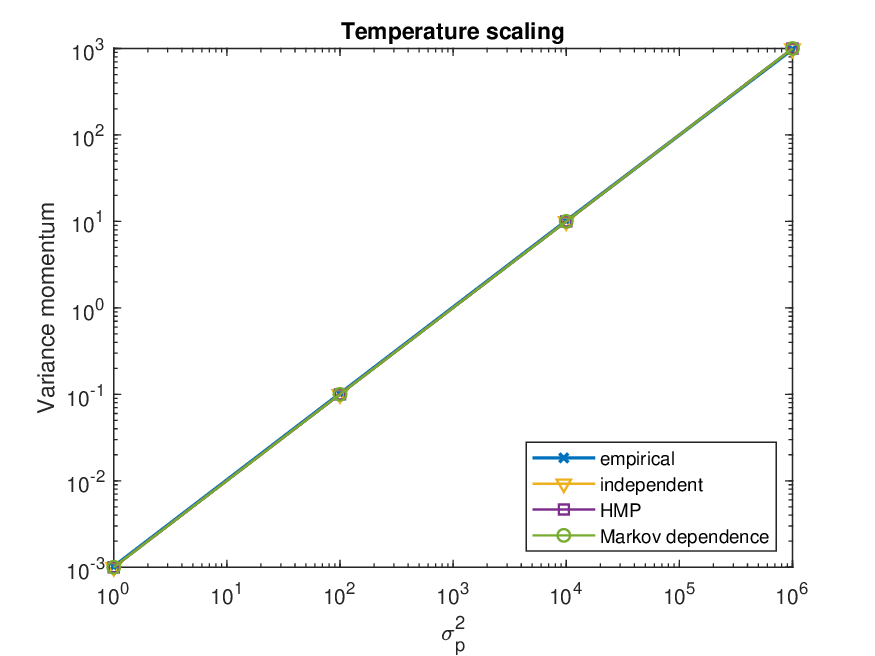}
\end{subfigure}
\begin{subfigure}{0.32\linewidth}
\centering
\includegraphics[width=\linewidth]{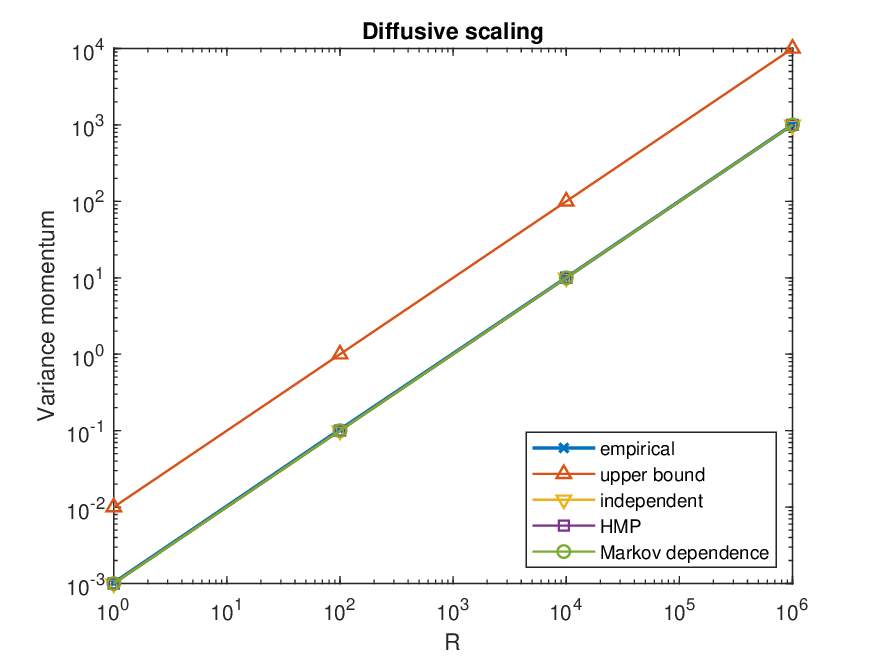}
\end{subfigure}
\begin{subfigure}{0.32\linewidth}
\centering
\includegraphics[width=\linewidth]{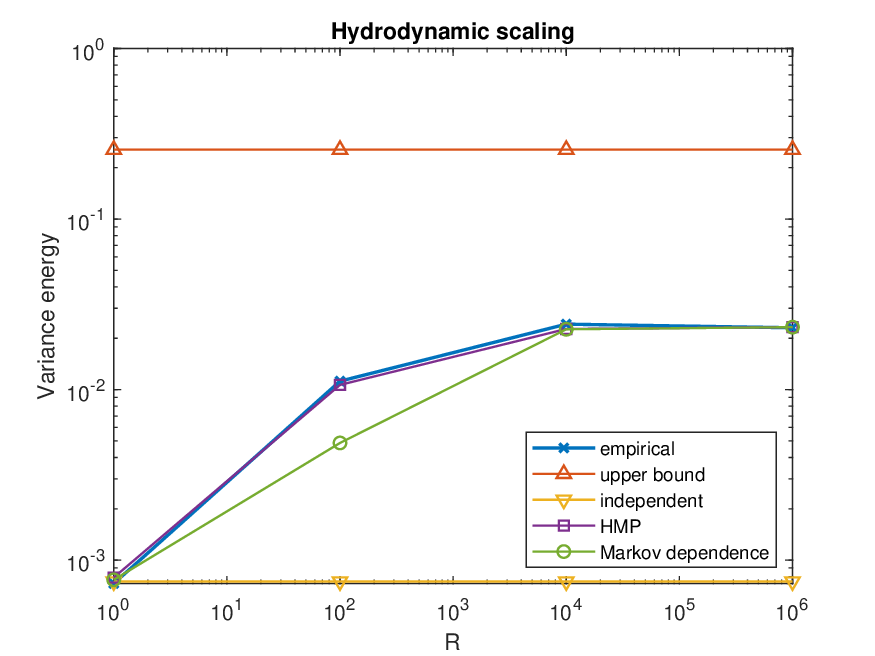}
\end{subfigure}
\begin{subfigure}{0.32\linewidth}
\centering
\includegraphics[width=\linewidth]{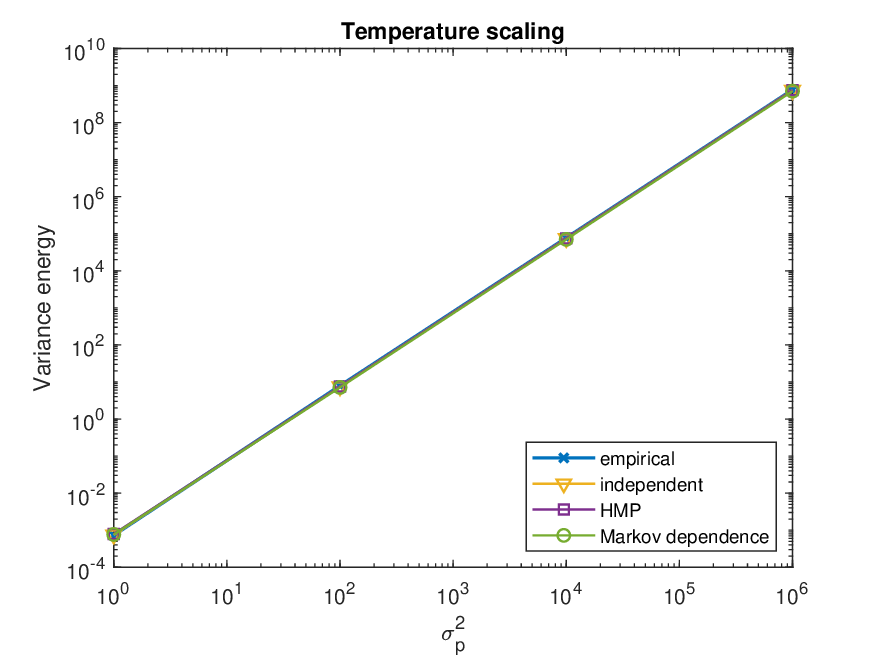}
\end{subfigure}
\begin{subfigure}{0.32\linewidth}
\centering
\includegraphics[width=\linewidth]{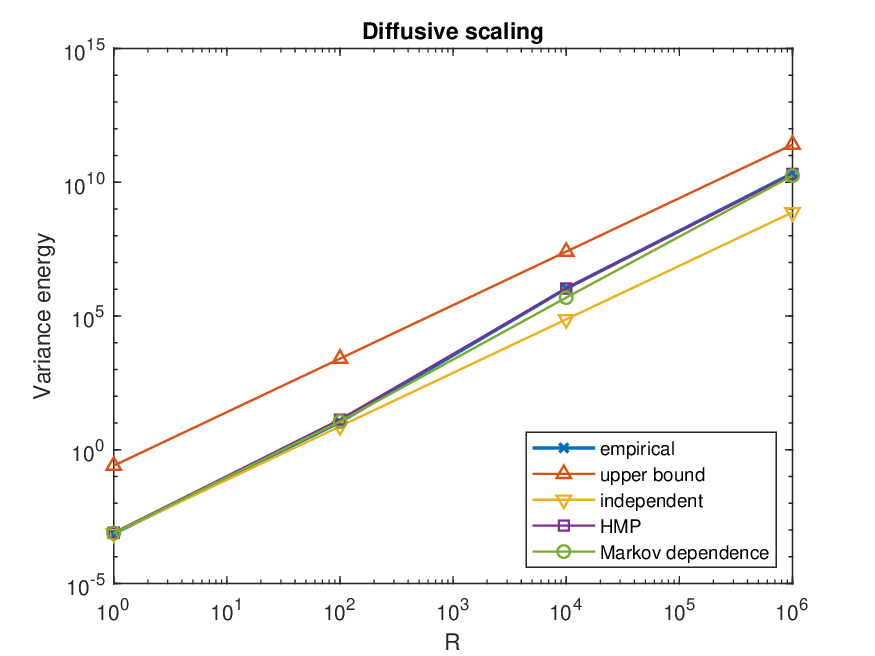}
\end{subfigure}
\caption{Analog particle tracing Monte Carlo with analog estimator. Particles originate from an uniform initial condition, there is no ionization sink ($R_i = 0$). Top to bottom: variance on density, momentum and energy. Left to right: hydrodynamic, temperature and diffusive scaling.}
\label{fig:analog_estimators_noRi_u0}
\end{figure}

\begin{figure}
\centering
\begin{subfigure}{0.32\linewidth}
\centering
\includegraphics[width=\linewidth]{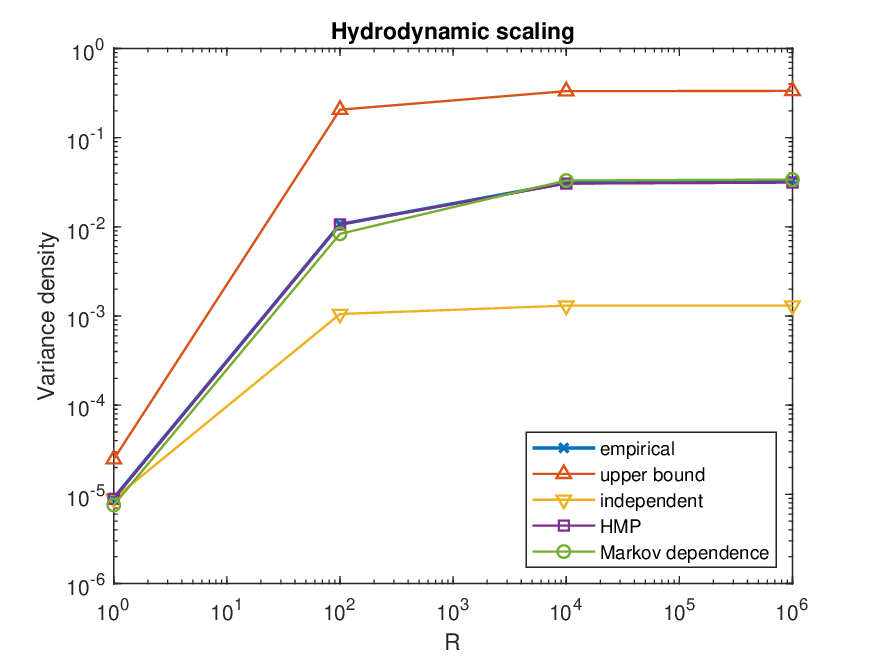}
\end{subfigure}
\begin{subfigure}{0.32\linewidth}
\centering
\includegraphics[width=\linewidth]{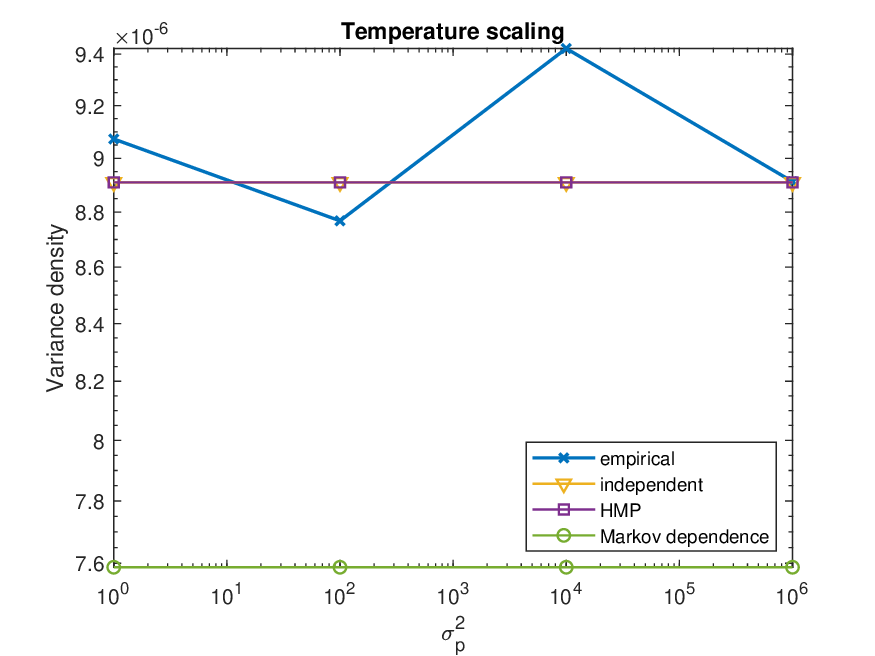}
\end{subfigure}
\begin{subfigure}{0.32\linewidth}
\centering
\includegraphics[width=\linewidth]{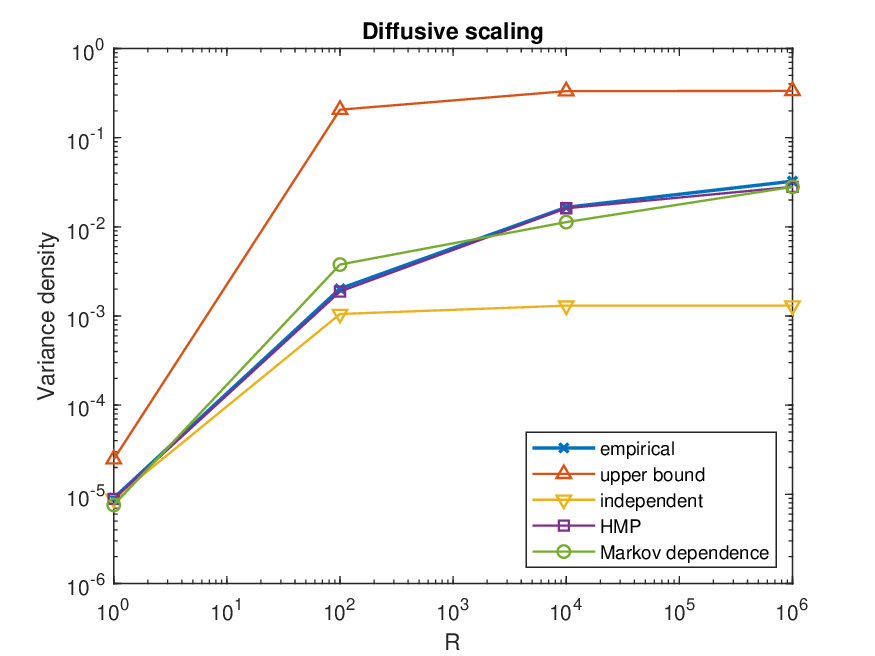}
\end{subfigure}
\begin{subfigure}{0.32\linewidth}
\centering
\includegraphics[width=\linewidth]{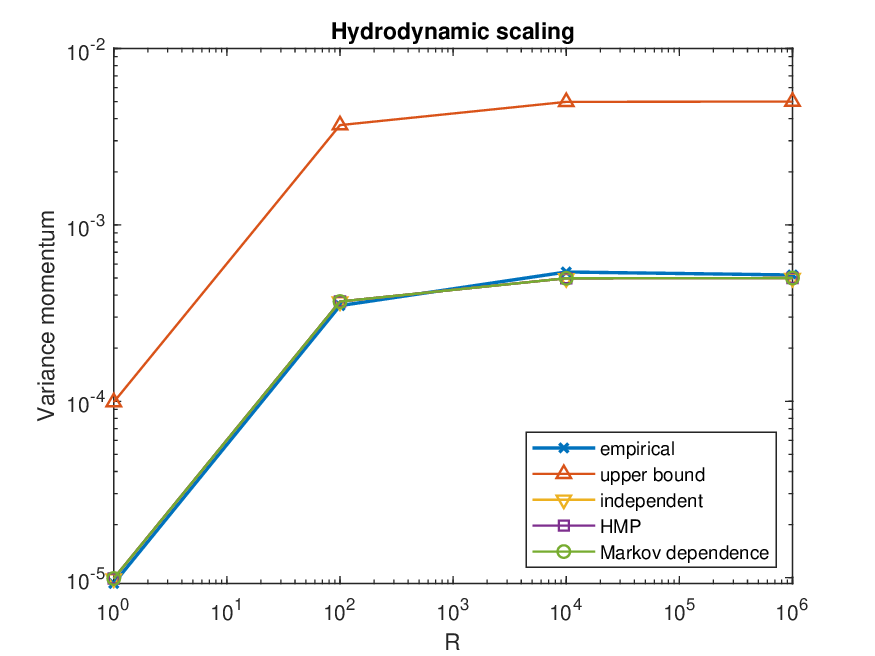}
\end{subfigure}
\begin{subfigure}{0.32\linewidth}
\centering
\includegraphics[width=\linewidth]{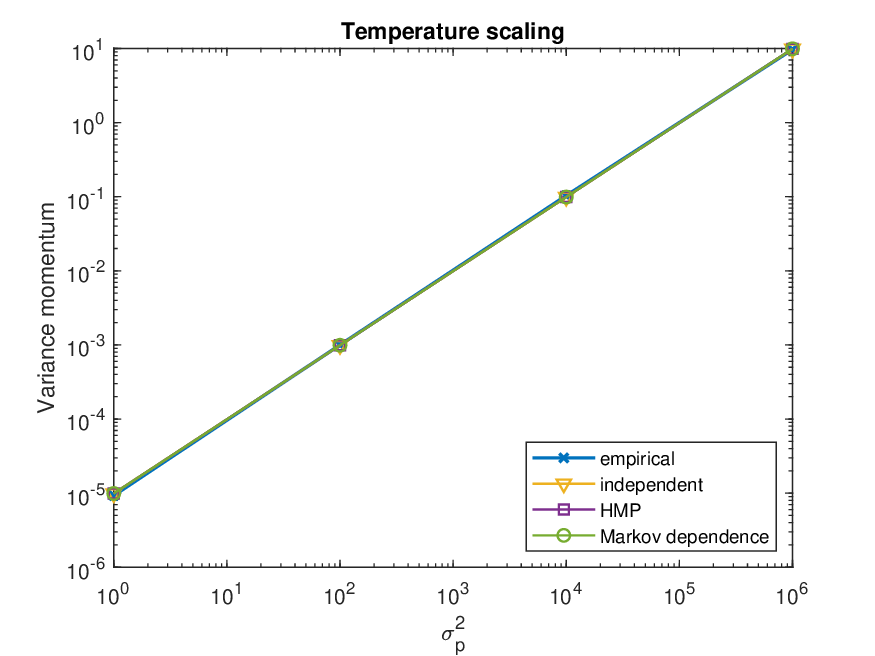}
\end{subfigure}
\begin{subfigure}{0.32\linewidth}
\centering
\includegraphics[width=\linewidth]{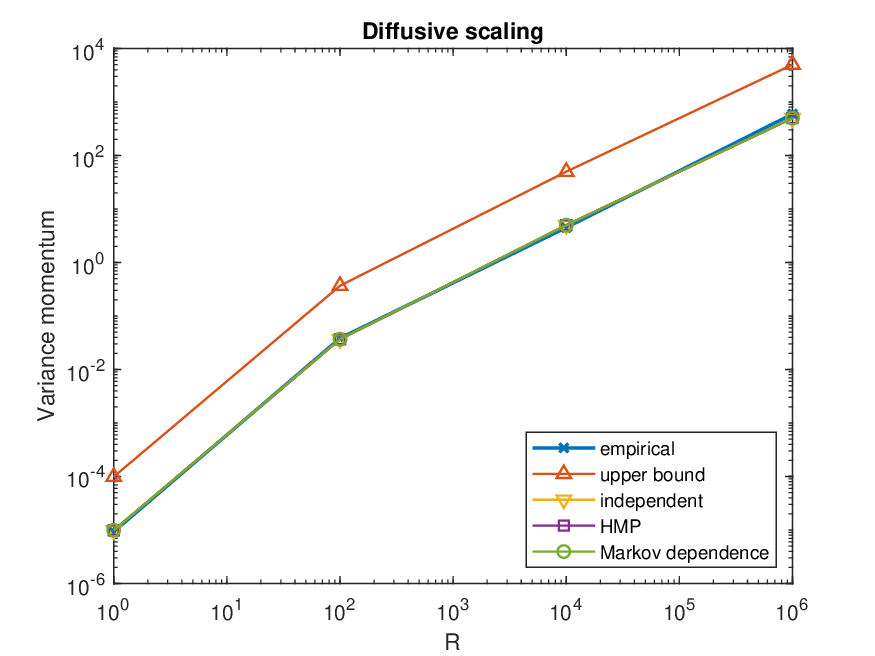}
\end{subfigure}
\begin{subfigure}{0.32\linewidth}
\centering
\includegraphics[width=\linewidth]{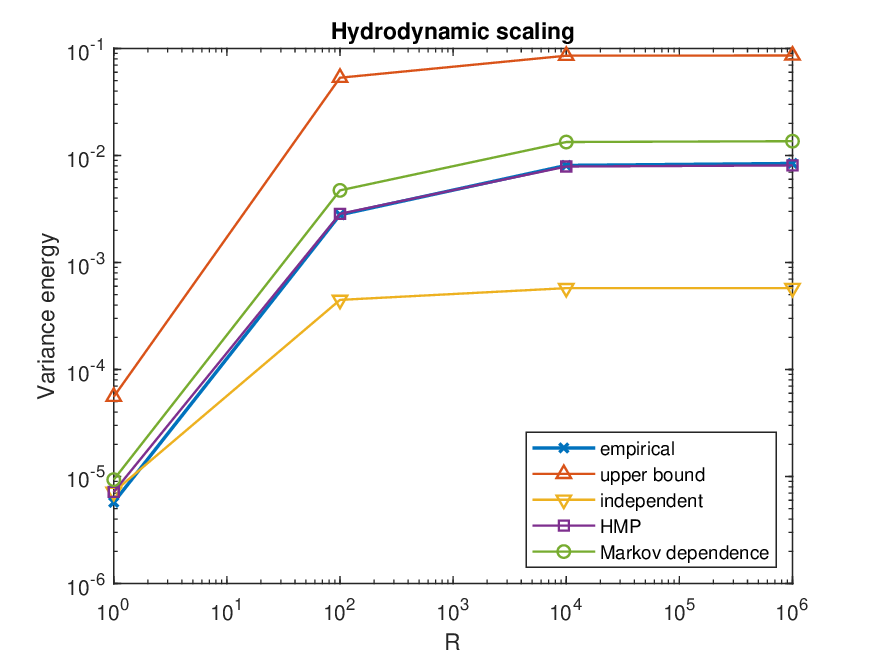}
\end{subfigure}
\begin{subfigure}{0.32\linewidth}
\centering
\includegraphics[width=\linewidth]{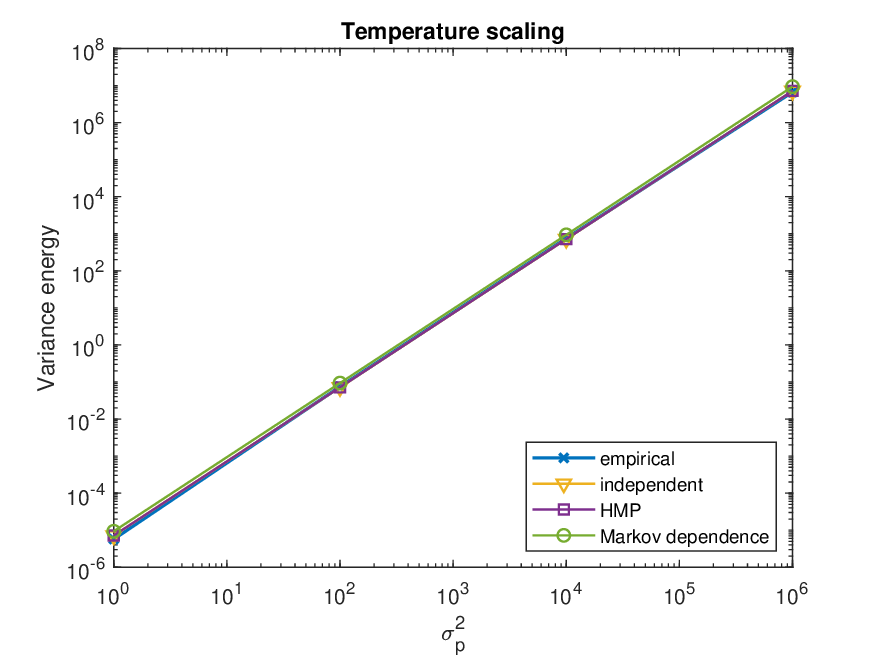}
\end{subfigure}
\begin{subfigure}{0.32\linewidth}
\centering
\includegraphics[width=\linewidth]{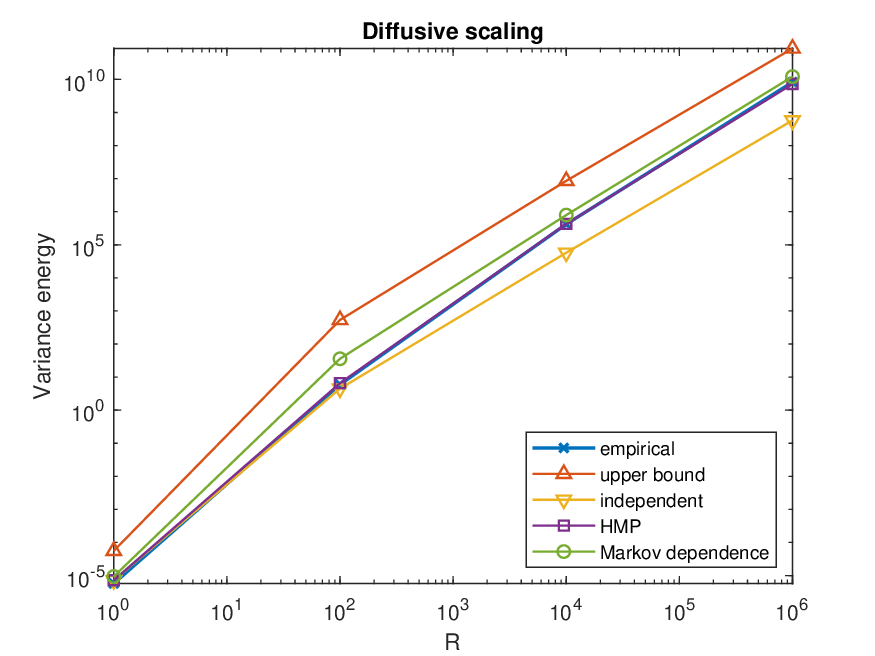}
\end{subfigure}
\caption{Analog particle tracing Monte Carlo with analog estimator. Particles are generated from a random source that is uniform in time, there is an ionization sink with $R_i = 1$. Top to bottom: variance on density, momentum and energy. Left to right: hydrodynamic, temperature and diffusive scaling.}
\label{fig:analog_estimators_Ri_u0}
\end{figure}

We consider the same three experiments as in Section~\ref{sec:verification_binomial_experiments}: the hydrodynamic scaling experiment, the temperature scaling experiment, and the diffusive scaling experiment. In each scaling, we estimate the variance on the three lowest order moments: particle density, momentum, and energy. For the variance on the binomial experiments, we insert the upper bound~\eqref{eq:binomial_variance_simple_upperbound}, the variance expression for independent trials~\eqref{eq:binomial_variance_independent_stationary}, the Markov process based variance expression~\eqref{eq:binomial_variance_markovdependence_klotz}, and the discretized hidden Markov process based variance expression. The hidden Markov process is discretized on an equidistant grid consisting of 100 cells. The conditional probability $\lambda_j$ for the Markov process and transition probabilities for the discretized hidden Markov process are calculated as a function of the model parameters as described in Ref.~\cite{ingelaere_prep_2023}. As a reference solution, we determine for each set of parameters the empirical variance based on $10^4$ realizations of the experiment. We plot the different variance predictors divided by $R t_2 = 100$ for these three scalings in Figure~\ref{fig:analog_estimators_noRi_u0} (no stationary particle source or ionization sink) and Figure~\ref{fig:analog_estimators_Ri_u0} (active stationary particle source and ionization sink), where we each time choose $u_p = 0$, and use $\sigma_p^2 = 1$ in the hydrodynamic scaling and $R=1$ in the temperature scaling. The upper bound for the variance is omitted in the temperature scaling plots, to zoom in on the other variance expressions. In all cases, the discretized hidden Markov process based expression corresponds well to the measured empirical values. The Markov process based variance predictor achieves the correct order of magnitude, which is valuable for error prediction where knowledge of the order of magnitude typically is sufficient.

\newpage
\section{Important remarks}
\label{sec: discussion}
The focus of this work is on predicting the variance on QoI estimators. For point estimators, the expressions based on independent contributions can easily be adopted. For analog estimators, the Markov process based expressions constitute easy-to-compute a priori variance predictors. This section deals with several aspects of statistical error prediction that need some more elaboration. The first aspect is how to choose the success probability $p_j$ and the conditional probability $\lambda_j$ in the Markov process based variance expressions. The second aspect is the performance of the different variance expressions in higher dimensions. Finally, the variance on a QoI estimator only gives a view of the absolute squared error. A better representation of the accuracy of the QoI estimates is given by the relative statistical error.
\subsection{Choice of $p_j$ and $\lambda_j$}
\label{subsec:pandlambda}
The probability for success $p_j$ scales with the particle density $\rho_j$. E.g., for the density point estimator, we have that 
\begin{equation}
\rho_j(T) = \langle \hat{\rho}_j(T) \rangle = wNp_j,
\label{eq:unbiased_point}
\end{equation}
meaning that $p_j = \frac{\rho_j(T)}{wN}$, i.e., the probability for success depends on the solution itself. The solution, of course, is not available a priori. To get a value for $p_j$, one can use a cheap (stationary) approximative solution or simply assume that $p_j$ is uniform: $p_j = \frac{1}{J}$. 

The conditional probability $\lambda_j$, introduced in~\eqref{eq:binomial_markovdependence_conditionalprobabilities}, depends on local model parameters in cell $D_j$ and the size of that cell, but unlike the probability for success $p_j$, it does not depend on the particle density $\rho_j$. Ref.~\cite{ingelaere_prep_2023} describes how $\lambda_j$ can be calculated when assuming an infinite domain without boundary conditions. As the calculation only involves local model parameters, it can be used to obtain local error predictors for each cell separately. Alternatively, one can only consider the cell with the worst-case model parameters to predict an upper bound on the variance.  The effect of boundary conditions and non-homogeneities between different cells in a real simulation still has to be investigated. For boundary cells, it might be possible to take the boundary conditions into account in the calculation of $\lambda_j$.

\subsection{Additional experiments and higher dimensions}
In addition to the experiments shown in this work, there are experiments with different parameter settings and in higher dimensions (2D and 3D) available in \reponame. The higher dimensional experiments are computationally expensive for the discretized hidden Markov process based variance expressions. The Markov process based expressions remain cheap and easily generalize to higher dimensions, because the generalization only requires the calculation of the conditional probability $\lambda_j$ for a higher dimensional bin of interest, which can be done as explained in Ref.~\cite{ingelaere_prep_2023}. The variance predictors also perform satisfactorily in these higher dimensional test cases.

\subsection{Relative statistical errors}
Using the predicted variances on QoI estimates $\mathbb{V}[\hat{Q}_j]$, we can also predict relative statistical errors, which we define as follows:
\begin{equation}
e_{j,x} = \frac{\sqrt{\mathbb{V}[\hat{Q}_j]}}{\langle \hat{Q}_j \rangle} = \sqrt{\frac{\mathbb{V}[\hat{Q}_j]}{\langle \hat{Q}_j \rangle^2}}.
\label{eq:relative_statistical_error}
\end{equation}
As an illustration, we will elaborate~\eqref{eq:relative_statistical_error} for the point estimator expression~\eqref{eq:var_predictor_point_estimators_general}. Using~\eqref{eq:unbiased_point}, we obtain
\begin{equation}
\begin{split}
e_{j,x} &= \sqrt{\frac{w^2 N p_j \mathbb{V}[q_x(v_n(T))] + w^2 N p_j(1-p_j) \langle q_x(v_n(T))\rangle^2}{w^2 \langle q_x(v_n(T)) \rangle^2 N^2 p_j^2}} = \sqrt{\frac{\mathbb{V}[q_x(v_n(T))] + (1-p_j) \langle q_x(v_n(T))\rangle^2}{ \langle q_x(v_n(T)) \rangle^2 N p_j}}\\
&= \sqrt{\frac{\mathbb{V}[q_x(v_n(T))] + (1-p_j) \langle q_x(v_n(T))\rangle^2}{ \langle q_x(v_n(T)) \rangle^2}} \cdot \sqrt{\frac{w}{\rho_j(T)}}.
\end{split}
\label{eq:statistical_error_general_expression}
\end{equation}

If we now, for example, write out the relative statistical error on the density, we obtain
\begin{equation}
e_{j,\rho} = \sqrt{\frac{1-p_j}{Np_j}} = \sqrt{1-p_j} \cdot \sqrt{\frac{w}{\rho_j(T)}}.
\end{equation}
This expression can also be found in Section 3.4.9 of the book~\cite{spanier_monte_1969} and a more generalized version for the so-called delta-f method can be found in Ref.~\cite{brunner_collisional_1999}. Note how the relative statistical errors scale with the square root of $\frac{w}{\rho_j(T)}$. If the particle weight $w$ is small compared to the density $\rho_j(T)$, then the relative statistical error will be small. If the particle weight $w$ is large compared to the density $\rho_j(T)$, then the relative statistical error is high. Increasing the number of particles $N$, results in smaller particle weights $w$ (see~\eqref{eq:particle_weight}), resulting in smaller relative statistical errors.

\paragraph{Remark 5} For the momentum estimator, we have that $\mathbb{V}[q_m(v_n(T))] = \sigma_p^2$ and $\langle q_m(v_n(T)) \rangle = |u_p|$. If $|u_p|^2 \ll \sigma_p^2$, then the momentum estimator has a large relative statistical error (see~\eqref{eq:statistical_error_general_expression}), making it difficult to get an accurate QoI estimate.

%\newpage
\section{Conclusions}
\label{sec: conclusion}
In this work, variance predictors are devised for analog particle tracing Monte Carlo methods that simulate linear kinetic equations. The expressions are illustrated in 1D test cases but can easily be applied to higher dimensional problems as well~\cite{ingelaere_prep_2023}. The variance predictors are built on the idea that constructing a histogram can be interpreted as performing a (correlated) binomial experiment. Using this insight, we first derived an upper bound, an independent trials based, Markov process based, and hidden Markov process based variance expressions for a (correlated) binomial experiment. Next, these expressions were used to construct variance predictors for different quantity of interest estimators. For point estimators, the independent trials based estimators suffice. For analog estimators, the hidden Markov process based expressions capture the variance exactly, but at a high cost, making them unattractive for a priori variance prediction. The Markov process based expressions are cheap to compute and typically capture the order of magnitude of the variance correctly, which is satisfactory from an error prediction point of view. The derived expressions give a prediction of the variance in a given bin of the histogram, based on the bin size, the used number of particles $N$, and the local model parameters. One can then choose to either predict the variance for all bins, or to select the bin with worst-case parameters and predict a worst-case upper bound for the variance over the whole simulation domain. The resulting variance predictors can be used in multiple ways: (1) to predict the relative statistical error, i.e., the accuracy, on an estimate of a quantity of interest; (2) to characterize the performance of a Monte Carlo method as a function of model parameters; (3) for Monte Carlo method selection and optimization, also, e.g., if the Monte Carlo method is a component of a more complex hybrid code.

In future work, the variance predictors can perhaps be extended to nonlinear (or coupled) Monte Carlo methods such as DSMC and PIC, or other methods with interacting particles, see, e.g., Ref.~\cite{ArneToon} and references therein. They can be applied to different kinetic equations and other (nonanalog) simulation and estimation procedures, enabling selection and optimization of nonanalog particle tracing Monte Carlo methods. The error analysis can be used in the design of hybrid simulation methods that contain a Monte Carlo component. The theory has to be tested on real application cases to assess the influence of the neglected $x,t$ dependencies of the model parameters. Finally, some work can still be done on taking boundary conditions into account, e.g., in the calculation of the conditional probability $\lambda_j$ in the Markov dependence approximation.

\section*{Acknowledgements}

The authors would like to acknowledge the financial support of the CogniGron research center and the Ubbo Emmius Funds (University of Groningen). The first author is a SB PhD fellow of the Research Foundation Flanders
(FWO), funded by grant 1S64723N. The second author has received funding from the European High-Performance Computing Joint Undertaking (JU) under grant agreement No 955701. The JU receives support from the European Union’s Horizon 2020 research and innovation programme, and Belgium, France, Germany, and Switzerland. Parts of the work are supported by the Research Foundation Flanders (FWO) under grant 1179820N for fundamental research and project grant G085922N.
This work has been carried out within the framework of the EUROfusion Consortium, funded by the European Union via the Euratom Research and Training Programme (Grant Agreement No 101052200 — EUROfusion). Views and opinions expressed are, however, those of the author(s) only and do not necessarily reflect those of the European Union or the European Commission. Neither the European Union nor the European Commission can be held responsible for them.

\newpage
\appendix

\section{Derivation of mean and variance of $\kappa$ events happening in a time interval with sinks}
\label{app:meanvark}
In this appendix, we derive the mean and variance~\eqref{eq:probability:collisions_sink_case_MV} of $K(n)$ with respect to the probability distribution $\mathcal{P}(K(n) = \kappa)$ given by~\eqref{eq:probability:collisions_sink_case}. 
\subsection{The mean} 
We start by inserting~\eqref{eq:probability:collisions_sink_case} in the definition of the mean:
\begin{equation}
\begin{split}
\langle K(n) \rangle &= \sum_{\kappa=0}^\infty \kappa \cdot \mathcal{P}(K(n) = \kappa)\\
&= \sum_{\kappa=1}^\infty \kappa \cdot \biggl( \left(\frac{R_{cx}}{R} \right)^{\kappa-1}  \cdot \frac{(R\Delta t)^\kappa \exp(-R\Delta t)}{\kappa!}\\
&+ \sum_{l = \kappa+1}^\infty \left(\frac{R_{cx}}{R} \right)^{\kappa-1} \left(1 - \frac{R_{cx}}{R} \right) \cdot \frac{(R\Delta t)^l \exp(-R\Delta t)}{l!} \biggr).\\
\end{split}
\label{eq:appendixA:fullmean}
\end{equation}
We tackle this big equation in multiple parts. For the first term we get
\begin{equation}
\begin{split}
&\sum_{\kappa=1}^\infty \kappa \cdot  \left(\frac{R_{cx}}{R} \right)^{\kappa-1}  \cdot \frac{(R\Delta t)^\kappa \exp(-R\Delta t)}{\kappa!}\\
&= R \Delta t \exp(-R\Delta t) \sum_{\kappa=1}^\infty \frac{(R_{cx}\Delta t)^{\kappa-1}}{(\kappa-1)!}\\ 
&= R \Delta t \exp(-R\Delta t) \exp(R_{cx} \Delta t)\\
&= R \Delta t \exp(-R_i\Delta t).
\end{split}
\end{equation}

For the remaining term of~\eqref{eq:appendixA:fullmean}, we use (in order) property~\eqref{eq:AppendixC:incompleteGammaProperty} and the definition~\eqref{eq:AppendixC:defIncompleteGamma} of the incomplete Gamma function (see Appendix~\ref{app:gamma}), and the Taylor expansion of the exponential function:
\begin{equation}
\begin{split}
&\sum_{\kappa=1}^\infty \kappa \cdot \sum_{l = \kappa+1}^\infty \left(\frac{R_{cx}}{R} \right)^{\kappa-1} \left(1 - \frac{R_{cx}}{R} \right) \cdot \frac{(R\Delta t)^l \exp(-R\Delta t)}{l!}\\
&= \sum_{\kappa=1}^\infty \kappa \cdot \left( \frac{R_{cx}}{R} \right)^{\kappa-1} \left(1 - \frac{R_{cx}}{R} \right) \cdot \exp(-R\Delta t) \left( \exp(R\Delta t) - \sum_{l = 0}^\kappa \frac{(R\Delta t)^l}{l!} \right)\\
&= \sum_{\kappa=1}^\infty \kappa \cdot \left( \frac{R_{cx}}{R} \right)^{\kappa-1} \left(1 - \frac{R_{cx}}{R} \right) \cdot \left( 1 - \frac{\Gamma(\kappa+1,R \Delta t)}{\kappa!} \right)\\
&= \left( \sum_{\kappa=1}^\infty \kappa \cdot \left( \frac{R_{cx}}{R} \right)^{\kappa-1} \left(1 - \frac{R_{cx}}{R} \right) \right) - \left( \sum_{\kappa=1}^\infty \kappa \cdot \left( \frac{R_{cx}}{R} \right)^{\kappa-1} \left(1 - \frac{R_{cx}}{R} \right) \cdot \frac{\Gamma(\kappa+1,R \Delta t)}{\kappa!} \right)\\
&= \frac{R}{R_i} - \left(1 - \frac{R_{cx}}{R} \right) \sum_{\kappa=1}^\infty \left( \frac{R_{cx}}{R} \right)^{\kappa-1}  \cdot \frac{\Gamma(\kappa+1,R \Delta t)}{(\kappa-1)!}\\
&= \frac{R}{R_i} - \left( 1 - \frac{R_{cx}}{R}\right) \int_{R \Delta t}^\infty t \exp(-t) \sum_{\kappa=1}^\infty \frac{\left(\frac{R_{cx}t}{R}\right)^{\kappa-1}}{(\kappa-1)!}dt\\
&= \frac{R}{R_i} - \left( 1 - \frac{R_{cx}}{R}\right) \int_{R \Delta t}^\infty t \exp(-t) \exp\left(\frac{R_{cx}t}{R} \right) dt\\
&= \frac{R}{R_i} - \frac{R (R_i \Delta t + 1)}{R_i} \exp(-R_i \Delta t).
\end{split}
\end{equation}

Combining the two parts, we finally obtain
\begin{equation}
\begin{split}
\langle K(n) \rangle &= R \Delta t \exp(-R_i\Delta t) + \frac{R}{R_i} - \frac{R (R_i \Delta t + 1)}{R_i} \exp(-R_i \Delta t)\\
&= \frac{R}{R_i} (1-\exp(-R_i \Delta t)).
\end{split}
\end{equation}

\subsection{The variance}
We start from the definition of the variance:
\begin{equation}
\begin{split}
\mathbb{V}[K(n)] &= \sum_{\kappa=0}^\infty (\kappa- \langle K(n) \rangle)^2 \cdot \mathcal{P}(K(n) = k)\\
&= \left( \sum_{\kappa=1}^\infty \kappa^2 \cdot \mathcal{P}(K(n) = \kappa)\right) - \langle K(n) \rangle^2.
\end{split}
\label{eq:AppendixA:fullvariance}
\end{equation}

As the second term is known from above, we focus on the first term and insert~\eqref{eq:probability:collisions_sink_case}:
\begin{equation}
\begin{split}
&\sum_{\kappa=1}^\infty \kappa^2 \cdot \mathcal{P}(K(n) = \kappa)\\
&= \sum_{\kappa=1}^\infty \kappa^2 \cdot \biggl( \left(\frac{R_{cx}}{R} \right)^{\kappa-1}  \cdot \frac{(R\Delta t)^\kappa \exp(-R\Delta t)}{\kappa!}\\
&+ \sum_{l = \kappa+1}^\infty \left(\frac{R_{cx}}{R} \right)^{\kappa-1} \left(1 - \frac{R_{cx}}{R} \right) \cdot \frac{(R\Delta t)^l \exp(-R\Delta t)}{l!} \biggr).
\end{split}
\end{equation}
We can work this out part by part as we did for the expected value of $K(n)$. We provide some intermediate results. For the first part of the expression, we obtain:
\begin{equation}
\begin{split}
&\sum_{\kappa=1}^\infty \kappa^2 \cdot \left(\frac{R_{cx}}{R} \right)^{\kappa-1}  \cdot \frac{(R\Delta t)^\kappa \exp(-R\Delta t)}{\kappa!}\\
&= \left(\sum_{\kappa=1}^\infty (\kappa-1) \left(\frac{R_{cx}}{R}\right)^{\kappa-1} \cdot \frac{(R \Delta t)^\kappa \exp(-R \Delta t)}{(\kappa-1)!} \right) + \left( \sum_{\kappa=1}^\infty \left(\frac{R_{cx}}{R}\right)^{\kappa-1} \frac{(R \Delta t)^\kappa \exp(-R \Delta t)}{(\kappa-1)!} \right)\\
&= (R \Delta t)^2 \left(\frac{R_{cx}}{R}\right)\exp(-R \Delta t)\exp(R_{cx} \Delta t) + R \Delta t \exp(-R \Delta t) \exp(R_{cx} \Delta t)\\
&= R \Delta t (1+ R_{cx} \Delta t) \exp(-R_i \Delta t).
\end{split}
\end{equation}

Again using property~\eqref{eq:AppendixC:incompleteGammaProperty} and the definition~\eqref{eq:AppendixC:defIncompleteGamma} of the incomplete Gamma function, we can write the second part of the expression as follows:
\begin{equation}
\begin{split}
&\sum_{\kappa=1}^\infty \kappa^2 \cdot \sum_{l = \kappa+1}^\infty \left(\frac{R_{cx}}{R} \right)^{\kappa-1} \left(1 - \frac{R_{cx}}{R} \right) \cdot \frac{(R\Delta t)^l \exp(-R\Delta t)}{l!}\\
&= \sum_{\kappa=1}^\infty \kappa^2 \left(\frac{R_{cx}}{R}\right)^{\kappa-1} \left( 1 - \frac{R_{cx}}{R}\right) \left( 1 - \frac{\Gamma(\kappa+1, R \Delta t)}{\kappa!} \right)\\
&= \frac{R(2R - R_i)}{R_i^2} - \sum_{\kappa=1}^\infty \kappa^2 \left(\frac{R_{cx}}{R}\right)^{\kappa-1} \left( 1 - \frac{R_{cx}}{R}\right) \frac{\Gamma(\kappa+1, R \Delta t)}{\kappa!}\\
&= \frac{R(2R - R_i)}{R_i^2}\\ 
&- \left( 1 - \frac{R_{cx}}{R}\right) \int_{R \Delta t}^\infty \exp(-t) \left( t^2 \left(\frac{R_{cx}}{R} \right) \exp\left( \left(1-\frac{R_i}{R}\right) t \right) + t \exp\left( \left(1-\frac{R_i}{R}\right) t \right) \right)dt\\
&= \frac{R(2R - R_i)}{R_i^2}\\ 
&- \frac{R_i}{R} \left(\frac{R^3 \exp(-R_i \Delta t) \left( (R_i \Delta t)^2 + 2 R_i \Delta t + 2 \right)}{R_i^3} \left( 1 - \frac{R_i}{R} \right) + \frac{R^2 \exp(-R_i \Delta t) (R_i \Delta t + 1 )}{R_i^2} \right).
\end{split}
\end{equation}
Combining the two parts, we finally obtain
\begin{equation}
\begin{split}
&\sum_{\kappa=1}^\infty \kappa^2 \cdot \mathcal{P}(K(n) = \kappa) = \frac{2R^2 - R_iR - (2R_i R_{cx} R \Delta t + 2 R^2 - R_i R) \exp(-R_i \Delta t)}{R_i^2}.
\end{split}
\end{equation}
Inserting this result in the expression for the variance~\eqref{eq:AppendixA:fullvariance}, we get the following result:
\begin{equation}
\begin{split}
\mathbb{V}[K(n)] &= \frac{2 R^2 - R_i R -(2R_i R_{cx} R \Delta t + 2 R^2 - R_i R) \exp(-R_i \Delta t)}{R_i^2} - \left(\frac{R}{R_i}(1-\exp(-R_i \Delta t)) \right)^2.
\end{split}
\end{equation}

\section{The law of total expectation and variance}
\label{app:total_exp_var}
The law of total expectation is as follows:
\begin{equation}
\langle x \rangle = \langle \langle x \mid y \rangle \rangle.
\end{equation}
The inner expectation is over the random variable $x$ conditioned on $y$. The outer expectation is over the random variable $y$. For example, if $x = \sum_{l=1}^L z_l$, with a random number of terms $L$, and $y=L$, then we have
\begin{equation}
\begin{split}
\langle \sum_{l=1}^L z_l \rangle &= \langle \langle \sum_{l=1}^L z_l \mid L = \lambda \rangle \rangle\\
&= \sum_{\lambda=0}^\infty \langle \sum_{l=1}^L z_l \mid L = \lambda \rangle \cdot \mathcal{P}(L=\lambda)\\
&= \sum_{\lambda=0}^\infty E(L=\lambda) \cdot \mathcal{P}(L=\lambda)\\
&= E(E(L=\lambda)),
\end{split}
\label{eq:law_of_total_expectation}
\end{equation}
where we define $E(L)$ as the expected value of the sum $\sum_{l=1}^L z_l$ for a fixed value of $L$. This expression is used, e.g., in~\eqref{eq:expected_value_with_random_k_collision} and~\eqref{eq:analog_estimators_expectedvalue_doublesum}.

The law of total variance is as follows:
\begin{equation}
\mathbb{V}[x] = \langle \mathbb{V}[x \mid y ] \rangle + \mathbb{V}[ \langle x \mid y \rangle ].
\end{equation}
In the first term, the variance is over $x$ conditioned on $y$ and the expected value is over $y$. In the second term, the expected value is over $x$ conditioned on $y$ and the variance is over $y$. Considering again the example with $x = \sum_{l=1}^L z_l$ and $y=L$, where for a fixed value of $L$ we have $\mathbb{V}[\sum_{l=1}^L z_l] = \mathcal{V}(L)$. We find
\begin{equation}
\begin{split}
\mathbb{V}[\sum_{l=1}^L z_l] &= \langle \mathbb{V}[\sum_{l=1}^L z_l \mid L=\lambda ] \rangle + \mathbb{V}[ \langle \sum_{l=1}^L z_l \mid L=\lambda \rangle ]\\
&= \langle \mathcal{V}(L=\lambda) \rangle + \mathbb{V}[ E(L=\lambda) ]\\
&= \sum_{\lambda=0}^\infty \mathcal{V}(L=\lambda) \cdot \mathcal{P}(L=\lambda) + \sum_{\lambda=0}^\infty (E(L=\lambda)-E(E(L=\lambda)))^2 \cdot \mathcal{P}(L=\lambda),
\end{split}
\end{equation}
where we defined $E(E(L=\lambda))$ in~\eqref{eq:law_of_total_expectation}. This expression is used, e.g., in~\eqref{eq:general_variance_with_random_k_collision} and~\eqref{eq:analog_estimators_variance_doublesum}.

\section{The incomplete Gamma function}
\label{app:gamma}
The Gamma function is defined as
\begin{equation}
\Gamma(z) = \int_0^\infty t^{z-1} \exp(-t) dt,
\end{equation}
with $\text{Re}(z) > 0$. The incomplete Gamma function is defined as
\begin{equation}
\Gamma(z,x) = \int_x^\infty t^{z-1} \exp(-t) dt,
\label{eq:AppendixC:defIncompleteGamma}
\end{equation}
with $\text{Re}(z) > 0$ and $x > 0$.

For the Poisson distributions used throughout this text, we are interested in the (incomplete) Gamma function where $z$ represents a number of collisions, i.e., an integer $z \in \mathbb{N}$. In that case, the Gamma function can be simplified using integration by parts as follows:
\begin{equation}
\begin{split}
\Gamma(z) &= \int_0^\infty t^{z-1} \exp(-t)dt\\
&= (z-1) \int_0^\infty t^{z-2} \exp(-t)dt\\
&= (z-1)(z-2) \int_0^\infty t^{z-3} \exp(-t)dt\\
& \vdots\\
&= (z-1)! \int_0^\infty \exp(-t)dt\\
&= (z-1)!.
\end{split}
\end{equation}
Analogously, we can simplify the incomplete Gamma function using integration by parts as follows:
\begin{equation}
\begin{split}
\Gamma(z,x) &= \int_x^\infty t^{z-1} \exp(-t) dt\\
&= x^{z-1} \exp(-x) + (z-1) \int_x^\infty t^{z-2} \exp(-t)dt\\
&= x^{z-1} \exp(-x) + (z-1) x^{z-2} \exp(-x) + (z-1)(z-2) \int_x^\infty t^{z-3} \exp(-t)dt\\
& \vdots\\
&=\exp(-x) (z-1)! \sum_{k=1}^{z-1} \left(\frac{x^k }{k!}\right) + (z-1)! \int_x^\infty \exp(-t)dt\\
&=\exp(-x) (z-1)! \sum_{k=1}^{z-1} \left(\frac{x^k }{k!}\right) + (z-1)! \exp(-x),\\
&=\exp(-x) (z-1)! \left(\sum_{k=1}^{z-1} \left(\frac{x^k }{k!}\right) + 1 \right),\\
\end{split}
\end{equation}
from which it follows that
\begin{equation}
\sum_{k=1}^{z-1} \left(\frac{x^k}{k!}\right) = \frac{\exp(x) \Gamma(z,x)}{(z-1)!} -1 = \frac{\exp(x) \Gamma(z,x)}{\Gamma(z)} -1
\end{equation}
and 
\begin{equation}
\sum_{k=0}^{z-1} \left(\frac{x^k}{k!}\right) = \frac{\exp(x) \Gamma(z,x)}{(z-1)!} = \frac{\exp(x) \Gamma(z,x)}{\Gamma(z)}.
\label{eq:AppendixC:incompleteGammaProperty}
\end{equation}
These expressions turn out to be useful when doing calculations with Poisson distributions where we can then replace sums by Gamma functions (integrals) that can be solved analytically, see, e.g., Appendix~\ref{app:meanvark}.

%\newpage
\bibliographystyle{abbrv}
\bibliography{Bib_20231005}

\end{document}